\documentclass[a4paper,10pt]{article}

\usepackage{color}
\usepackage{graphicx}
\usepackage{amsmath}
\usepackage{graphics}
\usepackage{amsthm}
\usepackage{amsfonts}
\usepackage{url}
\usepackage{multirow}
\usepackage{rotating}
\usepackage{wrapfig}
\usepackage{authblk}
\usepackage{url} 
\usepackage{pifont}
\usepackage{array}
\usepackage{subfig}
\usepackage{float}
\usepackage[section] {placeins}
\usepackage{array}
\usepackage[margin=0.75in]{geometry}

\usepackage{amssymb}
\usepackage{fancyhdr}

\newcolumntype{@}{>{\global\let\currentrowstyle\relax}}
\newcolumntype{^}{>{\currentrowstyle}}
\newcommand{\rowstyle}[1]{\gdef\currentrowstyle{#1}%
  #1\ignorespaces
}

\newcommand{\tick}{\ding{51}}
\newcommand{\cross}{\ding{55}}

\newtheorem{definition}{Definition}

\newcommand{\Keywords}[1]{\par\noindent
{\small{\em Keywords\/}: #1}}

\newcommand{\LV}{Low Voltage }
\newcommand{\MV}{Medium Voltage }
\newcommand{\HV}{High Voltage }
\newcommand{\CPL}{Characteristic Path Length }

\newcommand{\PG}{Power Grid }

\newcommand{\etal}{\textit{et al.}}
\newcommand{\G}{Grid }
\newcommand{\DG}{Distribution Grid }
\newcommand{\Gs}{Grids }
\newcommand{\SG}{Smart Grid }
\newcommand{\sw}{small-world }
\newcommand{\MLV}{Medium and Low Voltage }
\newcommand{\CNA}{Complex Network Analysis }

\newcommand{\cpl}{characteristic path length }
\newcommand{\cc}{clustering coefficient }
\newcommand{\apl}{average path length }
\newcommand{\Pl}{Power-law }

\title{The Complex Evolution of the Power Grid}
\title{Turning a Grid into a Denser and Smarter One}
\title{From the Grid to the Smart Grid, Topologically}

\author{Giuliano~Andrea~Pagani} 
\author{Marco~Aiello}

\affil{Distributed Systems Group\\Johann Bernoulli Institute for Mathematics and Computer Science
\\University of Groningen\\ Groningen, The Netherlands\\
\vspace{0.3cm}
email: \protect\url{{g.a.pagani,m.aiello}@rug.nl}\\
\protect\url{http://www.cs.rug.nl/ds/}
}

\begin{document}

\maketitle

\begin{abstract}  
  The \SG is not just about the digitalization of the Power Grid. In its more visionary acceptation, it is a model of energy management in which the users are engaged in producing energy as well as consuming it, while having information systems fully aware of the energy demand-response of the network and of dynamically varying prices. A natural question is then: to make the \SG a reality will the \DG have to be updated? We assume a positive answer to the question and we consider the lower layers of Medium and Low Voltage to be the most affected by the change. In our previous work, we have analyzed samples of the Dutch \DG~\cite{PaganiAielloTSG2011} and we have considered possible evolutions of these using synthetic topologies modeled after studies of complex systems in other technological domains~\cite{PaganiEvol2013}. In this paper, we take an extra important further step by defining a methodology for evolving any existing physical Power Grid to a good \SG model thus laying the foundations for a decision support system for utilities and governmental organizations. In doing so, we consider several possible evolution strategies and apply then to the Dutch Distribution Grid. We show how more connectivity is beneficial in realizing more efficient and reliable networks. Our proposal is topological in nature, and enhanced with economic considerations of the costs of such evolutions in terms of cabling expenses and economic benefits of evolving the Grid.
\end{abstract}


\section{Introduction}

The \PG traces its roots in the late 18$^{th}$ century. Then it involved large production facilities and a hierarchical \G composed of High, Medium and Low Voltage to transport and distribute the power to the end user. Such infrastructure and its principles are still valid today. As a sector it has traditionally being a (state-owned) monopoly. Nowadays there is a strong trend for innovation driven by the need of making the infrastructure more reliable, open and to accommodate for new technology, both in term of renewable energy generation and digital infrastructure. The technological innovation and the political pressure for a new Power Grid often go under the name of {\em Smart Grid}, which though has no unique definition~\cite{smartGrid09}.  One of the prominent aspects that the \SG will enable is the participation of small producers in the energy market. It is already possible today, for an end user to produce energy with small-scale production units such as solar panels, small wind turbines, and micro-CHP units and sell it back to the unique energy distributor. But more can be achieved if the producers would have access to an open energy market where they can auction their over-production and buy in a truly competitive market the energy for their needs. Local energy production and distribution will change the traditional way power systems have been considered so far: the \LV layer of the \G will go from being a passive end receiving energy from the upper layers, to an active segment of multi-directional energy flows thanks for the pervasive distribution of sustainable energy generators. But this trend will inevitably affect the physical structure of the \LV Grid, also topologically. The reason for this is that the cost of distribution will be of primary importance in enabling or repressing the local market of energy. We also consider that the design of the infrastructure will need to take into account the new paradigm of local energy energy generation and distribution.

\begin{figure}[htbp]
 %\begin{minipage}[htbp]{cm}
   \centering   
  \includegraphics[width=1\textwidth]{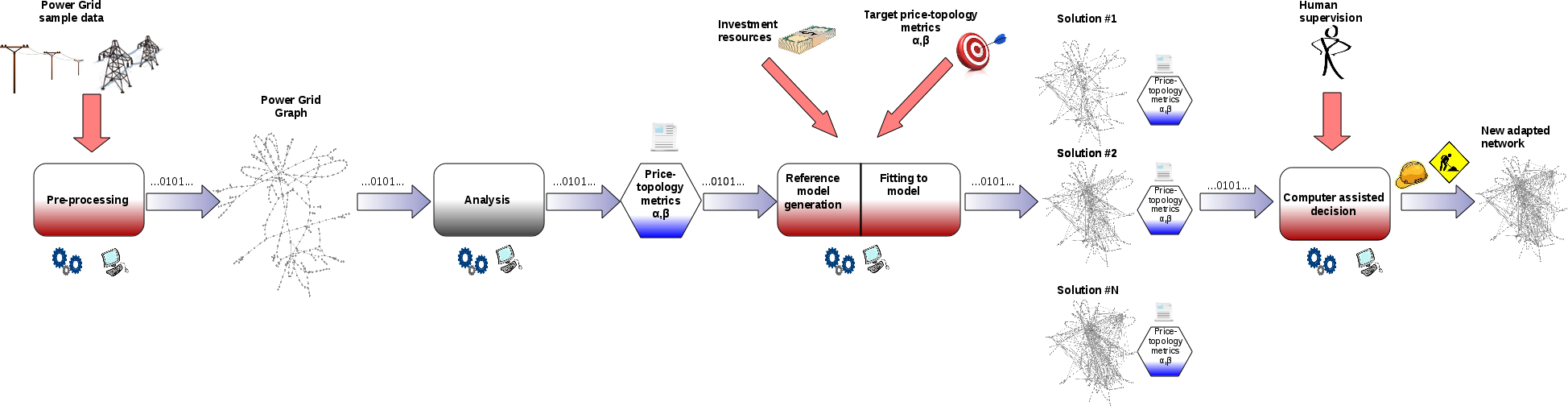}
   \caption{Engineering process for \MLV \G optimized for prosumer-based energy exchange.}
\label{fig:engProc}
\end{figure}

Assuming the need for an enhancement of the physical Low Voltage Grid, our aim is to realize a decision support system to guide the distribution operators, policy makers, and utilities to evaluate scenarios of network improvement and to realize distribution \Gs that are more efficient and that facilitate (from an economical point of view) the delocalized distribution of energy. We propose a process to analyze, design, and adapt distribution networks based on statistical models of the Power Grid as a weighted network.  A visual representation of the process we propose is presented in Figure~\ref{fig:engProc}.
In the figure, several phases and input are considered to plan the evolution of the infrastructure where a local energy exchange is the guiding goal. It starts with a pre-processing phase where the input data of the Grid is converted into a graph; the output of this initial phase is a \PG graph. The following phase consists of the analysis of the topological properties characterizing the graph. The output of this phase consists of a set of values representing the metrics related to the \PG that influence the price of electricity ($\alpha$ and $\beta$ metrics in the figure). The process continues with the generation of a network model. The number of nodes and edges of this reference model are provided according to the targets for the cost-related parameters ($\alpha$ and $\beta$) and the will to invest of the stakeholders. Based on the theoretical model identified, the physical network under assessment is then fitted to a topological structure similar to the one of the model. Several solutions are provided that differ in the topology and the $\alpha$ and $\beta$ metrics. All these solutions are then input to a computer assisted decision support system that presents the benefits/costs of the evolution of the network; an expert is involved in the selection of the evolution to be implemented among the most promising candidates built by the computer. Once the decision is made, the adaptation of the physical Grid can begin.

We covered the initial steps of the process in our previous works. In~\cite{PaganiAielloTSG2011}, we performed topological analysis of the Dutch \MLV Power Grid. We developed a set of metrics ($\alpha$ and $\beta$ metrics in the figure) based on weighted topological properties to assess the influence on the cost of electricity distribution for a set of real Dutch \MLV Grids. In the following~\cite{PaganiEvol2013}, we have evaluated known (reference) models of technological networks (such as the Web, the Internet, etc.) to evaluate how they would perform for local energy exchange.  We found that an increase in connectivity from the typical current value of average degree of two to higher values such as four and higher is beneficial in improving the efficiency and reliability of the network. We also found that the \sw model coming from the social sciences with average degree $\langle k \rangle\approx4$ provides the right compromise between performance improvement and the thrift in the realization of this connectivity (i.e., cost for cabling). Although the generation from scratch of a network topology is interesting from a modeling and theoretical perspective, it is not common to design a \DG from scratch.  In this paper, we take the a practical step in evaluating how to evolve the current \DG to a more interconnected network taking into account the existing network and the physical constraints. We keep the basis of our statistical approach to weighted network evolution and apply it to Dutch samples of the \MLV networks. We not only assess the benefits in terms of pure topology, but also the costs for cabling the evolved networks, and the benefit in terms of costs of electricity distribution. The are several novelties with this approach. First, we use statistical graph models as a tool for designing networks and not just for analyzing the existing; second, we focus on the \DG rather than the \HV \G that is usually the target of the analysis in the literature; third, we provide a decision support system that works in the large for the network and not for individual subcomponents of the Grid; fourth, we provide a tool that can help electricity Distribution companies in the design and expansion of \PG networks suited for local-scale energy exchange.

The rest of the paper is organized as follows. We open by analyzing the motivations for a new energy landscape and the required changes to the current Grid, Section~\ref{sec:motivations}. The basic Graph Theory background is presented in Section~\ref{sec:graphBasics}. Section~\ref{sec:evolMec} describes the evolution strategies followed in upgrading the samples of the Distribution Grid.  The analysis of the results is presented in Section~\ref{sec:results}, while an overall discussion comparing the evolution strategies is provided in Section~\ref{sec:discussion}.  Section~\ref{sec:economics} takes into account benefits and costs of evolution of the Dutch Grid samples.  Section~\ref{sec:relWks} reviews the main approaches to Electrical \G and System design and evolution, while Section~\ref{sec:conclusion} provides a conclusion of the paper. \ref{sec:metrics} is included to provide extra details about the topological metrics that are assessed in the evolution process.

\section{The Need for a New Grid}\label{sec:motivations}

From the executive summary of  {\em Grid 2030---A National Vision for Electricity’s Second 100 Years:'} ``America’s electric system, “the supreme engineering achievement of the 20th century,'' is aging, inefficient, and congested, and incapable of meeting the future energy needs''; ``Unprecedented levels of risk and uncertainty about future conditions in the electric industry have raised concerns about the ability of the system to meet future needs''; ``There are several promising technologies on the horizon that could
help modernize and expand the Nation's electric delivery system, relieve transmission congestion, and address other problems in system planning and operations''; ``The revolution in information technologies that has transformed other ``network'' industries in America (e.g., telecommunications) has yet to transform the electric power business''; and finally ``It is becoming increasingly difficult to site new conventional overhead transmission lines, particularly in urban and suburban areas experiencing the greatest load growth''~\cite{us03}. Here are all the ingredients that already ten years ago set the stage for a \G that needed to evolve and become smarter. The quoted comments refers mainly to the \HV \PG, but the same conclusion can be drawn for the \LV Grid. Actually, the \HV Grid is already a quite ``Smart'' with SCADA and EMS systems dealing with the control and communication of the system. On the other hand, the Distribution infrastructure, that is usually neglected in the big picture, is the layer of the \G that has less ICT technology in place so far~\cite{Wissner2011}. A key aspect is that the \MLV \G will be the ones that will have to accommodate the renewable sources coming from the distributed generation paradigm~\cite{Lopes2007}. We therefore consider that the same problems and challenges that were envisioned 10 years ago for the Transmission \G now need to be considered for the Distribution infrastructure.

Such requirement for the evolution in the Distribution \G is driven by two inter-related aspects: on the one hand, the pressure for unbundling the energy sector, and, on the other hand, the availability and affordability of small-scale renewable-based energy production units. Unbundling proposes to get rid of the monopoly system that has dominated the energy world so far by providing the possibility of competition in the energy market for production, transmission, distribution, and retail. The underlying aim of unbundling is the one of providing better services and tariffs for the end-user and promoting innovation and new investment in a traditionally sluggish (by definition) sector of the economy dominated by the demand following mindset~\cite{cossent09}. At the extreme of the unbundling process is the idea where potentially everybody can produce energy and participate as a seller on a free energy market~\cite{Vaithe05}. This last aspect is the linking point with renewable energy production: nowadays photovoltaic panels, small-wind turbines are affordable for everybody and often incentivized by governments' policies~\cite{friedman08}. The step is small to envision a future Grid where everybody can sell the surplus of energy not used at home in a market where it is traded as a commodity by software agents embedded in the future generation of Smart Meters.  In such a context, with many small-scale producers and still without an efficient and cheap energy storage technology, a local energy exchange at the neighborhood or municipal level between end-users is foreseeable and desirable. Micro-grids increased performance in terms of reduced losses and power quality have been successfully tested~\cite{lasseter11,barri10}, but little attention has been devoted to the network topology of these type of Grids.

A future with plenty of prosumers that produce energy and sell or share it at the level of neighborhoods will affect the Distribution Grid. The change from a passive-only \G to a Smart Grid~\cite{brown08} will require to rethink the role of the \MLV Grid and the design principles and techniques that have guided its development so far.  In our study of considering how to evolve or adapt the current Distribution Grid to a Smart Grid, we resort to Complex Network Analysis not only to analyze the existing, but also to drive the design of the next generation Grid. {\em \CNA (CNA)} is a branch of Graph Theory taking its root in the early studies of Erd{\H{o}}s and R{\'{e}}nyi~\cite{Erdos1959} on random graphs and considering statistical structural properties of evolving very large graphs.  Taking its root in the past, \CNA is a relatively young field of research. The first systematic studies appeared in the late 1990s~\cite{Watts98,Strogatz2001, Barabasi1999,Albert2000} having the goal of looking at the properties of large networks with a complex systems behavior. Afterwards, \CNA has been used in many diverse fields of knowledge, from biology~\cite{Jeong2000} to chemistry~\cite{Doye2002}, from linguistics to social sciences~\cite{Milgram69}, from telephone call patterns~\cite{Aiello2000} to computer networks~\cite{Faloutsos1999} and Web~\cite{albert99,Donato2007} to virus spreading~\cite{kephart91,colizza07,Gautreau2008} to logistics~\cite{Latora2002,Guimer2004,Colizza2006} and also inter-banking systems~\cite{Nationalbank2003}. Men-made infrastructures are especially interesting to study under the \CNA lenses, especially when they are large scale and grow in a decentralized and independent fashion, thus not being the result of a global, but rather of many local autonomous designs. The \PG is a prominent example.  In this work we consider a novel approach both in considering \CNA tools as a design instrument (i.e., CNA-related metrics are used in finding the most suited \MLV Grid for local energy exchange) and in focusing on the \MLV layers of the Power Grid. In fact, traditionally, \CNA studies applied to the \PG only evaluate reliability issues and disruption behavior of the \G when nodes or edges of the \HV layer are compromised.

\section{Graph Theory Background}\label{sec:graphBasics}

The approach used in this work to model the Power Grid and its evolution is based on Graph Theory and Complex Networks. Here we recall the basic definitions that we use throughout the paper and refer to standard textbooks such as~\cite{Bollobas79,Bollobas1998} for a broader introduction. First, we define a graph for the Power Grid~\cite{pag:preprint1102}.
\begin{definition}[Power Grid graph] A
{\em Power Grid graph} is a graph $G(V,E)$ such that each element $v_i \in V$ is either a substation, transformer, or consuming unit of a physical Power Grid. There is an edge $e_{i,j}=(v_i,v_j) \in E$ between two nodes if there is physical cable connecting directly the elements represented by $v_i$ and $v_j$.  \end{definition}
Next, we associate weights to the edges representing physical cable properties (e.g., resistance, voltage, supported current flow).
\begin{definition}[Weighted Power Grid graph]\label{def:wpgg}
A {\em Weighted Power Grid graph} is a Power Grid graph $G_w(V,E)$ with an additional function $f:E \to \mathbb{R}$ associating a real number to an edge representing the physical property of the corresponding cable (e.g., the resistance, expressed in Ohm, of the physical cable).
\end{definition}
A first classification of graphs is expressed in terms of their size. 
\begin{definition}[Order and size of a  graph]
Given the graph $G$ the order is given by $N=|V|$, while the
size is given by $M=|E|$.
\end{definition}
From \emph{order} and \emph{size} it is possible to have a global value for the connectivity of the vertexes of the graph, known as {\em average node degree} . That is $<\!k\!>=\frac{2M}{N}$. To characterize the relationship between a node and the others it is connected to, the following properties provide an indication of the bond between them.
\begin{definition}[Adjacency, neighborhood and degree]
  If $e_{x,y} \in E$ is an edge in graph $G$, then $x$ and $y$ are {\em adjacent,} or {\em neighboring,} vertexes, and the vertexes $x$ and $y$ are \emph{incident} with the edge $e_{x,y}$.  The set of vertexes adjacent to a vertex $x \in V$, called the \emph{neighborhood} of $x$, is denoted by $\Gamma_x$. The number $d(x)=|\Gamma_x|$ is the \emph{degree} of $x$.  \end{definition}

A measure of the average `density' of the graph is given by the
clustering coefficient, characterizing the extent to which vertexes
adjacent to any vertex $v$ are adjacent to each other.
\begin{definition}[Clustering coefficient (CC)]
The {\em clustering coefficient} $\gamma_v$ of $\Gamma_v$ is
\[
\gamma_v=\frac{|E(\Gamma_v)|}{\binom{k_v}{2}} 
\]
where $|E(\Gamma_v)|$ is the number of edges in the neighborhood of
$v$ and $\binom{k_v}{2}$ is the total number of $possible$ edges in
$\Gamma_v$.
\end{definition}
This local property of a node can be extended to an entire graph by
averaging over all nodes.

Another important property is how much any two nodes are far apart from each other, in particular the minimal distance between them or shortest path.
The concepts of \emph{path} and \emph{path length} are crucial to understand the way two vertexes are connected.
\begin{definition}[Path and  path length]
A \emph{path} of G is a subgraph $P$ of the form:
\[V(P)=\{x_0,x_1,\ldots,x_l\}, \hspace{10mm}
  E(P)=\{(x_0,x_1),(x_1,x_2),\ldots,(x_{l-1},x_l)\}.
\]
such that $V(P)\subseteq V\textrm{ and }E(P)\subseteq E$. The vertexes
$x_0$ and $x_l$ are \emph{end-vertexes} of $P$ and $l=|E(P)|$ is the
\emph{length} of $P$. A graph is {\em connected} if for any two
distinct vertexes $v_i,v_j\in V$ there is a finite path from $v_i$ to $v_j$.\\
\end{definition}

\begin{definition}[Distance]
Given a graph $G$ and vertexes $v_i$ and $v_j$, their \emph{distance} $d(v_i,v_j)$ is
the minimal length of any $v_i-v_j$ path in the graph. If there is no
$v_i-v_j$ path then it is conventionally  set to $d(v_i,v_j)=\infty$.\\
\end{definition}
\begin{definition}[Shortest path]
Given a graph $G$ and vertexes $v_i$ and $v_j$ the {\em shortest path} is
the the path corresponding to the minimum of to the set $\{|P_1|,|P_2|,\ldots,|P_k|\}$ containing the
lengths of all paths for which $v_i$ and $v_j$ are the end-vertexes.
\end{definition}
A global measure for a graph is given by its average distance among
any two nodes.
\begin{definition}[Average path length (APL)]
Let $v_i \in V$ be a vertex in graph $G$. The \emph{average path length} for $G$ $L_{av}$ is:
\[
 L_{av} = \frac{1}{N \cdot (N-1)} \sum_{i\neq j}d(v_i,v_j)
\]
where $d(v_i,v_j)$ is the finite distance between $v_i$ and $v_j$ and $N$ is the order of $G$.
\end{definition}
\begin{definition}[Characteristic path length (CPL)]\label{def:cpl}
%\begin{cpl}
Let $v_i \in V$ be a vertex in graph $G$, the \emph{characteristic path length} for $G$, $L_{cp}$ is defined as the median of ${d_{v_i}}$ where:
\[
 d_{v_i} = \frac{1}{(N-1)} \sum_{i\neq j}d(v_i,v_j)
\]
is the mean of the distances connecting $v_i$ to any other vertex $v_j$ in $G$ and $N$ is the order of $G$.\\
%\end{cpl}
\end{definition}
To describe the importance of a node with respect to minimal paths in the graph, the concept of betweenness helps. Betweenness (sometimes also referred as {\em load}) for a given vertex is the number of shortest paths between any other nodes that traverse it.
\begin{definition}[Betweenness]
 The {\em betweenness} $b(v)$ of vertex $v \in V$ is
\[
 b(v)=\sum_{v\neq s \neq t}\frac{\sigma_{st}(v)}{\sigma_{st}}
\]
where $\sigma_{st}(v)$ is 1 if the shortest path between vertex s and vertex t goes through vertex v, 0 otherwise and $\sigma_{st}$ is the number of shortest paths between vertex s and vertex t.
\end{definition}

Looking at large graphs, one is usually interested in global statistical measures rather than the properties of a specific node. A typical example is the node degree, where one measures the node degree probability distribution.
\begin{definition}[Node degree distribution]\label{def:ndd}
Consider the degree $k$ of a node in a graph as a random variable. The
function 
\[
N_k=\{v\in G:\: d(v)=k\}
\]
is called {\em probability node degree distribution}.
\end{definition}
The shape of the distribution is a salient characteristic of the network. For the Power Grid, the shape is typically either exponential or a Power-law~\cite{Barabasi1999,Amaral2000,PaganiAielloTSG2011,casals07}.
More precisely, an exponential node degree ($k$) distribution has a fast decay in the probability of having nodes with relative high node degree. The relation:
\[
 P(k)=\alpha e^{\beta k}
\]
follows, where $\alpha$ and $\beta$ are parameters of the specific network considered. On the contrary, a \Pl distribution has a slower decay with higher probability of having nodes with high node degree. It is expressed by the relation:
\[
 P(k)=\alpha k^{-\gamma}
\]
where $\alpha$ and $\gamma$ are parameters of the specific network considered. We remark that the graphs considered in the \PG domain are usually large, although finite, in terms of \textit{order} and \textit{size} thus providing limited and finite probability distributions.

A Graph can also be represented as a matrix, typically an adjacency matrix.  
\begin{definition}[Adjacency matrix] 
The adjacency matrix $A=A(G)=(a_{i,j})$ of a graph $G$ of order $N$ is the $N \times N$ matrix given by \[
 a_{ij} =
  \begin{cases}
   1      & \text{if } (v_i,v_j) \in E, \\
   0       & \text{otherwise.}
  \end{cases}
\]
\end{definition} 
We have now provided the basic definitions needed to present the modeling tools for the \PG evolutions.

\section{Evolving the Current \PG}\label{sec:evolMec}

We start by taking samples of \LV and \MV from the Dutch Power Grid. Tables~\ref{tab:uLow} and~\ref{tab:uMed} summarize the main facts of these samples (described in greater detail in~\cite{PaganiAielloTSG2011}). The first column of each table represents the identifier of the sample, column two and three provide the \textit{order} and \textit{size} of the sample; these two values are used to compute the average degree that is shown in the fourth column. Column five and six give an impression on the the effort to reach one node from any other in the network through the average path length and the characteristic path length. A measure of local clustering is given in column seven with the clustering coefficient metric. One notes a low average node degree $\langle k \rangle\approx2$ both for the \LV networks and the \MV networks. Besides the difference in the \textit{order} and \textit{size} between the two types of networks (generally the \MV network samples are bigger), one sees that the \LV samples have a mostly a null clustering coefficient, while the \MV networks present a small, but at least significant, value. This difference is explained in the different topology and purpose of the networks: a radial structure with no clustering to distribute electricity to the end-users (Low Voltage) and a more meshed structure for the \MV that shows small clustering values.

\begin{table*}[htb]
\begin{center}
\begin{small}
    \begin{tabular}{| p{0.4cm} || p{0.7cm} | p{0.5cm}  | p{0.9cm} | p{0.9cm}  | p{0.9cm} |p{1.0cm} || p{1.0cm}| p{1.0cm} |p{1.0cm} |}
    \hline
& \multicolumn{6}{|c||}{\sc{Present study}} &  \multicolumn{3}{|c|}{\sc{Random Graph}}\\
\hline
    ID & Order & Size & Avg. $d$ & APL & CPL & $\gamma$ &  APL &  CPL &  $\gamma$\\ \hline\hline
    1 & 17 & 18 & 2.118 & 3.398 & 3.313 & 0.00000 & 1.427 & 1.688 & 0.13726\\ \hline
    2 & 15 & 16 & 2.133 & 3.086	& 3.000&0.00000 & 2.319 & 2.358 & 0.00000\\ \hline
    3 & 24 & 23 & 2.087 & 4.499 & 4.228&0.00000 & 3.127 & 3.091 & 0.05508\\ \hline
    4 & 30 & 29 & 1.933 & 4.545 & 4.449&0.00000& 1.860 & 2.242 & 0.05778\\ \hline
    5 & 188 & 191 & 2.032 & 17.726 & 17.878&0.00000& 3.846 & 4.345 & 0.00532\\ \hline
    6 & 10 & 9 & 1.800 & 2.423 & 2.223&0.00000& 0.978 & 1.167 & 0.26667\\ \hline
    7 & 63 & 62 & 1.968 & 5.204 & 5.404&0.00000& 2.514&	2.904&	0.03175\\ \hline	
    8 & 28 & 27 & 1.929 &	4.784 &	5.000&0.00000& 2.553&	2.945&	0.04762\\ \hline
    9 & 133 & 140 & 2.105 &	11.543&	11.366&0.01112& 3.702 & 4.172 & 0.01482\\ \hline
    10 & 124 & 138 & 2.226&	8.053&	7.070&0.00869& 3.010 & 3.540 & 0.02914\\ \hline
    11 & 31 & 30 & 1.935&	4.353&	4.357&0.00000& 1.590 & 1.969 & 0.07475\\ \hline
    \end{tabular}
    \end{small}
\end{center}

\caption{Low Voltage samples from the northern Netherlands \PG compared with Random graphs of the same size.\label{tab:uLow}}

\end{table*}

\begin{center}
\begin{table}[htbp]
\centering
\begin{footnotesize}
\begin{tabular}{|@p{1.7cm}|^p{0.9cm}|^p{0.6cm}|^p{1.5cm}|^p{1.5cm}|^p{1.2cm}|}
\hline
\rowstyle{\bfseries}
Network sample  & \textit{Order} & \textit{Size} & {Avg. betweenness} & {Avg. betw/order} & {Coeff. variation} \\ \hline
\hline
1   & 17 & 18 & 32.933 & 0.074 & 0.773 \\ \hline
2   & 15 & 16 & 25.231 & 0.053 & 0.887 \\ \hline
3   & 24 & 23 & 70.286 & 0.295 & 0.643 \\ \hline
4  & 30 & 29 & 81.167 & 0.309 & 1.153 \\ \hline
5   & 188 & 191 & 2928.227 & 13.494 & 1.207 \\ \hline
6    & 10 & 9 & 9.000 & 0.047 & 1.291 \\ \hline
7    & 63 & 62 & 255.016 & 0.288 & 2.091 \\ \hline
8   & 28 & 27 & 102.143 & 0.279 & 1.301 \\ \hline
9    & 133 & 140 & 1355.953 & 6.220 & 1.534 \\ \hline
10   & 124 & 138 & 771.911 & 3.840 & 1.351 \\ \hline
11   & 31 & 30 & 139.677 & 0.691 & 1.265 \\ \hline
\end{tabular}
\caption{Betweenness for Dutch \LV samples.}\label{tab:samplesBetLv}
\end{footnotesize}
\end{table}
\end{center}

\begin{table*}[htb]
\begin{center}
\begin{small}
    \begin{tabular}{| p{0.4cm} || p{0.7cm} | p{0.5cm}  | p{0.9cm} | p{0.9cm}  | p{0.9cm} |p{1.0cm} || p{1.0cm}| p{1.0cm} |p{1.0cm} |}
    \hline
& \multicolumn{6}{|c||}{\sc{Present study}} & \multicolumn{3}{|c|}{\sc{Random Graph}}\\
\hline
    ID & Order & Size & Avg. $d$ & APL & CPL & $\gamma$ &  APL &  CPL &  $\gamma$\\ \hline\hline
 
1 & 444	& 486 & 2.189 &	11.033	&10.858&0.00537& 5.547 & 6.163 & 0.00333\\ \hline
2 & 472	& 506 & 2.144 & 17.095 & 17.174&0.01360& 5.039 & 5.700 & 0.00106\\ \hline
3 & 238	&245		&2.059	&11.715	&11.580&0.00000& 3.558 & 4.234 & 0.00595\\ \hline
4 & 263	&288&		2.190	&12.775	&12.311&0.01118& 5.046 & 5.368 & 0.01080\\ \hline
5 & 217	&229&		2.111	&10.321	&10.241&0.00140& 4.894 & 5.391 & 0.00121\\ \hline
6 & 191 & 207 & 2.168&	9.288&	8.990&0.00296& 4.616 & 5.079 & 0.00225  \\ \hline
7 & 884 & 1059 & 2.396 & 9.817 & 9.527&0.00494& 5.440 & 6.010 & 0.00170\\ \hline
8 & 366	&382&		2.087	&15.113	&14.546&0.00000 & 4.691 & 5.249 & 0.00405\\ \hline
9 & 218	&232&		2.128	&10.850	&10.915&0.00000& 5.454 & 5.856 & 0.00539\\ \hline
10 & 201&	204		&2.030	&15.742	&15.257&0.00166& 4.898 & 5.503 & 0.00491\\ \hline
11 & 202&	213		&2.109	&13.504&	12.891&0.00140&4.801&	5.217&	0.08750\\ \hline
%12 & 25	&24	&	1.920	&5.781	&5.500&0.00000& 4.924&	5.084&	0.00000\\ \hline
12 & 464	&499		&2.151	&13.144	&12.703&0.00036& 4.718 & 5.390 & 0.00209\\ \hline
    \end{tabular}
    \end{small}
\end{center}

\caption{Medium Voltage samples from the northern Netherlands \PG compared with Random graphs of the same size.\label{tab:uMed}}
\end{table*}

\begin{center}
\begin{table}[htbp]
\centering
\begin{footnotesize}
\begin{tabular}{|@p{1.7cm}|^p{0.9cm}|^p{0.6cm}|^p{1.5cm}|^p{1.5cm}|^p{1.2cm}|}
\hline
\rowstyle{\bfseries}
Network sample  & \textit{Order} & \textit{Size} & {Avg. betweenness} & {Avg. betw/order} & {Coeff. variation} \\ \hline
\hline
1   & 444 & 486 & 4329.054 & 9.750 & 2.050 \\ \hline
2  & 472 & 506 & 5087.728 & 10.779 & 1.704 \\ \hline
3  & 238 & 245 & 1910.757 & 8.028 & 1.566 \\ \hline
4   & 263 & 288 & 1237.711 & 4.706 & 1.517 \\ \hline
5   & 217 & 229 & 3169.571 & 14.606 & 1.743 \\ \hline
6   & 191 & 207 & 3870.640 & 20.265 & 1.432 \\ \hline
7   & 884 & 1059 & 7755.542 & 8.773 & 2.875 \\ \hline
8   & 366 & 382 & 5136.520 & 14.034 & 1.691 \\ \hline
9   & 218 & 232 & 1244.663 & 5.709 & 1.544 \\ \hline
10  & 201 & 204 & 3613.691 & 17.979 & 1.173 \\ \hline
11   & 202 & 213 & 2690.183 & 13.318 & 1.331 \\ \hline
%12   & 25 & 24 & 114.720 & 4.589 & 0.954 \\ \hline
12   & 464 & 499 & 3424.602 & 7.381 & 1.687 \\ \hline
\end{tabular}
\caption{Betweenness for Dutch \MV samples.}\label{tab:samplesBetMv}
\end{footnotesize}
\end{table}
\end{center}

Next, we consider evolutions starting from the Dutch samples, that is, adding cables according to several strategies of network growth. We break the evolutions into four groups of edge growth: increments of 25\%, 50\%, 75\%, and 100\%. The choice of stopping at 100\% is performed based on the results of~\cite{PaganiEvol2013}, where it is shown that an average node degree of 4 has the right balance of improved network qualities and costs of network evolution. We consider several strategies for evolving the graph by adding more links, namely:
\begin{itemize}

\item \textbf{Assortativity.} A network is assortative if nodes having similar characteristics or properties are connected one another~\cite{newman02}. The property we consider is that of node degree and then take two strategies:
\begin{itemize}
\item \textbf{High degree} nodes are connected one another. The process starts with considering a set of nodes with the highest equal node degree and connect them together.  The process goes on considering the next set of nodes with equal high degree in the order of rank and so on.
\item \textbf{Low degree} nodes are connected one another. The process goes on as for the the high degree strategy, but node are linked starting from the couples with lowest degree.
\end{itemize}
\item \textbf{Dissortativity} is the opposite of assortativity, that is, a network is dissortative if nodes having different characteristics or properties are connected together. Following this strategy, nodes with highest node degree are linked to nodes with lowest node degree.
\item \textbf{Triangle closure} is based on the principle of increasing the clustering coefficient of the network. At each step, a node is selected at random and for each pair of its neighbors an edge is added between them, if not already present.
\item \textbf{Least distance} gives priority to the connection of nodes that are geographically closer to each other. This strategy can minimize the costs of cabling since such costs are directly proportional to the length of cables.
\item  \textbf{Random} is based on the random selection of nodes to attach edges. At each step of the growth process, a pair of distinct nodes are randomly selected and an edge between them is added.
\end{itemize}
For every strategy, if two nodes already have an edge that connects them the edge is not added and the evolution strategy continues. In fact, in the graph models we only allow a single edge between a pair of nodes, if not already present.

\section{From the Current Distribution Infrastructure to the \SG}\label{sec:results}

We adapt current physical networks according the strategies described in Section~\ref{sec:evolMec} and we analyze the obtained graphs according to a set of metrics that provide a view of efficiency of the whole network and its adequacy for local energy distribution.  Such metrics consider the path length properties of the graph, the presence of cliques at local scale (i.e., clustering coefficient), the presence of critical nodes that manage the majority of paths (i.e., betweenness). Robustness of the network to failures concerning its connectivity ($Rob_N$) is evaluated by computing the the \textit{order} of the maximal connected
component (MCC) of the graph when nodes (20\% of the initial \textit{order} of the graph) are removed randomly or in a targeted way by focusing on the nodes with higher degree first, then taking the average  between the two values to have an overall estimation of the disruption. The last metric we analyze deals with the redundancy of paths ($APL_{10^{th}}$); which takes into account the increase in the average path length when the 10$^{th}$ shortest path is computed, therefore this metric provides a measure of the additional effort required to benefit of alternative paths than the optimal one. For a more detailed dissertation over the metric analyzed we refer to~\ref{sec:metrics}.
To study the effects of the strategy, we implement the strategies described in a software based on the JAVA graph library JGraphT (\protect\url{http://www.jgrapht.org/}). 
The same library suite has been used to compute the metrics just described. The only metric computed with a different software library is the `betweennees' one. For this computation the Stanford Network Analysis Project (SNAP - {\protect\url{http://snap.stanford.edu/}}) software library has been used since it leverages on the the algorithm developed by Brandes~\cite{Brandes2001} with optimal performance.  To perform the generation and computation of the metrics we used a PC with Intel Core2 Quad CPU Q9400 2.66GHz with 4GB RAM. The Operating system is based on the Linux kernel 2.6.32 with a 4.4.3 GCC compiler and JAVA framework 1.6. The versions of JGraphT and SNAP software libraries used are respectively v10.10.01 and v0.8.1. Next we present the results of the generation and metrics evaluation.

\subsection{Evolution of \MV Distribution \Gs}

We start by considering the \MV Grid and apply the evolution strategies presented in the previous section one at the time.

\subsubsection*{Assortative high node degree evolution}

\begin{figure}
 \captionsetup{type=figure}
    \centering
    \subfloat[Random node removal robustness.]{\label{fig:rndHD}\includegraphics[scale=.4]{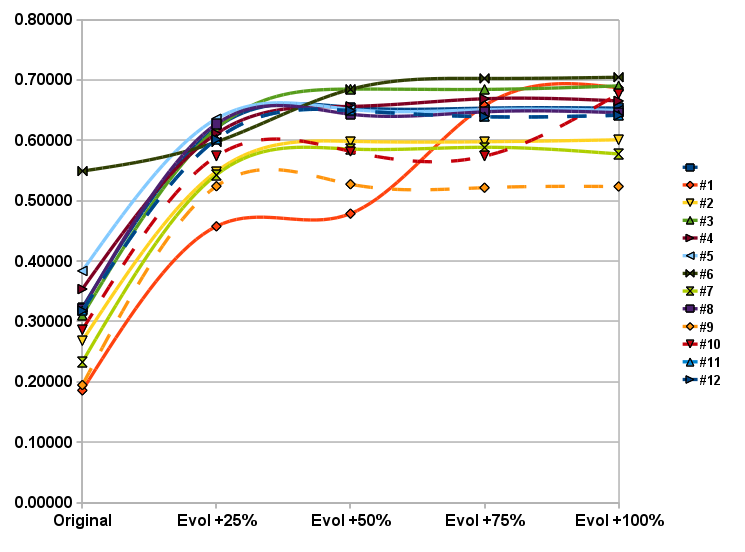}} 
    \subfloat[Targeted node removal robustness.]{\label{fig:tarHD}\includegraphics[scale=.4]{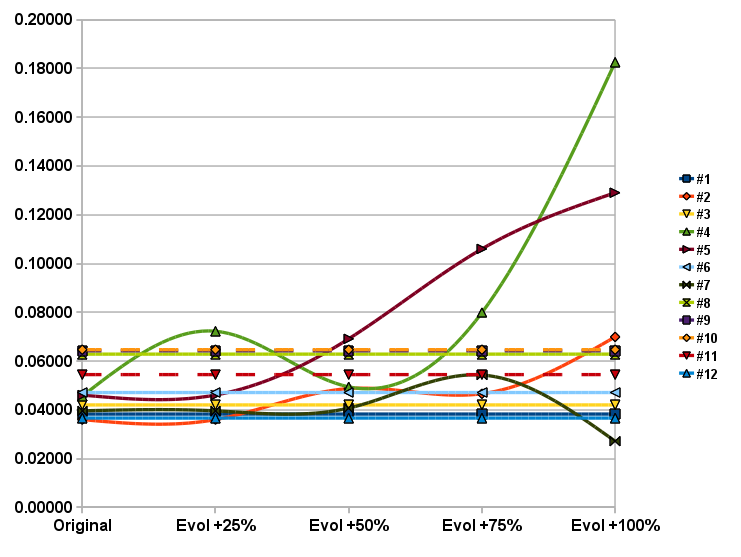}} %\\
    \caption{Evolution for robustness.}
    \label{fig:assHDRob}
\end{figure}

\begin{figure}[htbp]
 %\begin{minipage}[htbp]{7cm}
   \centering   
  \includegraphics[width=0.7\textwidth]{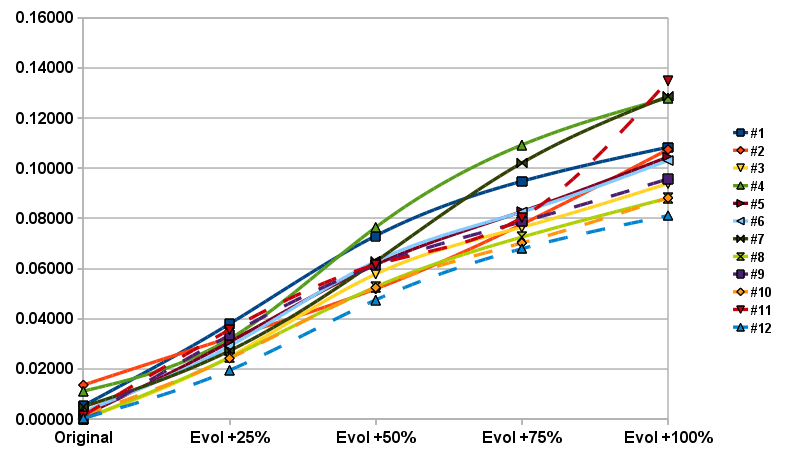}
   \caption{Evolution for the clustering coefficient.}
\label{fig:ccAssEvolMV}
\end{figure}

The results for evolution according of the assortative strategy is summarized in Table~\ref{tab:mvAss}. For each sample of the \MV \G (column one) the values for the main topological quantities: \textit{order} and \textit{size} in columns three and four, respectively; average node degree in the fifth column; the \cpl is reported in column six; the clustering coefficient follows in column seven; robustness is shown in eight column; the cost in term of redundant path length closes the data series (column nine). For the path length, we note that the increase in the connectivity is extremely beneficial. This is notable already with the addition of 25\% of links, where on average about 50\% of the path length is reduced. The improvement is then minor when more link are added, in fact, when the number of links is doubled the reduction of the \cpl is around 60\%, on average. The increment in the transition +75\%/+100\% of the number of links brings just a gain in the reduction of the \cpl of 3\%. This tendency of saturation in the reduction of the path length is explained by the small-world phenomenon that arises when a sufficient connectivity threshold is reached. The improvement for the clustering coefficient is significant, reaching values that are even two orders of magnitude higher. In general, the \cc has a linear improvement in the evolution steps considered, some of the samples show however a behavior that is similar to a logistic shape,\footnote{\protect\url{http://en.wikipedia.org/wiki/Logistic_curve}} as shown in Figure~\ref{fig:ccAssEvolMV}. The improvement in robustness tends to double in values. An exception is sample \#7 that had already a good initial value for robustness. The addition of edges according to this strategy is mainly beneficial in the random node attack case, while the fraction of the metric that considers targeted attacks against the most connected nodes is marginally affected. In fact, the ranking of the nodes is almost untouched reinforcing the node degree of those nodes that had already an high degree at the beginning of the evolution process.  As explained by Newman~\cite{newman02}, assortative networks that link high degree nodes have a sort of redundancy in the main cluster. The redundancy of new connections is not particularly helpful in improving the resilience of the network since the nodes with high connectivity already form a cluster structure. The addition of edges between the high degree nodes does not increase the number of new potential target nodes. This is empirically evident by the results of the robustness metric presented in Figures~\ref{fig:rndHD} and~\ref{fig:tarHD}, showing the evolution of the robustness metric in the random and targeted node removal situations, respectively. One sees that the increase in connectivity is beneficial to contrast random attacks and the first step of link addition is the most beneficial. Then the disruption of the network does not benefit from the additional connectivity anymore. The lack of benefit from the additional connectivity is emphasized in the resilience against targeted attacks (cf. Figure~\ref{fig:rndHD}) where almost all samples do not have improvements in their performance with just two exceptions.

Considering the values for the redundant path robustness, the same considerations on the \cpl apply. There is a reduction of more than 50\% already when the networks are evolved with a 25\% increase in the \textit{size}. An exception is sample \#2 that shows a consistent improvement also in the later stages of evolution, especially between step two and step three where the average redundant path length from about ten, declines to a value slightly higher than six.

Table~\ref{tab:mvAssBet} shows for each sample of the \MV \G (column one) the values of metrics related to betweenness. Columns three and four show \textit{order} and \textit{size}, then average betweenness is provided in column five, while a value of average betweenness normalized by the \textit{order} of the graph is presented in column six as it provides a normalized value. The statistical coefficient of variation which is shown in the seventh column.

The assortative strategy involving the nodes with highest node degree is beneficial in reducing the average betweenness of all the \MV samples. Already from the first step of the network evolution the betweenness on average reduces to almost 50\% of the original value. Some samples (i.e., samples \#5 and \#6) reach even higher reduction of up to 80\%. The same trend is followed by the normalized value of average betweenness divided by the \textit{order} of the network. Already in the first step no sample exceeds seven for the betweenness to order ratio. Both metrics improve of up to 60\% (on average) in the last step of the evolution. Considering the variability of betweenness in each of the samples we note a general increase. In particular, only four of the twelve samples show a decrease when the connectivity in the network is double of the original \textit{size} of the network. Such behavior is due to a slower decrease of the standard deviation of betweenness compared to its average. This strategy of adding the connection between the nodes that already have the highest connectivity, that are usually the nodes that also have highest betweenness, reduces the number of ``bottleneck'' nodes. However, the strategy of adding links is not helpful for substantially reducing the variability of betweenness. Evidence to this claim comes from the fact that the median of the betweenness in each of the samples at every state of the evolution process is zero, that is, the majority of nodes are terminal nodes that are not involved in any path between other nodes.

\begin{table}[htbp]
\begin{footnotesize}
\begin{center}
\begin{tabular}{|@p{1.7cm}|^p{1.7cm}|^p{0.9cm}|^p{0.6cm}|^p{1cm}|^p{1cm}|^p{1.2cm}|^p{1.5cm}|^p{2cm}|}
\hline
\rowstyle{\bfseries} Sample ID &Network type  & \textit{Order} & \textit{Size} & {Avg. deg.} & {CPL} & {CC} & {Removal robustness ($Rob_N$)} & {Redundancy cost ($APL_{10^{th}}$)} \\ \hline
\hline

\multicolumn{9}{|c|}{\textbf{original \textit{order} +25\%}} \\ \hline
1  & MV & 444 & 607 & 2.734 & 4.784 & 0.03797 & 0.330 & 7.871 \\ \hline
2 & MV & 472 & 632 & 2.678 & 9.203 & 0.03242 & 0.247 & 11.476 \\ \hline
3  & MV & 238 & 306 & 2.571 & 6.243 & 0.02486 & 0.295 & 9.549 \\ \hline
4  & MV & 263 & 360 & 2.738 & 5.973 & 0.03196 & 0.346 & 8.388 \\ \hline
5  & MV & 217 & 286 & 2.636 & 4.704 & 0.03072 & 0.329 & 7.855 \\ \hline
6  & MV & 191 & 258 & 2.702 & 4.889 & 0.02907 & 0.341 & 7.455 \\ \hline
7  & MV & 884 & 1323 & 2.993 & 7.312 & 0.02723 & 0.318 & 9.094 \\ \hline
8  & MV & 366 & 477 & 2.607 & 6.532 & 0.02451 & 0.303 & 9.120 \\ \hline
9  & MV & 218 & 290 & 2.661 & 4.806 & 0.03351 & 0.346 & 7.452 \\ \hline
10  & MV & 201 & 255 & 2.537 & 6.360 & 0.02425 & 0.294 & 8.943 \\ \hline
11  & MV & 202 & 266 & 2.634 & 5.669 & 0.03563 & 0.314 & 8.449 \\ \hline
%12  &MV & 25 & 30 & 2.400 & 3.500 & 0.13200 & 0.458 & 7.615 \\ \hline
12 & MV & 464 & 623 & 2.685 & 4.871 & 0.01950 & 0.319 & 7.884 \\ \hline \hline
\multicolumn{9}{|c|}{\textbf{original \textit{order} +50\%}} \\ \hline
1  & MV & 444 & 729 & 3.284 & 4.734 & 0.07307 & 0.346 & 6.305 \\ \hline
2 & MV & 472 & 759 & 3.216 & 8.304 & 0.05193 & 0.264 & 9.975 \\ \hline
3  & MV & 238 & 367 & 3.084 & 5.405 & 0.05780 & 0.320 & 6.995 \\ \hline
4  & MV & 263 & 432 & 3.285 & 4.676 & 0.07645 & 0.367 & 6.000 \\ \hline
5  & MV & 217 & 343 & 3.161 & 4.486 & 0.06157 & 0.363 & 6.228 \\ \hline
6  & MV & 191 & 310 & 3.246 & 4.805 & 0.06290 & 0.349 & 6.425 \\ \hline
7  & MV & 884 & 1588 & 3.593 & 4.496 & 0.06287 & 0.363 & 6.848 \\ \hline
8  & MV & 366 & 573 & 3.131 & 5.638 & 0.05272 & 0.324 & 7.972 \\ \hline
9  & MV & 218 & 348 & 3.193 & 4.749 & 0.06138 & 0.354 & 6.732 \\ \hline
10  & MV & 201 & 306 & 3.045 & 6.145 & 0.05243 & 0.296 & 8.195 \\ \hline
11  & MV & 202 & 319 & 3.158 & 5.502 & 0.06155 & 0.318 & 8.171 \\ \hline
%12  & MV & 25 & 36 & 2.880 & 2.792 & 0.15133 & 0.471 & 5.660 \\ \hline
12  & MV & 464 & 748 & 3.224 & 4.847 & 0.04752 & 0.343 & 6.728 \\ \hline \hline
\multicolumn{9}{|c|}{\textbf{original \textit{order} +75\%}} \\ \hline
1  & MV & 444 & 850 & 3.829 & 4.705 & 0.09481 & 0.346 & 6.267 \\ \hline
2  & MV & 472 & 885 & 3.750 & 4.327 & 0.07763 & 0.352 & 6.311 \\ \hline
3  & MV & 238 & 428 & 3.597 & 5.344 & 0.07648 & 0.320 & 6.980 \\ \hline
4  & MV & 263 & 504 & 3.833 & 4.622 & 0.10929 & 0.382 & 6.018 \\ \hline
5  & MV & 217 & 400 & 3.687 & 4.449 & 0.08276 & 0.387 & 5.833 \\ \hline
6  & MV & 191 & 362 & 3.791 & 4.532 & 0.08257 & 0.349 & 6.049 \\ \hline
7  & MV & 884 & 1853 & 4.192 & 4.384 & 0.10213 & 0.378 & 5.738 \\ \hline
8  & MV & 366 & 668 & 3.650 & 5.573 & 0.07256 & 0.326 & 7.627 \\ \hline
9  & MV & 218 & 406 & 3.725 & 4.712 & 0.07867 & 0.356 & 6.740 \\ \hline
10  & MV & 201 & 357 & 3.552 & 5.935 & 0.07022 & 0.293 & 8.019 \\ \hline
11  & MV & 202 & 372 & 3.683 & 5.435 & 0.08035 & 0.314 & 7.380 \\ \hline
%12  & MV & 25 & 42 & 3.360 & 2.583 & 0.27846 & 0.547 & 4.808 \\ \hline
12  & MV & 464 & 873 & 3.763 & 4.809 & 0.06792 & 0.338 & 6.641 \\ \hline \hline
\multicolumn{9}{|c|}{\textbf{original \textit{order} +100\%}} \\ \hline
1  & MV & 444 & 972 & 4.378 & 4.646 & 0.10837 & 0.346 & 6.195 \\ \hline
2  & MV & 472 & 1012 & 4.288 & 4.297 & 0.10749 & 0.378 & 5.684 \\ \hline
3  & MV & 238 & 490 & 4.118 & 5.135 & 0.09390 & 0.321 & 6.773 \\ \hline
4  & MV & 263 & 576 & 4.380 & 4.595 & 0.12805 & 0.436 & 5.476 \\ \hline
5  & MV & 217 & 458 & 4.221 & 4.347 & 0.10451 & 0.397 & 5.580 \\ \hline
6  & MV & 191 & 414 & 4.335 & 4.268 & 0.10310 & 0.348 & 5.947 \\ \hline
7  & MV & 884 & 2118 & 4.792 & 4.375 & 0.12869 & 0.366 & 5.410 \\ \hline
8  & MV & 366 & 764 & 4.175 & 5.538 & 0.08809 & 0.320 & 7.673 \\ \hline
9  & MV & 218 & 464 & 4.257 & 4.664 & 0.09572 & 0.355 & 6.227 \\ \hline
10  & MV & 201 & 408 & 4.060 & 5.610 & 0.08821 & 0.294 & 6.741 \\ \hline
11  & MV & 202 & 426 & 4.218 & 4.391 & 0.13484 & 0.366 & 6.249 \\ \hline
%12  & MV & 25 & 48 & 3.840 & 2.583 & 0.31899 & 0.502 & 4.487 \\ \hline
12  & MV & 464 & 998 & 4.302 & 4.798 & 0.08111 & 0.339 & 6.488 \\ \hline

\end{tabular}
\caption{Metrics for assortative high node degree strategy \MV samples evolution.}\label{tab:mvAss}
\end{center}
\end{footnotesize}
\end{table}

\begin{center}
\begin{table}[htbp]
\centering
\begin{footnotesize}
\begin{tabular}{|@p{1.7cm}|^p{1.5cm}|^p{0.8cm}|^p{0.8cm}|^p{1.5cm}|^p{1.5cm}|^p{1.2cm}|}
\hline
\rowstyle{\bfseries}
Sample  ID  &Network type &\textit{Order} & \textit{Size} & {Avg. betweenness} & {Avg. betw/order} & {Coeff. variation} \\ \hline
\hline
\multicolumn{7}{|c|}{\textbf{original \textit{order} +25\%}} \\ \hline
1 & MV & 444 & 607 & 1696.993 & 3.822 & 4.047 \\ \hline
2 & MV & 472 & 632 & 3290.807 & 6.972 & 1.651 \\ \hline
3 & MV & 238 & 306 & 1048.901 & 4.407 & 1.803 \\ \hline
4 & MV & 263 & 360 & 1236.901 & 4.703 & 1.909 \\ \hline
5 & MV & 217 & 286 & 857.71 & 3.953 & 2.937 \\ \hline
6 & MV & 191 & 258 & 750.095 & 3.927 & 2.181 \\ \hline
7 & MV & 884 & 1323 & 5826.819 & 6.591 & 2.067 \\ \hline
8 & MV & 366 & 477 & 1989.602 & 5.436 & 1.901 \\ \hline
9 & MV & 218 & 290 & 905.139 & 4.152 & 2.4 \\ \hline
10 & MV & 201 & 255 & 1144.505 & 5.694 & 1.524 \\ \hline
11 & MV & 202 & 266 & 1026.579 & 5.082 & 1.756 \\ \hline
%12 & MV & 25 & 30 & 63.6 & 2.544 & 1.107 \\ \hline
12 & MV & 464 & 623 & 2039.336 & 4.395 & 4.074 \\ \hline \hline
\multicolumn{7}{|c|}{\textbf{original \textit{order} +50\%}} \\ \hline

1 & MV & 444 & 729 & 1673.842 & 3.77 & 2.707 \\ \hline
2 & MV & 472 & 759 & 2827.063 & 5.99 & 1.707 \\ \hline
3 & MV & 238 & 367 & 886.658 & 3.725 & 1.67 \\ \hline
4 & MV & 263 & 432 & 886.347 & 3.37 & 2.294 \\ \hline
5 & MV & 217 & 343 & 789.065 & 3.636 & 2.236 \\ \hline
6 & MV & 191 & 310 & 732.148 & 3.833 & 1.643 \\ \hline
7 & MV & 884 & 1588 & 3243.961 & 3.67 & 6.472 \\ \hline
8 & MV & 366 & 573 & 1748.53 & 4.777 & 1.85 \\ \hline
9 & MV & 218 & 348 & 884 & 4.055 & 1.817 \\ \hline
10 & MV & 201 & 306 & 1097.948 & 5.462 & 1.221 \\ \hline
11 & MV & 202 & 319 & 965.401 & 4.779 & 1.724 \\ \hline
%12 & MV & 25 & 36 & 47.12 & 1.885 & 1.475 \\ \hline
12 & MV & 464 & 748 & 2020.101 & 4.354 & 2.556 \\ \hline
\hline
\multicolumn{7}{|c|}{\textbf{original \textit{order} +75\%}} \\ \hline
1 & MV & 444 & 850 & 1659.788 & 3.738 & 2.195 \\ \hline
2 & MV & 472 & 885 & 1577.847 & 3.343 & 4.268 \\ \hline
3 & MV & 238 & 428 & 867.261 & 3.644 & 1.499 \\ \hline
4 & MV & 263 & 504 & 870.669 & 3.311 & 1.825 \\ \hline
5 & MV & 217 & 400 & 775.355 & 3.573 & 1.782 \\ \hline
6 & MV & 191 & 362 & 688.18 & 3.603 & 1.624 \\ \hline
7 & MV & 884 & 1853 & 3111.479 & 3.52 & 4.838 \\ \hline
8 & MV & 366 & 668 & 1734.622 & 4.739 & 1.674 \\ \hline
9 & MV & 218 & 406 & 869.273 & 3.987 & 1.571 \\ \hline
10 & MV & 201 & 357 & 1061.907 & 5.283 & 1.135 \\ \hline
11 & MV & 202 & 372 & 940.802 & 4.657 & 1.683 \\ \hline
%12 & MV & 25 & 42 & 41.92 & 1.677 & 1.706 \\ \hline
12 & MV & 464 & 873 & 1998.422 & 4.307 & 2.111 \\ \hline
\hline
\multicolumn{7}{|c|}{\textbf{original \textit{order} +100\%}} \\ \hline
1 & MV & 444 & 972 & 1639.753 & 3.693 & 1.964 \\ \hline
2 & MV & 472 & 1012 & 1544.805 & 3.273 & 3.358 \\ \hline
3 & MV & 238 & 490 & 827.586 & 3.477 & 1.618 \\ \hline
4 & MV & 263 & 576 & 858.372 & 3.264 & 1.625 \\ \hline
5 & MV & 217 & 458 & 754.514 & 3.477 & 1.615 \\ \hline
6 & MV & 191 & 414 & 657.291 & 3.441 & 1.592 \\ \hline
7 & MV & 884 & 2118 & 3099.314 & 3.506 & 3.815 \\ \hline
8 & MV & 366 & 764 & 1719.62 & 4.698 & 1.606 \\ \hline
9 & MV & 218 & 464 & 857.522 & 3.934 & 1.458 \\ \hline
10 & MV & 201 & 408 & 1015.278 & 5.051 & 1.11 \\ \hline
11 & MV & 202 & 426 & 935.381 & 4.631 & 1.679 \\ \hline
%12 & MV & 25 & 48 & 39.68 & 1.587 & 1.44 \\ \hline
12 & MV & 464 & 998 & 1989.684 & 4.288 & 1.875 \\ \hline

\end{tabular}
\caption{Betweenness for assortative high node degree strategy \MV samples evolution.}\label{tab:mvAssBet}
\end{footnotesize}
\end{table}
\end{center}

\subsubsection*{Assortative low node degree evolution}

Table~\ref{tab:mvAssLD} contains for each sample of the \MV \G (column one) the values for main topological quantities: \textit{order} and \textit{size} in columns three and four, respectively; average node degree in the fifth column; the \cpl is reported in column six; the clustering coefficient follows in column seven; robustness is shown in eight column; and the cost in term of redundant path length closes the data series (column nine).

\begin{figure}
 \captionsetup{type=figure}
    \centering
    \subfloat[Random node removal robustness.]{\label{fig:rndld}\includegraphics[scale=.4]{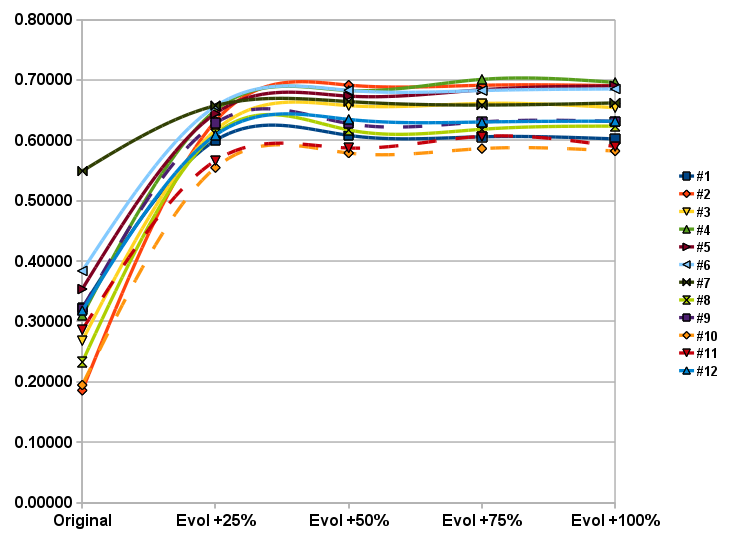}} 
    \subfloat[Targeted node removal robustness.]{\label{fig:tarld}\includegraphics[scale=.4]{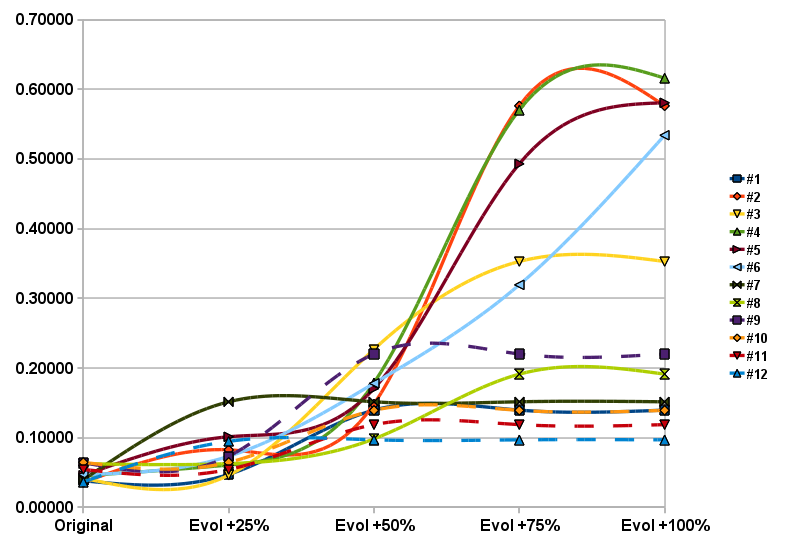}} %\\
    \caption{Evolution of robustness sub-metrics.}
    \label{fig:assldRob}
\end{figure}

\begin{figure}[htbp]
 %\begin{minipage}[htbp]{7cm}
   \centering   
  \includegraphics[width=0.7\textwidth]{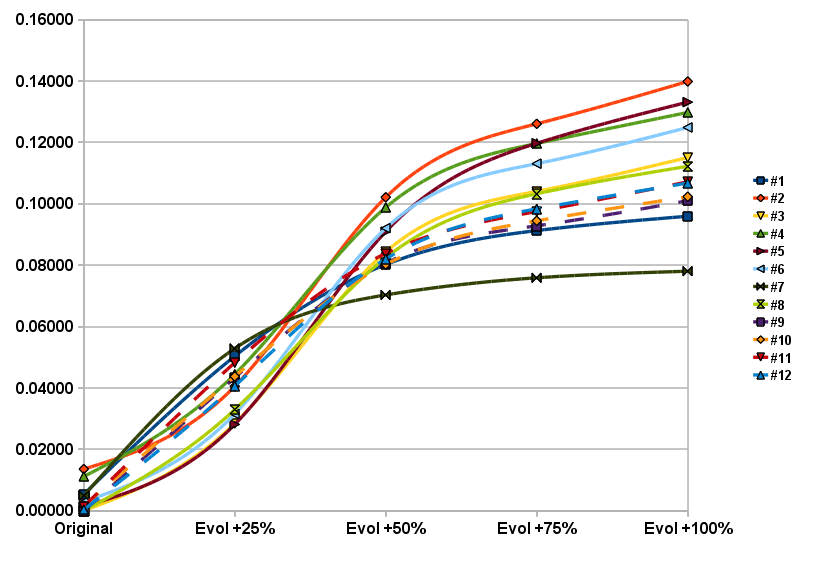}
   \caption{Evolution of the \cc metric.}
\label{fig:ccAssLDEvolMV}
\end{figure}

The increase in the connectivity with the assortative low node degree strategy is beneficial for the path length already with the addition of 25\% of links reducing about 45\% the initial value of the original graph. The improvement is lower when more and more links are added. In fact, when the number of link is doubled compared to the original \textit{size}, the reduction of the \cpl is around 47\% on average. In particular, the improvement comparing all the evolution steps is around 0.9, which shows a limit in the benefits achievable by the increased connectivity. This effect is explained by the small-world phenomenon that arises when a sufficient connectivity threshold is obtained. Once this threshold is obtained, the subsequent addition of edges following the same strategy have a reduced effect on the improvement of the property. Following this evolution strategy, we see that a saturation effect arises after the first two steps of the growth process since the \cpl then has no more significant improvement. The improvement for the \cc increases substantially compared to the initial values. It reaches values that are even three order of magnitude higher than the initial situation. In general, the improvement in the \cc tends to have a logistic trend in the evolution step considered, as shown in Figure~\ref{fig:ccAssLDEvolMV}. The improvement in robustness is in general three times higher compared to the initial value for the same samples. Three out of the twelve samples have values for robustness higher than 0.6. Even in this evolution scenario, sample \#7 poses an exception whose improvement are limited compared to the other samples. The addition of edges according to this strategy is beneficial in contrasting the effects of both random attacks and targeted attacks against the most connected nodes. This evolution strategy tends to change the hierarchy (in terms of node degree) of the nodes, adding more connections between the nodes that are less connected. Therefore, these nodes assume more importance in terms of node degree compared to the initial situation, giving more homogeneity in the degree. We see in this strategy how the new connectivity is particularly beneficial in contrasting targeted attacks. The new established connections tend to bond nodes with low degree (that anyway become small hubs of the network) thus improving the resilience of the network. This is empirically evident by the results of the robustness metric (Figures~\ref{fig:rndld} and~\ref{fig:tarld}) that show the evolution of the robustness metric in the random and targeted node removal situations,  respectively. One sees that the increase in connectivity is beneficial to contrast random attacks that reach a plateau around 0.6-0.7 after the second evolution step and more beneficial against the attacks that target high degree nodes: the majority of the samples experiences an improvement in the metric about one order of magnitude.

Considering the values for the redundant path robustness the same considerations done for the \cpl apply. There is already a reduction of 45\% compared to the initial value of the samples already when the networks are evolved with a 25\% increase in the \textit{size} of the original graph.

Table~\ref{tab:mvAssBetLD} shows for each sample of the \MV \G (column one) the values of metrics related to betweenness. In addition to \textit{order} and \textit{size} (columns three and four), average betweenness is provided in column five, while a value of average betweenness normalized by the \textit{order} of the graph is computed in column six. A measure of the statistical variation of betweenness is the coefficient of variation which is shown in the seventh column.

Considering betweenness, the assortative strategy involving the nodes with lowest node degree is beneficial in reducing the average betweenness of all the \MV samples. Already from the first step of the network evolution the betweenness on average reduces more than 38\% of the original value. However samples \#4 and \#9 have in the first step a slight increment in the average betweenness. The same reduction trend is followed by the normalized value of average betweenness divided by the \textit{order} of the network; already in the first step no sample exceeds 7.5 for the betweenness to order ratio. The benefits for both these metrics improve up to 41\% (on average) in the last step of the evolution. Considering the variability of betweenness in each of the samples, we note a general increase for this metric in the first two steps of the evolution, while the tendency is inverted in the last two steps. In particular, only sample \#8 shows a slight increase when the connectivity in the network is double the original \textit{size} of the network. Such behavior is due to an initial increase in the standard deviation of betweenness when only few edges are added, then after the second step of the evolution the tendency is inverted and the standard deviation decreases. This strategy of adding the connection, in fact, provides more connections between the nodes that have small connectivity. These are usually the nodes at the periphery of the network and do not have paths traversing them. A proof is that the median of the betweenness for each sample in the first two stages of evolution is zero, then in the later stages some samples (four out of fourteen) have a non-zero median that is a sign that all nodes are more evenly involved in the paths between other nodes.

\begin{table}[htbp]
\begin{footnotesize}
\begin{center}
\begin{tabular}{|@p{1.7cm}|^p{1.7cm}|^p{0.9cm}|^p{0.6cm}|^p{1cm}|^p{1cm}|^p{1.2cm}|^p{1.5cm}|^p{2cm}|}
\hline
\rowstyle{\bfseries} Sample ID &Network type  & \textit{Order} & \textit{Size} & {Avg. deg.} & {CPL} & {CC} & {Removal robustness ($Rob_N$)} & {Redundancy cost ($APL_{10^{th}}$)} \\ \hline
\hline

\multicolumn{9}{|c|}{\textbf{original \textit{order} +25\%}} \\ \hline
1 & MV & 444 & 607 & 2.734 & 8.037 & 0.05039 & 0.324 & 10.170 \\ \hline
2 & MV & 472 & 632 & 2.678 & 6.146 & 0.04055 & 0.357 & 9.194 \\ \hline
3 & MV & 238 & 306 & 2.571 & 6.171 & 0.02847 & 0.331 & 9.713 \\ \hline
4 & MV & 263 & 360 & 2.738 & 6.115 & 0.04448 & 0.355 & 8.514 \\ \hline
5 & MV & 217 & 286 & 2.636 & 5.667 & 0.02814 & 0.373 & 8.608 \\ \hline
6 & MV & 191 & 258 & 2.702 & 5.379 & 0.03156 & 0.365 & 8.331 \\ \hline
7 & MV & 884 & 1323 & 2.993 & 7.439 & 0.05290 & 0.404 & 8.827 \\ \hline
8 & MV & 366 & 477 & 2.607 & 7.130 & 0.03298 & 0.337 & 10.249 \\ \hline
9 & MV & 218 & 290 & 2.661 & 6.661 & 0.04338 & 0.351 & 10.134 \\ \hline
10 & MV & 201 & 255 & 2.537 & 6.650 & 0.04379 & 0.310 & 9.466 \\ \hline
11 & MV & 202 & 266 & 2.634 & 6.774 & 0.04825 & 0.310 & 9.734 \\ \hline
%12 & MV & 25 & 30 & 2.400 & 3.833 & 0.09714 & 0.487 & 10.814 \\ \hline
12 & MV & 464 & 623 & 2.685 & 7.254 & 0.04071 & 0.352 & 9.779 \\ \hline \hline
\multicolumn{9}{|c|}{\textbf{original \textit{order} +50\%}} \\ \hline
1 & MV & 444 & 729 & 3.284 & 7.860 & 0.08023 & 0.374 & 9.941 \\ \hline
2 & MV & 472 & 759 & 3.216 & 6.142 & 0.10215 & 0.420 & 7.791 \\ \hline
3 & MV & 238 & 367 & 3.084 & 6.105 & 0.08438 & 0.442 & 8.317 \\ \hline
4 & MV & 263 & 432 & 3.285 & 6.103 & 0.09884 & 0.431 & 7.649 \\ \hline
5 & MV & 217 & 343 & 3.161 & 5.620 & 0.09096 & 0.422 & 7.236 \\ \hline
6 & MV & 191 & 310 & 3.246 & 5.332 & 0.09208 & 0.430 & 7.121 \\ \hline
7 & MV & 884 & 1588 & 3.593 & 7.376 & 0.07033 & 0.408 & 8.576 \\ \hline
8 & MV & 366 & 573 & 3.131 & 7.068 & 0.08246 & 0.358 & 9.747 \\ \hline
9 & MV & 218 & 348 & 3.193 & 6.569 & 0.08068 & 0.424 & 10.086 \\ \hline
10 & MV & 201 & 306 & 3.045 & 6.565 & 0.08028 & 0.359 & 8.689 \\ \hline
11 & MV & 202 & 319 & 3.158 & 6.714 & 0.08388 & 0.353 & 8.946 \\ \hline
%12 & MV & 25 & 36 & 2.880 & 3.500 & 0.13667 & 0.514 & 6.750 \\ \hline
12 & MV & 464 & 748 & 3.224 & 7.211 & 0.08205 & 0.366 & 9.542 \\ \hline \hline
\multicolumn{9}{|c|}{\textbf{original \textit{order} +75\%}} \\ \hline

1 & MV & 444 & 850 & 3.829 & 7.795 & 0.09130 & 0.373 & 10.521 \\ \hline
2 & MV & 472 & 885 & 3.750 & 6.125 & 0.12608 & 0.634 & 7.455 \\ \hline
3 & MV & 238 & 428 & 3.597 & 6.046 & 0.10408 & 0.507 & 8.178 \\ \hline
4 & MV & 263 & 504 & 3.833 & 6.069 & 0.11968 & 0.636 & 7.619 \\ \hline
5 & MV & 217 & 400 & 3.687 & 5.593 & 0.11974 & 0.588 & 6.776 \\ \hline
6 & MV & 191 & 362 & 3.791 & 5.279 & 0.11311 & 0.501 & 7.030 \\ \hline
7 & MV & 884 & 1853 & 4.192 & 7.328 & 0.07592 & 0.405 & 8.261 \\ \hline
8 & MV & 366 & 668 & 3.650 & 7.049 & 0.10329 & 0.405 & 9.180 \\ \hline
9 & MV & 218 & 406 & 3.725 & 6.498 & 0.09285 & 0.425 & 8.571 \\ \hline
10 & MV & 201 & 357 & 3.552 & 6.430 & 0.09449 & 0.363 & 8.551 \\ \hline
11 & MV & 202 & 372 & 3.683 & 6.654 & 0.09751 & 0.363 & 8.521 \\ \hline
%12 & MV & 25 & 42 & 3.360 & 3.333 & 0.18619 & 0.708 & 5.897 \\ \hline
12 & MV & 464 & 873 & 3.763 & 7.195 & 0.09838 & 0.364 & 9.375 \\ \hline \hline
\multicolumn{9}{|c|}{\textbf{original \textit{order} +100\%}} \\ \hline

1 & MV & 444 & 972 & 4.378 & 7.761 & 0.09597 & 0.371 & 9.426 \\ \hline
2 & MV & 472 & 1012 & 4.288 & 6.113 & 0.13987 & 0.633 & 6.909 \\ \hline
3 & MV & 238 & 490 & 4.118 & 5.977 & 0.11510 & 0.503 & 7.432 \\ \hline
4 & MV & 263 & 576 & 4.380 & 5.966 & 0.12978 & 0.656 & 7.346 \\ \hline
5 & MV & 217 & 458 & 4.221 & 5.560 & 0.13315 & 0.635 & 7.478 \\ \hline
6 & MV & 191 & 414 & 4.335 & 5.200 & 0.12489 & 0.609 & 7.037 \\ \hline
7 & MV & 884 & 2118 & 4.792 & 7.258 & 0.07806 & 0.407 & 8.759 \\ \hline
8 & MV & 366 & 764 & 4.175 & 6.758 & 0.11224 & 0.407 & 8.902 \\ \hline
9 & MV & 218 & 464 & 4.257 & 6.394 & 0.10102 & 0.426 & 8.791 \\ \hline
10 & MV & 201 & 408 & 4.060 & 6.340 & 0.10233 & 0.361 & 8.074 \\ \hline
11 & MV & 202 & 426 & 4.218 & 6.604 & 0.10718 & 0.354 & 9.461 \\ \hline
%12 & MV & 25 & 48 & 3.840 & 3.125 & 0.23067 & 0.721 & 5.769 \\ \hline
12 & MV & 464 & 998 & 4.302 & 7.188 & 0.10677 & 0.364 & 8.800 \\ \hline
\end{tabular}
\caption{Metrics for assortative low node degree strategy \MV samples evolution.}\label{tab:mvAssLD}
\end{center}
\end{footnotesize}
\end{table}

\begin{center}
\begin{table}[htbp]
\centering
\begin{footnotesize}
\begin{tabular}{|@p{1.7cm}|^p{1.5cm}|^p{0.8cm}|^p{0.8cm}|^p{1.5cm}|^p{1.5cm}|^p{1.2cm}|}
\hline
\rowstyle{\bfseries}
Sample  ID  &Network type &\textit{Order} & \textit{Size} & {Avg. betweenness} & {Avg. betw/order} & {Coeff. variation} \\ \hline
\hline
\multicolumn{7}{|c|}{\textbf{original \textit{order} +25\%}} \\ \hline
1 & MV & 444 & 607 & 3329.321 & 7.498 & 1.943 \\ \hline
2 & MV & 472 & 632 & 2506.855 & 5.311 & 3.427 \\ \hline
3 & MV & 238 & 306 & 1184.601 & 4.977 & 2.264 \\ \hline
4 & MV & 263 & 360 & 1321.642 & 5.025 & 2.091 \\ \hline
5 & MV & 217 & 286 & 1053.607 & 4.855 & 2.147 \\ \hline
6 & MV & 191 & 258 & 887.095 & 4.644 & 2.001 \\ \hline
7 & MV & 884 & 1323 & 5840.286 & 6.607 & 2.351 \\ \hline
8 & MV & 366 & 477 & 2393.787 & 6.54 & 2.241 \\ \hline
9 & MV & 218 & 290 & 1251.694 & 5.742 & 1.708 \\ \hline
10 & MV & 201 & 255 & 1212.602 & 6.033 & 1.527 \\ \hline
11 & MV & 202 & 266 & 1232.495 & 6.101 & 1.63 \\ \hline
%12 & MV & 25 & 30 & 67.12 & 2.685 & 0.941 \\ \hline
12 & MV & 464 & 623 & 3134.5 & 6.755 & 2.236 \\ \hline \hline
\multicolumn{7}{|c|}{\textbf{original \textit{order} +50\%}} \\ \hline

1 & MV & 444 & 729 & 3260.014 & 7.342 & 1.992 \\ \hline
2 & MV & 472 & 759 & 2501.373 & 5.3 & 2.294 \\ \hline
3 & MV & 238 & 367 & 1165.202 & 4.896 & 1.785 \\ \hline
4 & MV & 263 & 432 & 1314.276 & 4.997 & 1.556 \\ \hline
5 & MV & 217 & 343 & 1044.607 & 4.814 & 1.555 \\ \hline
6 & MV & 191 & 310 & 874.358 & 4.578 & 1.604 \\ \hline
7 & MV & 884 & 1588 & 5781.723 & 6.54 & 2.091 \\ \hline
8 & MV & 366 & 573 & 2362.72 & 6.456 & 1.936 \\ \hline
9 & MV & 218 & 348 & 1228.775 & 5.637 & 1.43 \\ \hline
10 & MV & 201 & 306 & 1189.092 & 5.916 & 1.239 \\ \hline
11 & MV & 202 & 319 & 1218.99 & 6.035 & 1.437 \\ \hline
%12 & MV & 25 & 36 & 60.24 & 2.41 & 0.695 \\ \hline
12 & MV & 464 & 748 & 3102.316 & 6.686 & 1.853 \\ \hline \hline
\multicolumn{7}{|c|}{\textbf{original \textit{order} +75\%}} \\ \hline

1 & MV & 444 & 850 & 3228.822 & 7.272 & 1.905 \\ \hline
2 & MV & 472 & 885 & 2489.82 & 5.275 & 1.978 \\ \hline
3 & MV & 238 & 428 & 1151.462 & 4.838 & 1.623 \\ \hline
4 & MV & 263 & 504 & 1301.374 & 4.948 & 1.316 \\ \hline
5 & MV & 217 & 400 & 1036.047 & 4.774 & 1.343 \\ \hline
6 & MV & 191 & 362 & 861.737 & 4.512 & 1.462 \\ \hline
7 & MV & 884 & 1853 & 5734.982 & 6.488 & 1.994 \\ \hline
8 & MV & 366 & 668 & 2350.599 & 6.422 & 1.821 \\ \hline
9 & MV & 218 & 406 & 1209.014 & 5.546 & 1.311 \\ \hline
10 & MV & 201 & 357 & 1164.48 & 5.793 & 1.13 \\ \hline
11 & MV & 202 & 372 & 1201.455 & 5.948 & 1.349 \\ \hline
%12 & MV & 25 & 42 & 56.32 & 2.253 & 0.613 \\ \hline
12 & MV & 464 & 873 & 3088.206 & 6.656 & 1.701 \\ \hline
\hline
\multicolumn{7}{|c|}{\textbf{original \textit{order} +100\%}} \\ \hline
1 & MV & 444 & 972 & 3213.273 & 7.237 & 1.864 \\ \hline
2 & MV & 472 & 1012 & 2481.614 & 5.258 & 1.746 \\ \hline
3 & MV & 238 & 490 & 1137.444 & 4.779 & 1.516 \\ \hline
4 & MV & 263 & 576 & 1283.073 & 4.879 & 1.207 \\ \hline
5 & MV & 217 & 458 & 1025.907 & 4.728 & 1.215 \\ \hline
6 & MV & 191 & 414 & 844.295 & 4.42 & 1.406 \\ \hline
7 & MV & 884 & 2118 & 5673.277 & 6.418 & 1.941 \\ \hline
8 & MV & 366 & 764 & 2240.144 & 6.121 & 1.985 \\ \hline
9 & MV & 218 & 464 & 1186.383 & 5.442 & 1.278 \\ \hline
10 & MV & 201 & 408 & 1140.622 & 5.675 & 1.072 \\ \hline
11 & MV & 202 & 426 & 1188.545 & 5.884 & 1.305 \\ \hline
%12 & MV & 25 & 48 & 53.84 & 2.154 & 0.582 \\ \hline
12 & MV & 464 & 998 & 3080.978 & 6.64 & 1.619 \\ \hline
\end{tabular}
\caption{Betweenness for assortative low node degree strategy \MV samples evolution.}\label{tab:mvAssBetLD}
\end{footnotesize}
\end{table}
\end{center}

\subsubsection*{Triangle closure evolution}

Table~\ref{tab:mvCC} contains for each sample of the \MV \G (column one) the values for the main topological quantities: \textit{order} and \textit{size} in columns three and four, respectively; average node degree in the fifth column; the \cpl is reported in column six; the clustering coefficient follows in column seven; robustness is shown in eight column; the cost in term of redundant path length closes the data series (column nine).

\begin{figure}[htbp]
 %\begin{minipage}[htbp]{7cm}
   \centering   
  \includegraphics[width=0.7\textwidth]{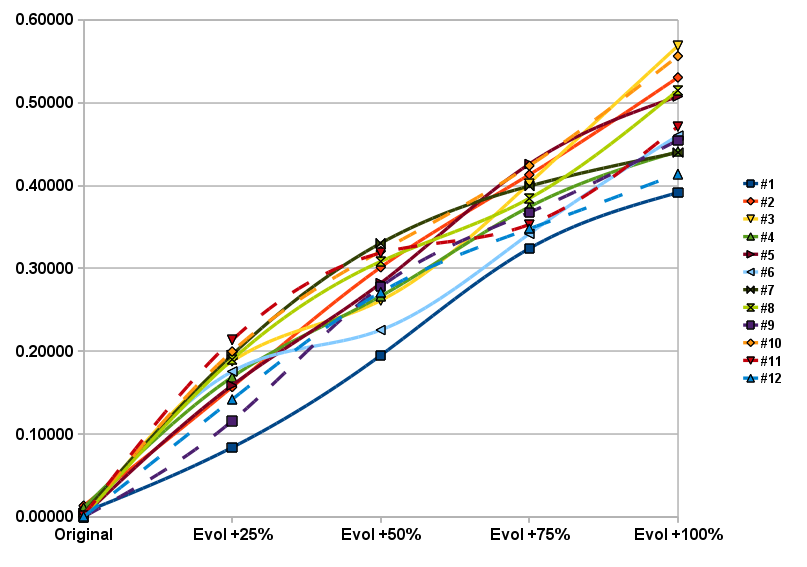}
   \caption{Evolution of the \cc metric.}
\label{fig:ccCCEvolMV}
\end{figure}

\begin{figure}
 \captionsetup{type=figure}
    \centering
    \subfloat[Random node removal robustness.]{\label{fig:rndCC}\includegraphics[scale=.4]{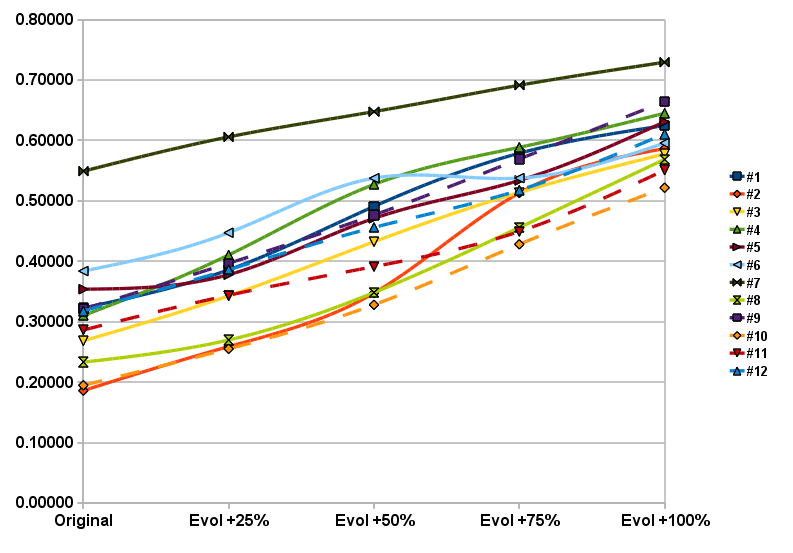}} \subfloat[Targeted node removal robustness.]{\label{fig:tarCC}\includegraphics[scale=.4]{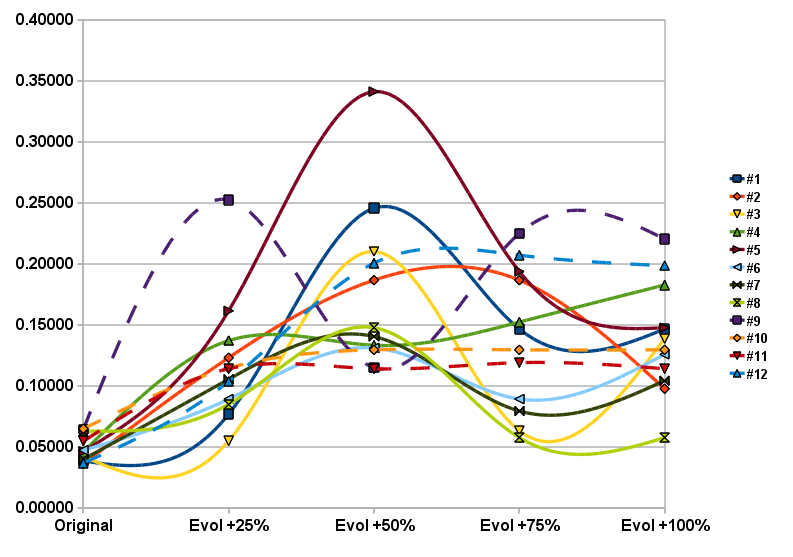}} %\\
    \caption{Evolution of robustness sub-metrics.}
    \label{fig:ccRob}
\end{figure}

The primary focus of the triangle closure strategy is to improve the \cc of the network and, as a side effect, the other metrics benefit from the additional `local' links (the links added connect the neighbors of a node). The increase in the connectivity with such a strategy is beneficial for the path length: the addition of 25\% of links reduces the path length around 12\% on average, the trend is slightly sub-linear and the final step of the evolution provides a shrinking of the \cpl measure to about 38\%. The best improvement is naturally obtained for the clustering coefficient. The average of the \cc reaches almost 0.5 when the edges of the networks are doubled compared to the original \textit{size}; for some samples the improvement is more than three orders of magnitude. The improvement in the \cc tend to have a logistic trend in the evolution steps considered (Figure~\ref{fig:ccCCEvolMV}). In general, robustness doubles when the \textit{size} of the graph doubles, in particular, one notices a sharp increase in this metric between the first and the second evolution step. In addition, it is interesting to note that for some samples the highest value of robustness is reached not when the connectivity is the highest, but in intermediate steps of evolution. For example, sample \#1 has better robustness in evolution step two than step three with values of 0.368 and 0.363; for sample \#2, the highest value is reached in step three of the evolution process. An aspect that needs to be highlighted is  the effect of this evolution on robustness against random and targeted attacks. The addition of edges to contrast the first type of attacks is always beneficial, increasing sub-linearly while more edges are added (Figure~\ref{fig:rndCC}); on the other hand, to contrast the second type of attack the maximum robustness is obtained in the second step of edges addition (Figure~\ref{fig:tarCC}). For the redundant path robustness, the same considerations done for the \cpl apply. There is a reduction of 28\%, compared to the initial value of the samples already when the networks are evolved with a 25\% increase in the \textit{size}. The maximal reduction in the redundant \apl is obtained when the edges are doubled with the average $10^{th}$ path that is about 55\% less than the initial value.

Table~\ref{tab:mvCCBet} contains for each sample of the \MV \G (column one) the values of metrics related to betweenness. In addition to \textit{order} and \textit{size} (columns three and four), average betweenness is provided in columns five, while a value of average betweenness normalized by the \textit{order} of the graph is computed in column six in order to compare the different samples. A measure of the statistical variation of betweenness is the coefficient of variation which is shown in the seventh column.

Considering betweenness, the triangle closure strategy involving the nodes with highest node degree is beneficial in reducing the average betweenness of all the \MV samples. The reduction in the average betweenness is around 45\% compared to the original value. In the first step of the network evolution, the betweenness on average reduces about 20\%. Some samples (i.e., samples \#5 and\#6) reach even higher reduction already in the first phase of the evolution (45\% and 66\% respectively). The same trend is followed by the normalized value of average betweenness divided by the \textit{order} of the network; already in the first step no sample exceeds seven for the betweenness to order ratio. Considering the variability of betweenness in each of the samples we note a general increase for this metric. All the twelve samples present an increase in the coefficient of variation that increments at each stage of the evolution. Such behavior is due to a slower decrease of the standard deviation of betweenness compared to its average. This strategy of adding the connection, by providing more connections between neighbors of a node, has less effect on the nodes at the edge of the network which have a marginal role in providing shortest path management. A proof is that the median of the betweenness in each of the samples at every state of the evolution process is zero, that is the majority of nodes are terminal nodes that are not involved in any path between other nodes.

\begin{table}[htbp]
\begin{footnotesize}
\begin{center}
\begin{tabular}{|@p{1.7cm}|^p{1.7cm}|^p{0.9cm}|^p{0.6cm}|^p{1cm}|^p{1cm}|^p{1.2cm}|^p{1.5cm}|^p{2cm}|}
\hline
\rowstyle{\bfseries} Sample ID &Network type  & \textit{Order} & \textit{Size} & {Avg. deg.} & {CPL} & {CC} & {Removal robustness ($Rob_N$)} & {Redundancy cost ($APL_{10^{th}}$)} \\ \hline
\hline

\multicolumn{9}{|c|}{\textbf{original \textit{order} +25\%}} \\ \hline
1 & MV & 444 & 607 & 2.734 & 11.178 & 0.08388 & 0.231 & 14.166 \\ \hline
2 & MV & 472 & 632 & 2.678 & 15.929 & 0.15681 & 0.191 & 20.221 \\ \hline
3 & MV & 238 & 306 & 2.571 & 11.768 & 0.18787 & 0.199 & 15.655 \\ \hline
4 & MV & 263 & 360 & 2.738 & 11.271 & 0.16888 & 0.274 & 14.735 \\ \hline
5 & MV & 217 & 286 & 2.636 & 9.093 & 0.15887 & 0.269 & 14.314 \\ \hline
6 & MV & 191 & 258 & 2.702 & 7.763 & 0.17561 & 0.268 & 10.813 \\ \hline
7 & MV & 884 & 1323 & 2.993 & 8.552 & 0.19566 & 0.355 & 11.051 \\ \hline
8 & MV & 366 & 477 & 2.607 & 13.453 & 0.18984 & 0.177 & 16.282 \\ \hline
9 & MV & 218 & 290 & 2.661 & 9.933 & 0.11586 & 0.324 & 19.088 \\ \hline
10 & MV & 201 & 255 & 2.537 & 13.400 & 0.19964 & 0.185 & 15.747 \\ \hline
11 & MV & 202 & 266 & 2.634 & 11.376 & 0.21345 & 0.229 & 15.254 \\ \hline
%12 & MV & 25 & 30 & 2.400 & 4.333 & 0.15600 & 0.394 & 6.571 \\ \hline
12 & MV & 464 & 623 & 2.685 & 11.848 & 0.14181 & 0.245 & 15.371 \\ \hline \hline
\multicolumn{9}{|c|}{\textbf{original \textit{order} +50\%}} \\ \hline

1 & MV & 444 & 729 & 3.284 & 10.025 & 0.19503 & 0.368 & 11.951 \\ \hline
2 & MV & 472 & 759 & 3.216 & 13.900 & 0.30131 & 0.267 & 17.160 \\ \hline
3 & MV & 238 & 367 & 3.084 & 9.848 & 0.26126 & 0.321 & 12.735 \\ \hline
4 & MV & 263 & 432 & 3.285 & 9.538 & 0.26645 & 0.330 & 11.308 \\ \hline
5 & MV & 217 & 343 & 3.161 & 8.412 & 0.28220 & 0.406 & 11.819 \\ \hline
6 & MV & 191 & 310 & 3.246 & 7.116 & 0.22559 & 0.334 & 8.789 \\ \hline
7 & MV & 884 & 1588 & 3.593 & 7.871 & 0.33041 & 0.394 & 9.865 \\ \hline
8 & MV & 366 & 573 & 3.131 & 11.775 & 0.30838 & 0.248 & 13.960 \\ \hline
9 & MV & 218 & 348 & 3.193 & 8.323 & 0.27785 & 0.296 & 9.878 \\ \hline
10 & MV & 201 & 306 & 3.045 & 11.570 & 0.32079 & 0.229 & 12.207 \\ \hline
11 & MV & 202 & 319 & 3.158 & 10.634 & 0.31854 & 0.252 & 13.553 \\ \hline
%12 & MV & 25 & 36 & 2.880 & 3.792 & 0.38400 & 0.404 & 5.872 \\ \hline
12 & MV & 464 & 748 & 3.224 & 10.748 & 0.27124 & 0.328 & 12.741 \\ \hline \hline
\multicolumn{9}{|c|}{\textbf{original \textit{order} +75\%}} \\ \hline

1 & MV & 444 & 850 & 3.829 & 8.796 & 0.32400 & 0.363 & 10.642 \\ \hline
2 & MV & 472 & 885 & 3.750 & 12.008 & 0.41309 & 0.350 & 13.903 \\ \hline
3 & MV & 238 & 428 & 3.597 & 8.880 & 0.40270 & 0.288 & 10.186 \\ \hline
4 & MV & 263 & 504 & 3.833 & 8.950 & 0.37425 & 0.370 & 10.482 \\ \hline
5 & MV & 217 & 400 & 3.687 & 7.194 & 0.42601 & 0.364 & 8.353 \\ \hline
6 & MV & 191 & 362 & 3.791 & 6.126 & 0.34172 & 0.313 & 7.584 \\ \hline
7 & MV & 884 & 1853 & 4.192 & 7.323 & 0.39933 & 0.385 & 8.965 \\ \hline
8 & MV & 366 & 668 & 3.650 & 10.363 & 0.38451 & 0.257 & 11.336 \\ \hline
9 & MV & 218 & 406 & 3.725 & 7.297 & 0.36753 & 0.397 & 8.770 \\ \hline
10 & MV & 201 & 357 & 3.552 & 10.185 & 0.42392 & 0.279 & 11.888 \\ \hline
11 & MV & 202 & 372 & 3.683 & 9.963 & 0.35298 & 0.284 & 11.178 \\ \hline
%12 & MV & 25 & 42 & 3.360 & 3.708 & 0.48267 & 0.427 & 5.590 \\ \hline
12 & MV & 464 & 873 & 3.763 & 9.477 & 0.34785 & 0.362 & 11.815 \\ \hline \hline
\multicolumn{9}{|c|}{\textbf{original \textit{order} +100\%}} \\ \hline

1 & MV & 444 & 972 & 4.378 & 7.869 & 0.39146 & 0.385 & 9.194 \\ \hline
2 & MV & 472 & 1012 & 4.288 & 10.839 & 0.53045 & 0.342 & 11.835 \\ \hline
3 & MV & 238 & 490 & 4.118 & 8.086 & 0.56850 & 0.358 & 9.286 \\ \hline
4 & MV & 263 & 576 & 4.380 & 8.057 & 0.44110 & 0.414 & 9.389 \\ \hline
5 & MV & 217 & 458 & 4.221 & 6.597 & 0.50752 & 0.389 & 7.457 \\ \hline
6 & MV & 191 & 414 & 4.335 & 5.700 & 0.46025 & 0.361 & 7.443 \\ \hline
7 & MV & 884 & 2118 & 4.792 & 6.745 & 0.43981 & 0.417 & 8.210 \\ \hline
8 & MV & 366 & 764 & 4.175 & 9.393 & 0.51501 & 0.313 & 10.760 \\ \hline
9 & MV & 218 & 464 & 4.257 & 6.569 & 0.45466 & 0.442 & 7.729 \\ \hline
10 & MV & 201 & 408 & 4.060 & 9.190 & 0.55617 & 0.325 & 9.556 \\ \hline
11 & MV & 202 & 426 & 4.218 & 8.338 & 0.47119 & 0.332 & 11.194 \\ \hline
%12 & MV & 25 & 48 & 3.840 & 3.167 & 0.52667 & 0.507 & 5.038 \\ \hline
12 & MV & 464 & 998 & 4.302 & 8.486 & 0.41405 & 0.405 & 9.871 \\ \hline
\end{tabular}
\caption{Metrics for triangle closure strategy \MV samples evolution.}\label{tab:mvCC}
\end{center}
\end{footnotesize}
\end{table}

\begin{center}
\begin{table}[htbp]
\centering
\begin{footnotesize}
\begin{tabular}{|@p{1.7cm}|^p{1.5cm}|^p{0.8cm}|^p{0.8cm}|^p{1.5cm}|^p{1.5cm}|^p{1.2cm}|}
\hline
\rowstyle{\bfseries}
Sample  ID  &Network type &\textit{Order} & \textit{Size} & {Avg. betweenness} & {Avg. betw/order} & {Coeff. variation} \\ \hline
\hline
\multicolumn{7}{|c|}{\textbf{original \textit{order} +25\%}} \\ \hline
1 & MV & 444 & 607 & 3837.713 & 8.643 & 2.052 \\ \hline
2 & MV & 472 & 632 & 4671.382 & 9.897 & 1.849 \\ \hline
3 & MV & 238 & 306 & 1646.108 & 6.916 & 1.537 \\ \hline
4 & MV & 263 & 360 & 1079.628 & 4.105 & 1.614 \\ \hline
5 & MV & 217 & 286 & 1786.327 & 8.232 & 1.724 \\ \hline
6 & MV & 191 & 258 & 1297.746 & 6.794 & 1.956 \\ \hline
7 & MV & 884 & 1323 & 6924.908 & 7.834 & 3.154 \\ \hline
8 & MV & 366 & 477 & 4258.659 & 11.636 & 1.916 \\ \hline
9 & MV & 218 & 290 & 1100.51 & 5.048 & 1.67 \\ \hline
10 & MV & 201 & 255 & 2982.928 & 14.84 & 1.303 \\ \hline
11 & MV & 202 & 266 & 2334.081 & 11.555 & 1.56 \\ \hline
%12 & MV & 25 & 30 & 88.88 & 3.555 & 1.055 \\ \hline
12 & MV & 464 & 623 & 3076.954 & 6.631 & 1.826 \\ \hline \hline
\multicolumn{7}{|c|}{\textbf{original \textit{order} +50\%}} \\ \hline

1 & MV & 444 & 729 & 3183.386 & 7.17 & 2.213 \\ \hline
2 & MV & 472 & 759 & 4035.469 & 8.55 & 1.969 \\ \hline
3 & MV & 238 & 367 & 1387.793 & 5.831 & 1.791 \\ \hline
4 & MV & 263 & 432 & 950.802 & 3.615 & 1.731 \\ \hline
5 & MV & 217 & 343 & 1625.935 & 7.493 & 1.817 \\ \hline
6 & MV & 191 & 310 & 1180.635 & 6.181 & 1.73 \\ \hline
7 & MV & 884 & 1588 & 6298.27 & 7.125 & 3.405 \\ \hline
8 & MV & 366 & 573 & 3625.364 & 9.905 & 1.777 \\ \hline
9 & MV & 218 & 348 & 898.163 & 4.12 & 1.848 \\ \hline
10 & MV & 201 & 306 & 2562.557 & 12.749 & 1.319 \\ \hline
11 & MV & 202 & 319 & 2194.406 & 10.863 & 1.582 \\ \hline
%12 & MV & 25 & 36 & 68.64 & 2.746 & 1.251 \\ \hline
12 & MV & 464 & 748 & 2901.182 & 6.253 & 1.906 \\ \hline
\hline
\multicolumn{7}{|c|}{\textbf{original \textit{order} +75\%}} \\ \hline
1 & MV & 444 & 850 & 2782.5 & 6.267 & 2.158 \\ \hline
2 & MV & 472 & 885 & 3468.145 & 7.348 & 2.085 \\ \hline
3 & MV & 238 & 428 & 1179.631 & 4.956 & 1.828 \\ \hline
4 & MV & 263 & 504 & 873.116 & 3.32 & 1.741 \\ \hline
5 & MV & 217 & 400 & 1389.383 & 6.403 & 1.939 \\ \hline
6 & MV & 191 & 362 & 1030.677 & 5.396 & 1.827 \\ \hline
7 & MV & 884 & 1853 & 5736.492 & 6.489 & 3.495 \\ \hline
8 & MV & 366 & 668 & 3097.301 & 8.463 & 1.936 \\ \hline
9 & MV & 218 & 406 & 809.837 & 3.715 & 1.818 \\ \hline
10 & MV & 201 & 357 & 2130.351 & 10.599 & 1.54 \\ \hline
11 & MV & 202 & 372 & 2060 & 10.198 & 1.505 \\ \hline
%12 & MV & 25 & 42 & 67.92 & 2.717 & 1.2 \\ \hline
12 & MV & 464 & 873 & 2733.855 & 5.892 & 1.974 \\ \hline
\hline
\multicolumn{7}{|c|}{\textbf{original \textit{order} +100\%}} \\ \hline
1 & MV & 444 & 972 & 2540.609 & 5.722 & 2.093 \\ \hline
2 & MV & 472 & 1012 & 3034.596 & 6.429 & 2.323 \\ \hline
3 & MV & 238 & 490 & 1065.207 & 4.476 & 2.108 \\ \hline
4 & MV & 263 & 576 & 807.934 & 3.072 & 1.719 \\ \hline
5 & MV & 217 & 458 & 1223.467 & 5.638 & 2.064 \\ \hline
6 & MV & 191 & 414 & 941.228 & 4.928 & 1.969 \\ \hline
7 & MV & 884 & 2118 & 5292.311 & 5.987 & 3.439 \\ \hline
8 & MV & 366 & 764 & 2757.96 & 7.535 & 2.146 \\ \hline
9 & MV & 218 & 464 & 684.923 & 3.142 & 2.054 \\ \hline
10 & MV & 201 & 408 & 1837.454 & 9.142 & 1.703 \\ \hline
11 & MV & 202 & 426 & 1710.355 & 8.467 & 1.666 \\ \hline
%12 & MV & 25 & 48 & 57.36 & 2.294 & 1.11 \\ \hline
12 & MV & 464 & 998 & 2545.442 & 5.486 & 2.042 \\ \hline
\end{tabular}
\caption{Betweenness for triangle closure strategy \MV samples evolution.}\label{tab:mvCCBet}
\end{footnotesize}
\end{table}
\end{center}

\begin{figure}
 \captionsetup{type=figure}
    \centering
    \subfloat[Physical sample.]{\label{fig:rndCCevol}\includegraphics[scale=.23]{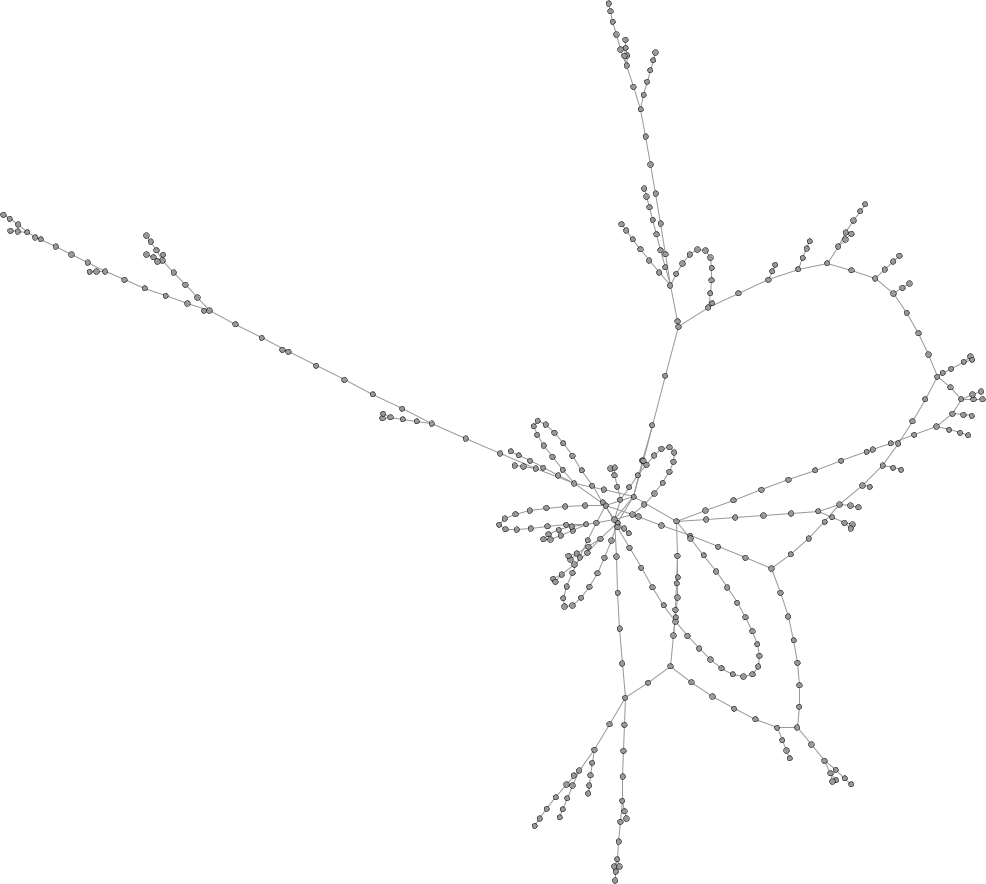}} 
    \subfloat[1$^{st}$ stage of evolution (i.e., +25\% edges).]{\label{fig:cc25}\includegraphics[scale=.23]{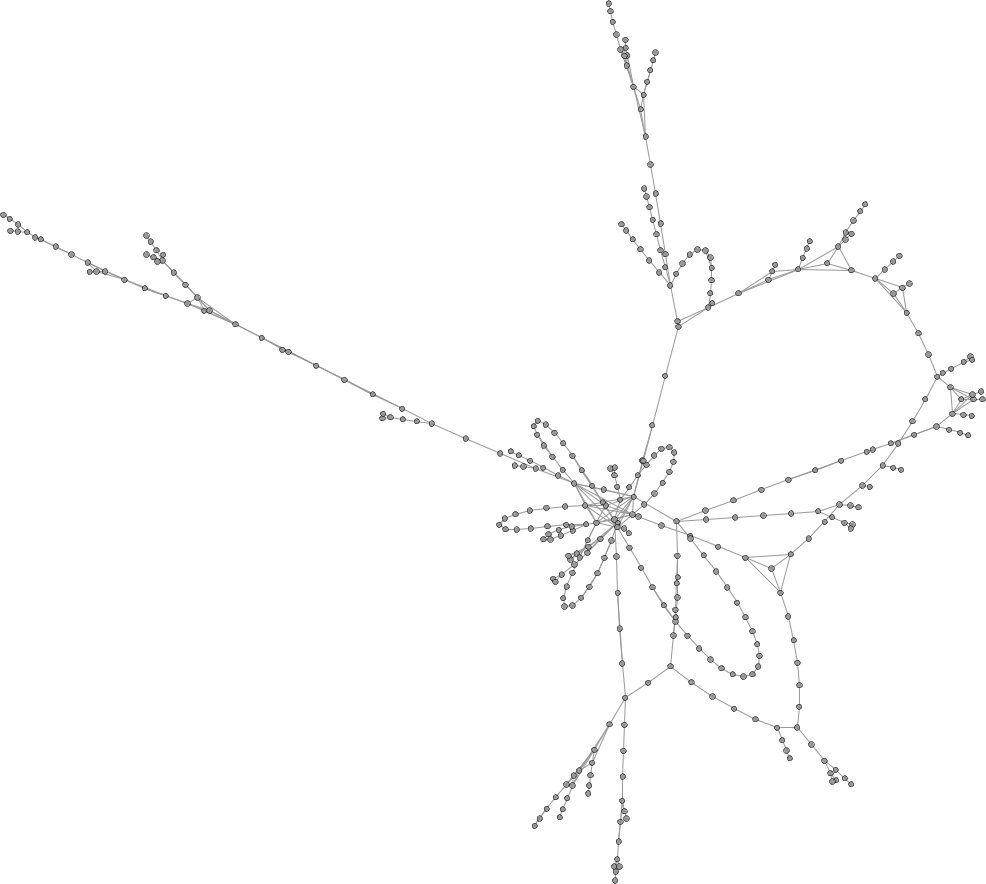}}\\
     \hspace*{-1.8cm}\subfloat[2$^{nd}$ stage of evolution (i.e., +50\% edges)]{\label{fig:cc50}\includegraphics[scale=.23]{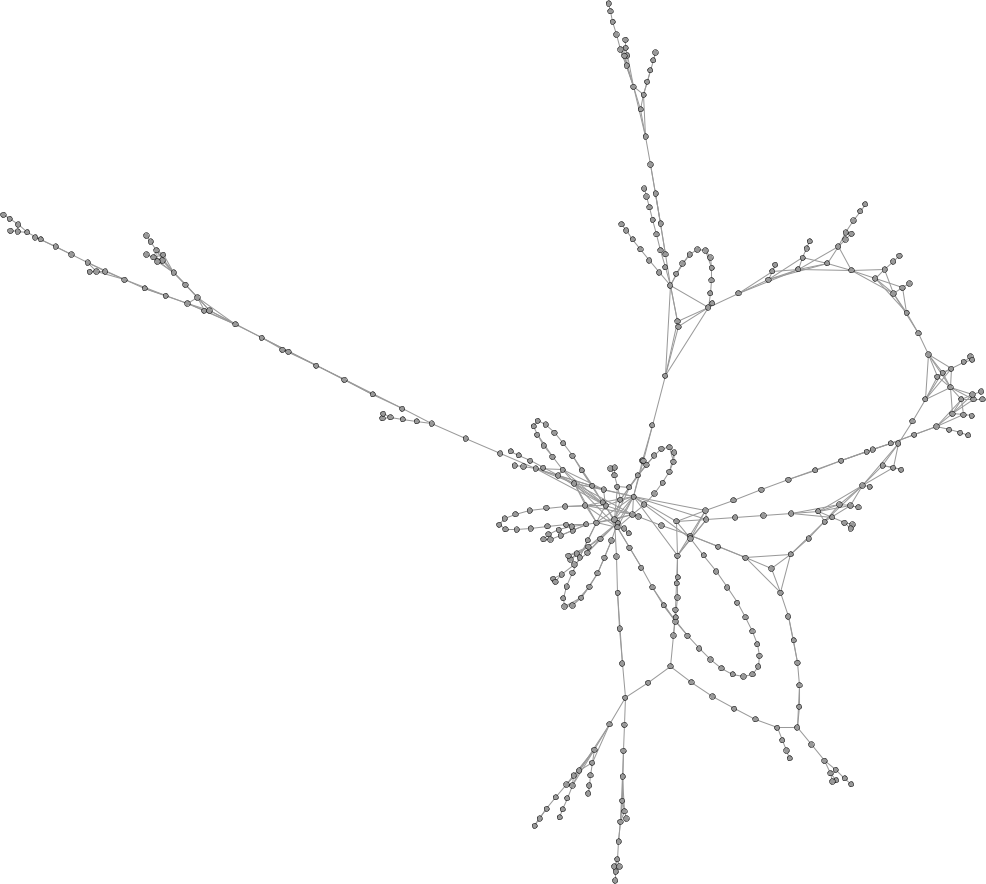}}  
      \subfloat[3$^{rd}$ stage of evolution (i.e., +75\% edges).]{\label{fig:cc75}\includegraphics[scale=.23]{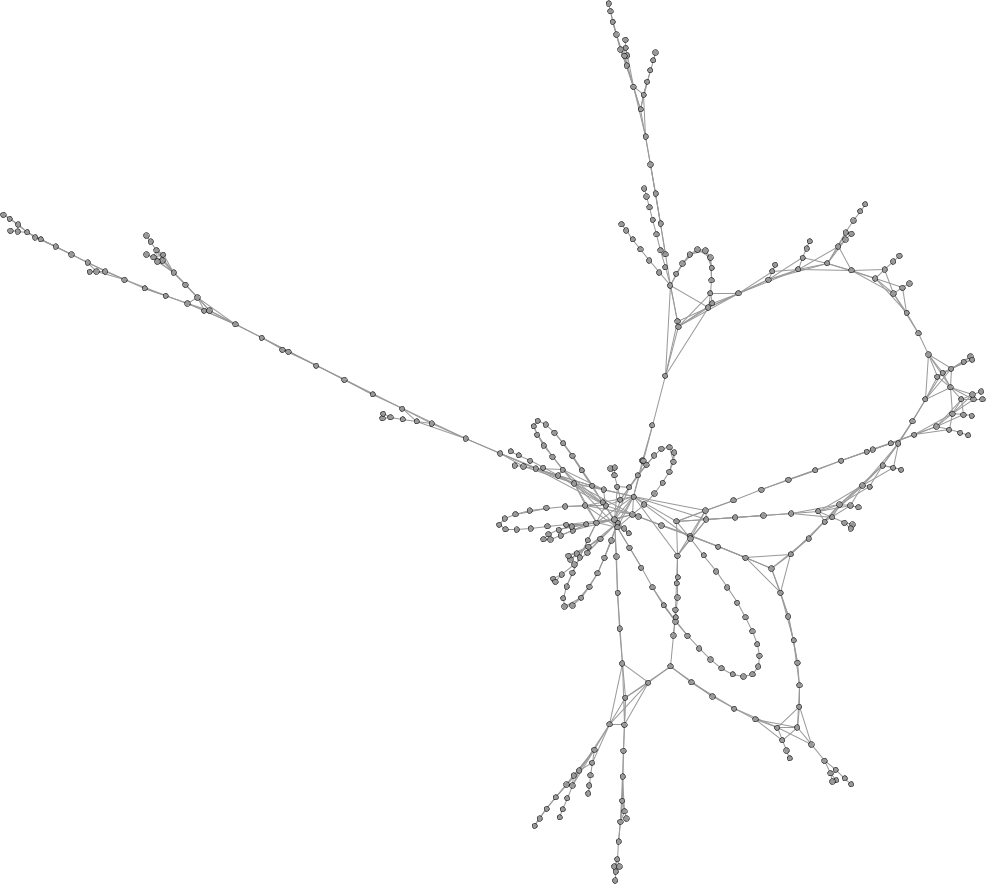}} \\
       \subfloat[4$^{th}$ stage of evolution (i.e., +100\% edges).]{\label{fig:cc100}\includegraphics[scale=.23]{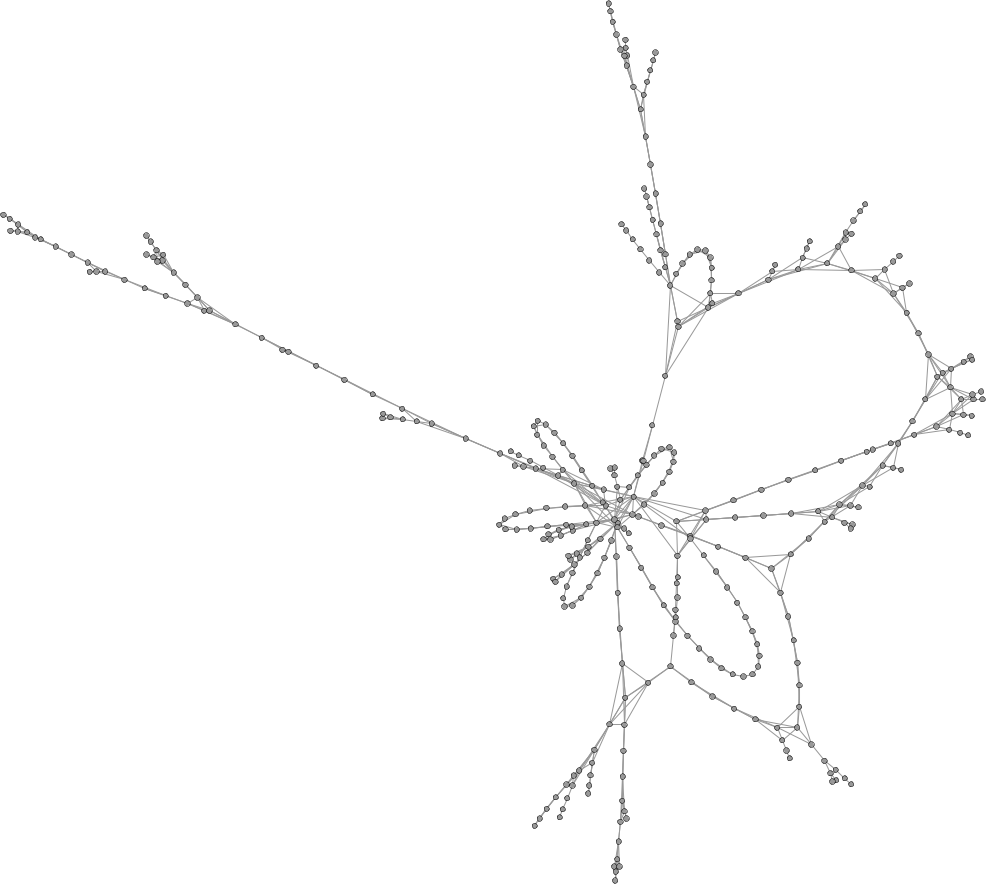}}  %\\
    \caption{Stages of evolution of \MV sample \#8 following the triangle closure strategy.}
    \label{fig:evolchartcc}
\end{figure}

\subsubsection*{Dissortative node degree evolution}

Table~\ref{tab:mvDiss} contains for each sample of the \MV \G (column one) the values for main topological quantities: \textit{order} and \textit{size} in columns three and four, respectively; average node degree in the fifth column; the \cpl is reported in column six; the clustering coefficient follows in column seven; robustness is shown in eight column; the cost in term of redundant path length closes the data series (column nine).

\begin{figure}
 \captionsetup{type=figure}
    \centering
    \subfloat[Random node removal robustness.]{\label{fig:rnddiss}\includegraphics[scale=.4]{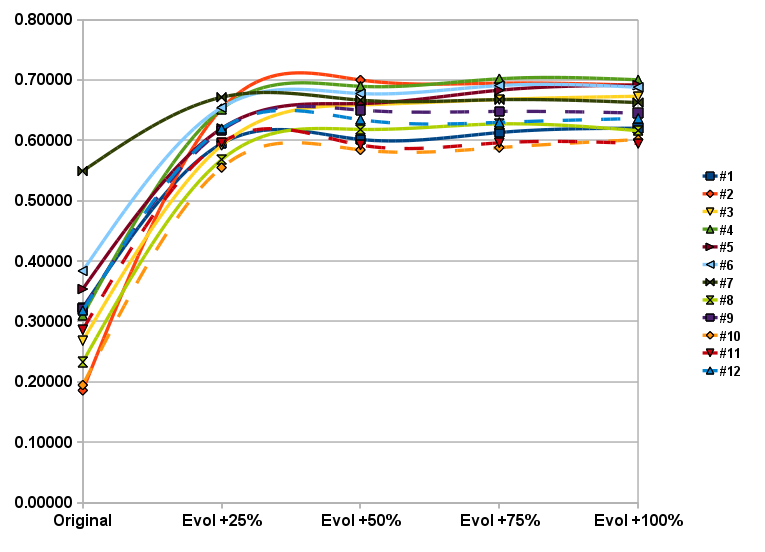}} 
    \subfloat[Targeted node removal robustness.]{\label{fig:tardiss}\includegraphics[scale=.4]{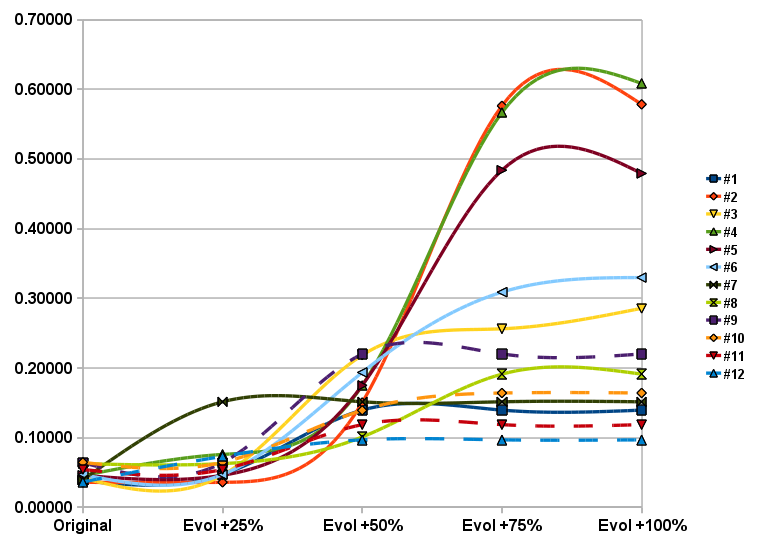}} %\\
    \caption{Evolution of robustness sub-metrics.}
    \label{fig:dissRob}
\end{figure}

\begin{figure}[htbp]
 %\begin{minipage}[htbp]{7cm}
   \centering   
  \includegraphics[width=0.7\textwidth]{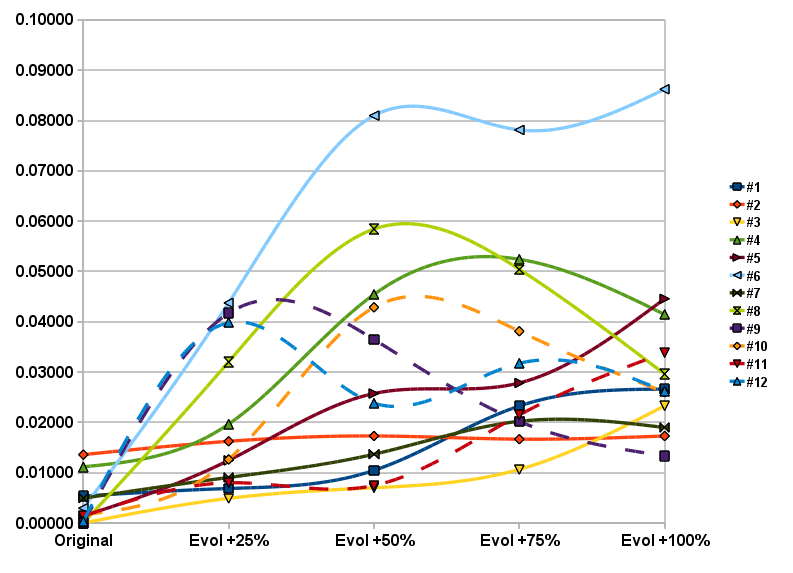}
   \caption{Evolution of the \cc metric.}
\label{fig:ccDissEvolMV}
\end{figure}

The primary focus of the dissortative node degree strategy is to connect nodes with small node degree with nodes with high nodes degree. A general consideration that can be made analyzing the data in Table~\ref{tab:mvDiss} is that the most of the improvement for all the metrics is obtained after the third step of evolution, the last increment in connectivity is scarcely beneficial. Considering the \cpl metric, the increase in the connectivity with such an evolution strategy is beneficial. The samples score for this metric all below 6.3 showing an improvement of 55\% in general when the number of edges is doubled. As already remarked, the final evolution step is slightly beneficial, providing a reduction in \cpl of just 1\% compared to step three. The clustering coefficient evolution for such  strategy provides benefits that are not more than two orders of magnitude. It is also interesting to remark that according to this evolution strategy for many samples 
the peak value of the \cc is not obtained when the maximum amount of edges are added, but in the first or second step of the evolution. 
In fact, more connectivity between heterogeneous nodes (in terms of node degree) leads to a situation where   nodes with a big neighborhood of nodes in which the new node is not likely have other connections, therefore reducing the \cc of the whole network.
%\footnote{sta frase e' da mal di testa :-) SPEZZATA E RISCRITTA UN PO} 
The graphical representation of the evolution of the \cc is shown in Figure~\ref{fig:ccDissEvolMV}. In general, robustness triples when the \textit{size} of the graph doubles, in particular a notable increase takes place until the third evolution step. As mentioned for other strategies, the addition of edges provides benefits in dealing with random attacks, but for some samples (i.e., samples \#2, \#5, \#6) the improvement is particularly high also for the targeted attacks having the two values that compose the robustness metric (cf.~\ref{sec:metrics}) that score almost equally around 0.5-0.6. 
Newman~\cite{newman02} explains that assortative networks that tight high node degree have a sort of redundancy in the main cluster that connects the nodes with high degree, while this is absent in a dissortative network.
%\footnote{sta frase non la capisco in inglese. Mi manca un verbo. MESSA A POSTO} 
In the evolution following this strategy we see that the new connectivity is particularly beneficial in contrasting the targeted attacks. The new established connections tend to bond nodes with low degree to the already highly connected nodes. On the one hand, this strategy reinforces the connectivity of already established hubs; on the other hand, it creates small hubs at the periphery of the network since the nodes with smallest connectivity tend to be more and more connected with the central nodes while more new edges are attached.  Therefore, these new external and redundant hubs are new targets for the high node removal policy once they have sufficient connectivity. This improvement in the reliability is empirically evident by the results of the robustness metric (Figures~\ref{fig:rnddiss} and~\ref{fig:tardiss}) that show the evolution in the random and targeted node removal situations, respectively. One sees that the increase in connectivity is beneficial to contrast random attacks. A plateau for random failures around 0.6-0.7 is reached after the second evolution step. More benefits are obtained against the attacks that target high degree nodes: about half of the samples experience an improvement in the metric about one order of magnitude.

Considering the values for the redundant path robustness, the same considerations for the \cpl apply. There is already a reduction of 50\%, compared to the initial value of the samples, already when the networks are evolved with a 25\% increase in the \textit{size} of the original graph. The additional connectivity that is provided in the last evolution step is marginally beneficial, as seen for the characteristic path length, providing just an additional 2\% reduction in the path length.

Table~\ref{tab:mvDissBet} contains for each sample of the \MV \G (column one) the values of metrics related to betweenness. In addition to \textit{order} and \textit{size} (columns three and four), average betweenness is provided in columns five, while a value of average betweenness normalized by the \textit{order} of the graph is computed in column six in order to compare the different samples. A measure of the statistical variation of betweenness is the coefficient of variation which is shown in the seventh column.

Considering betweenness, the dissortative strategy involving the nodes with lowest node degree is substantially beneficial in reducing the average betweenness of all the \MV samples just in the first step of the evolution. In fact, with the addition of just a quarter of the initial number of links, the average betweenness reduces around 45\% of the original value. The additional three evolution stages contribute modestly in further reducing this metric (just 5\%). The same trend is followed by the normalized value of average betweenness divided by the \textit{order} of the network; already in the first step just one sample exceeds 6 for the betweenness to order ratio. Considering the variability (i.e., coefficient of variation) of betweenness in each of the samples we note a general increase for this metric, only after the last stage of the evolution the coefficient of variation returns to values closer to the initial value.  Such behavior is due to the decrease in the standard deviation of betweenness that is slower compared to the average betweenness. However, this strategy of adding connections between the nodes that are at the opposite in their degree, tends to make them more evenly involved in the shortest paths. A proof is that the median of the betweenness for seven of the twelve samples is higher than zero.

\begin{table}[htbp]
\begin{footnotesize}
\begin{center}
\begin{tabular}{|@p{1.7cm}|^p{1.7cm}|^p{0.9cm}|^p{0.6cm}|^p{1cm}|^p{1cm}|^p{1.2cm}|^p{1.5cm}|^p{2cm}|}
\hline
\rowstyle{\bfseries} Sample ID &Network type  & \textit{Order} & \textit{Size} & {Avg. deg.} & {CPL} & {CC} & {Removal robustness ($Rob_N$)} & {Redundancy cost ($APL_{10^{th}}$)} \\ \hline
\hline

\multicolumn{9}{|c|}{\textbf{original \textit{order} +25\%}} \\ \hline
1 & MV & 444 & 607 & 2.734 & 6.800 & 0.00687 & 0.321 & 9.311 \\ \hline
2 & MV & 472 & 632 & 2.678 & 5.809 & 0.01625 & 0.344 & 8.604 \\ \hline
3 & MV & 238 & 306 & 2.571 & 5.842 & 0.00491 & 0.319 & 9.112 \\ \hline
4 & MV & 263 & 360 & 2.738 & 5.580 & 0.01966 & 0.364 & 8.399 \\ \hline
5 & MV & 217 & 286 & 2.636 & 4.903 & 0.01245 & 0.333 & 7.992 \\ \hline
6 & MV & 191 & 258 & 2.702 & 4.979 & 0.04375 & 0.351 & 7.648 \\ \hline
7 & MV & 884 & 1323 & 2.993 & 6.866 & 0.00906 & 0.411 & 8.235 \\ \hline
8 & MV & 366 & 477 & 2.607 & 6.314 & 0.03204 & 0.315 & 9.743 \\ \hline
9 & MV & 218 & 290 & 2.661 & 5.970 & 0.04170 & 0.340 & 8.548 \\ \hline
10 & MV & 201 & 255 & 2.537 & 6.290 & 0.01264 & 0.310 & 8.746 \\ \hline
11 & MV & 202 & 266 & 2.634 & 6.413 & 0.00800 & 0.326 & 9.951 \\ \hline
%12 & MV & 25 & 30 & 2.400 & 3.208 & 0.00000 & 0.459 & 9.994 \\ \hline
12 & MV & 464 & 623 & 2.685 & 6.470 & 0.03983 & 0.346 & 9.295 \\ \hline \hline
\multicolumn{9}{|c|}{\textbf{original \textit{order} +50\%}} \\ \hline 

1 & MV & 444 & 729 & 3.284 & 6.341 & 0.01046 & 0.370 & 9.079 \\ \hline
2 & MV & 472 & 759 & 3.216 & 5.645 & 0.01730 & 0.425 & 7.778 \\ \hline
3 & MV & 238 & 367 & 3.084 & 5.586 & 0.00701 & 0.439 & 7.748 \\ \hline
4 & MV & 263 & 432 & 3.285 & 5.477 & 0.04544 & 0.432 & 7.482 \\ \hline
5 & MV & 217 & 343 & 3.161 & 4.801 & 0.02574 & 0.418 & 7.258 \\ \hline
6 & MV & 191 & 310 & 3.246 & 4.647 & 0.08098 & 0.436 & 7.002 \\ \hline
7 & MV & 884 & 1588 & 3.593 & 6.577 & 0.01368 & 0.409 & 8.199 \\ \hline
8 & MV & 366 & 573 & 3.131 & 6.178 & 0.05844 & 0.360 & 8.085 \\ \hline
9 & MV & 218 & 348 & 3.193 & 5.751 & 0.03649 & 0.435 & 8.133 \\ \hline
10 & MV & 201 & 306 & 3.045 & 6.240 & 0.04287 & 0.362 & 8.330 \\ \hline
11 & MV & 202 & 319 & 3.158 & 5.995 & 0.00745 & 0.355 & 8.982 \\ \hline
%12 & MV & 25 & 36 & 2.880 & 3.125 & 0.00000 & 0.494 & 6.224 \\ \hline
12 & MV & 464 & 748 & 3.224 & 6.323 & 0.02381 & 0.366 & 8.452 \\ \hline \hline
\multicolumn{9}{|c|}{\textbf{original \textit{order} +75\%}} \\ \hline 

1 & MV & 444 & 850 & 3.829 & 6.246 & 0.02326 & 0.376 & 9.187 \\ \hline
2 & MV & 472 & 885 & 3.750 & 5.590 & 0.01667 & 0.635 & 7.656 \\ \hline
3 & MV & 238 & 428 & 3.597 & 5.308 & 0.01065 & 0.462 & 6.847 \\ \hline
4 & MV & 263 & 504 & 3.833 & 5.359 & 0.05242 & 0.634 & 7.009 \\ \hline
5 & MV & 217 & 400 & 3.687 & 4.671 & 0.02784 & 0.583 & 6.949 \\ \hline
6 & MV & 191 & 362 & 3.791 & 4.563 & 0.07813 & 0.500 & 6.212 \\ \hline
7 & MV & 884 & 1853 & 4.192 & 6.532 & 0.02022 & 0.409 & 8.163 \\ \hline
8 & MV & 366 & 668 & 3.650 & 6.012 & 0.05044 & 0.409 & 7.997 \\ \hline
9 & MV & 218 & 406 & 3.725 & 5.297 & 0.02018 & 0.434 & 7.492 \\ \hline
10 & MV & 201 & 357 & 3.552 & 6.000 & 0.03812 & 0.376 & 7.988 \\ \hline
11 & MV & 202 & 372 & 3.683 & 5.764 & 0.02148 & 0.357 & 7.971 \\ \hline
%12 & MV & 25 & 42 & 3.360 & 2.833 & 0.04778 & 0.484 & 5.397 \\ \hline
12 & MV & 464 & 873 & 3.763 & 6.056 & 0.03175 & 0.363 & 7.598 \\ \hline \hline
\multicolumn{9}{|c|}{\textbf{original \textit{order} +100\%}} \\ \hline

1 & MV & 444 & 972 & 4.378 & 6.089 & 0.02662 & 0.380 & 8.296 \\ \hline
2 & MV & 472 & 1012 & 4.288 & 5.089 & 0.01730 & 0.635 & 7.110 \\ \hline
3 & MV & 238 & 490 & 4.118 & 5.097 & 0.02329 & 0.479 & 6.761 \\ \hline
4 & MV & 263 & 576 & 4.380 & 5.011 & 0.04146 & 0.654 & 6.320 \\ \hline
5 & MV & 217 & 458 & 4.221 & 4.551 & 0.04456 & 0.585 & 6.222 \\ \hline
6 & MV & 191 & 414 & 4.335 & 4.547 & 0.08621 & 0.509 & 6.516 \\ \hline
7 & MV & 884 & 2118 & 4.792 & 6.278 & 0.01899 & 0.407 & 7.475 \\ \hline
8 & MV & 366 & 764 & 4.175 & 5.929 & 0.02961 & 0.404 & 8.028 \\ \hline
9 & MV & 218 & 464 & 4.257 & 5.189 & 0.01336 & 0.433 & 6.919 \\ \hline
10 & MV & 201 & 408 & 4.060 & 5.840 & 0.02601 & 0.383 & 7.391 \\ \hline
11 & MV & 202 & 426 & 4.218 & 5.637 & 0.03385 & 0.357 & 7.656 \\ \hline
%12 & MV & 25 & 48 & 3.840 & 2.583 & 0.04489 & 0.495 & 4.994 \\ \hline
12 & MV & 464 & 998 & 4.302 & 5.975 & 0.02619 & 0.367 & 8.007 \\ \hline
\end{tabular}
\caption{Metrics for dissortative node degree strategy \MV samples evolution.}\label{tab:mvDiss}
\end{center}
\end{footnotesize}
\end{table}

\begin{center}
\begin{table}[htbp]
\centering
\begin{footnotesize}
\begin{tabular}{|@p{1.7cm}|^p{1.5cm}|^p{0.8cm}|^p{0.8cm}|^p{1.5cm}|^p{1.5cm}|^p{1.2cm}|}
\hline
\rowstyle{\bfseries}
Sample  ID  &Network type &\textit{Order} & \textit{Size} & {Avg. betweenness} & {Avg. betw/order} & {Coeff. variation} \\ \hline
\hline
\multicolumn{7}{|c|}{\textbf{original \textit{order} +25\%}} \\ \hline
1 & MV & 444 & 607 & 2763.762 & 6.225 & 2.254 \\ \hline
2 & MV & 472 & 632 & 2298.132 & 4.869 & 3.515 \\ \hline
3 & MV & 238 & 306 & 1103.668 & 4.637 & 2.652 \\ \hline
4 & MV & 263 & 360 & 1161.52 & 4.416 & 2.695 \\ \hline
5 & MV & 217 & 286 & 926.215 & 4.268 & 2.897 \\ \hline
6 & MV & 191 & 258 & 749.916 & 3.926 & 2.783 \\ \hline
7 & MV & 884 & 1323 & 5258.973 & 5.949 & 2.815 \\ \hline
8 & MV & 366 & 477 & 2084.738 & 5.696 & 2.577 \\ \hline
9 & MV & 218 & 290 & 1067.579 & 4.897 & 2.061 \\ \hline
10 & MV & 201 & 255 & 1125.408 & 5.599 & 1.729 \\ \hline
11 & MV & 202 & 266 & 1149 & 5.688 & 1.739 \\ \hline
%12 & MV & 25 & 30 & 58 & 2.32 & 1.136 \\ \hline
12 & MV & 464 & 623 & 2713.298 & 5.848 & 2.77 \\ \hline \hline
\multicolumn{7}{|c|}{\textbf{original \textit{order} +50\%}} \\ \hline

1 & MV & 444 & 729 & 2630.233 & 5.924 & 1.979 \\ \hline
2 & MV & 472 & 759 & 2178.939 & 4.616 & 2.644 \\ \hline
3 & MV & 238 & 367 & 1011.731 & 4.251 & 2.063 \\ \hline
4 & MV & 263 & 432 & 1124.138 & 4.274 & 1.948 \\ \hline
5 & MV & 217 & 343 & 894.523 & 4.122 & 2.235 \\ \hline
6 & MV & 191 & 310 & 725.905 & 3.801 & 2.11 \\ \hline
7 & MV & 884 & 1588 & 4969.261 & 5.621 & 2.495 \\ \hline
8 & MV & 366 & 573 & 2002.795 & 5.472 & 1.96 \\ \hline
9 & MV & 218 & 348 & 1019.053 & 4.675 & 1.648 \\ \hline
10 & MV & 201 & 306 & 1104.204 & 5.494 & 1.396 \\ \hline
11 & MV & 202 & 319 & 1063.747 & 5.266 & 1.517 \\ \hline
%12 & MV & 25 & 36 & 51.6 & 2.064 & 1.113 \\ \hline
12 & MV & 464 & 748 & 2619.583 & 5.646 & 2.276 \\ \hline
\hline
\multicolumn{7}{|c|}{\textbf{original \textit{order} +50\%}} \\ \hline
1 & MV & 444 & 850 & 2588.741 & 5.83 & 1.842 \\ \hline
2 & MV & 472 & 885 & 2141.579 & 4.537 & 2.301 \\ \hline
3 & MV & 238 & 428 & 970.009 & 4.076 & 1.763 \\ \hline
4 & MV & 263 & 504 & 1108.724 & 4.216 & 1.667 \\ \hline
5 & MV & 217 & 400 & 856.327 & 3.946 & 1.858 \\ \hline
6 & MV & 191 & 362 & 704.274 & 3.687 & 1.775 \\ \hline
7 & MV & 884 & 1853 & 4906.982 & 5.551 & 2.34 \\ \hline
8 & MV & 366 & 668 & 1936.04 & 5.29 & 1.778 \\ \hline
9 & MV & 218 & 406 & 947.675 & 4.347 & 1.478 \\ \hline
10 & MV & 201 & 357 & 1052.286 & 5.235 & 1.278 \\ \hline
11 & MV & 202 & 372 & 1016.879 & 5.034 & 1.394 \\ \hline
%12 & MV & 25 & 42 & 47.2 & 1.888 & 1.055 \\ \hline
12 & MV & 464 & 873 & 2449.978 & 5.28 & 2.082 \\ \hline
\hline
\multicolumn{7}{|c|}{\textbf{original \textit{order} +50\%}} \\ \hline
1 & MV & 444 & 972 & 2510.314 & 5.654 & 1.668 \\ \hline
2 & MV & 472 & 1012 & 2109.969 & 4.47 & 2.068 \\ \hline
3 & MV & 238 & 490 & 944.942 & 3.97 & 1.574 \\ \hline
4 & MV & 263 & 576 & 1058.61 & 4.025 & 1.58 \\ \hline
5 & MV & 217 & 458 & 842.523 & 3.883 & 1.708 \\ \hline
6 & MV & 191 & 414 & 690.221 & 3.614 & 1.569 \\ \hline
7 & MV & 884 & 2118 & 4818.952 & 5.451 & 2.128 \\ \hline
8 & MV & 366 & 764 & 1905.977 & 5.208 & 1.682 \\ \hline
9 & MV & 218 & 464 & 925.225 & 4.244 & 1.394 \\ \hline
10 & MV & 201 & 408 & 1018.806 & 5.069 & 1.177 \\ \hline
11 & MV & 202 & 426 & 998.545 & 4.943 & 1.347 \\ \hline
%12 & MV & 25 & 48 & 42.64 & 1.706 & 1.083 \\ \hline
12 & MV & 464 & 998 & 2408.728 & 5.191 & 1.986 \\ \hline
\end{tabular}
\caption{Betweenness for dissortative node degree strategy \MV samples evolution.}\label{tab:mvDissBet}
\end{footnotesize}
\end{table}
\end{center}

\begin{figure}
 \captionsetup{type=figure}
    \centering
    \subfloat[Physical sample.]{\label{fig:dissOrig}\includegraphics[scale=.2]{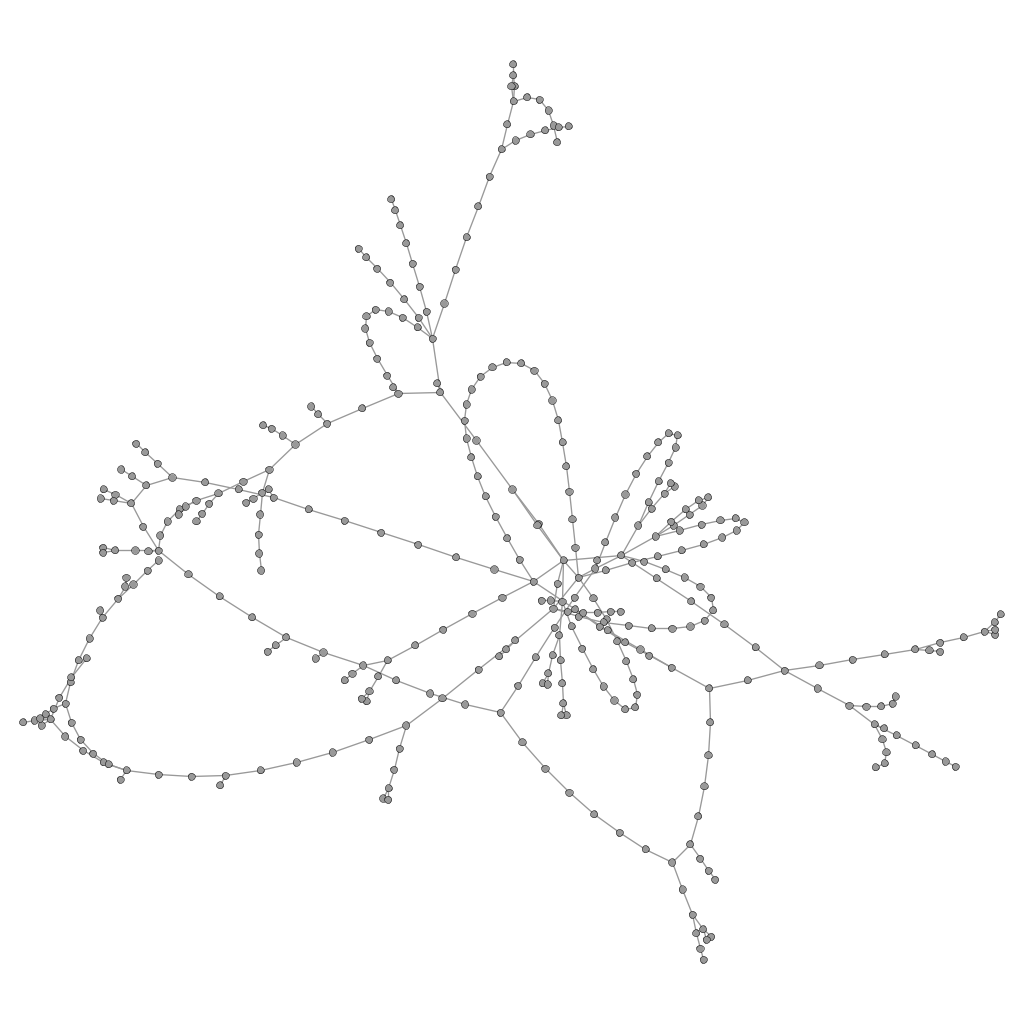}} 
    \subfloat[1$^{st}$ stage of evolution (i.e., +25\% edges).]{\label{fig:diss25}\includegraphics[scale=.2]{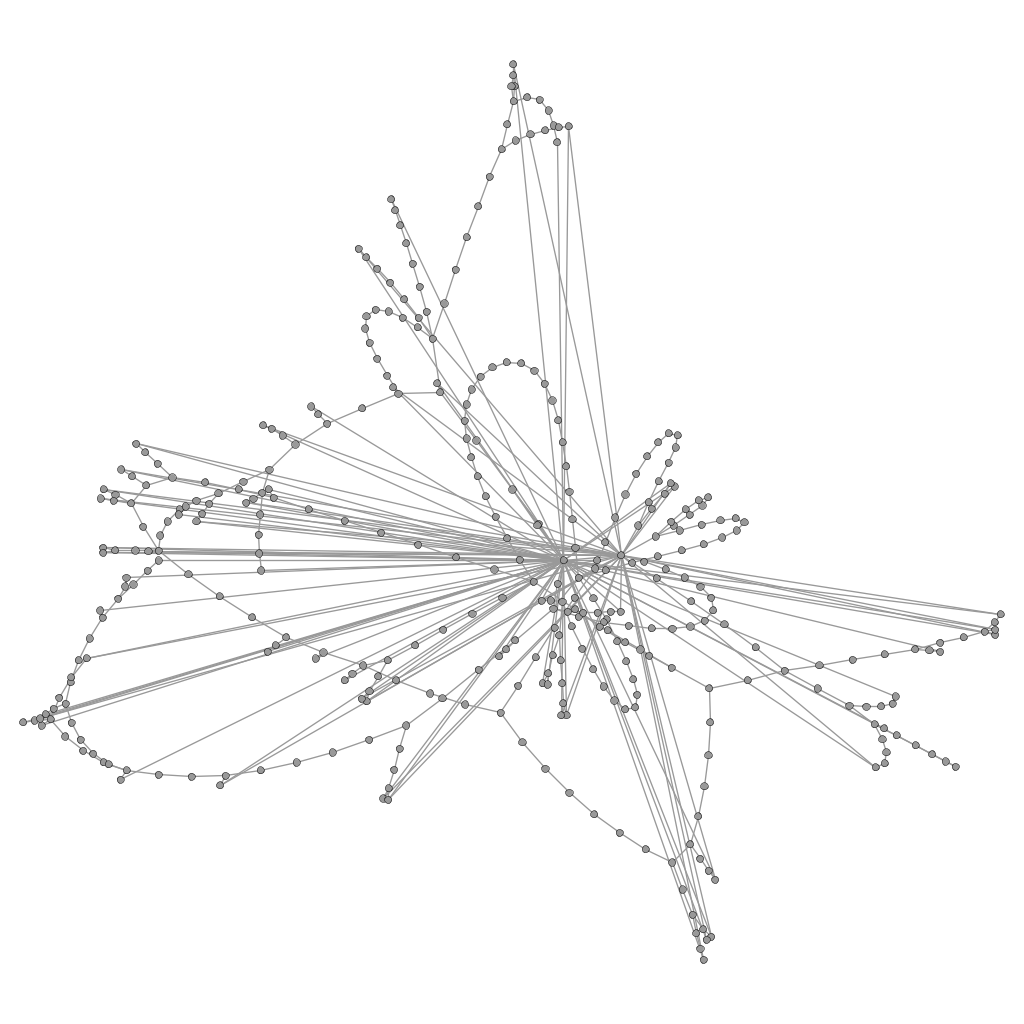}}\\
     \hspace*{-1.8cm}\subfloat[2$^{nd}$ stage of evolution (i.e., +50\% edges)]{\label{fig:diss50}\includegraphics[scale=.2]{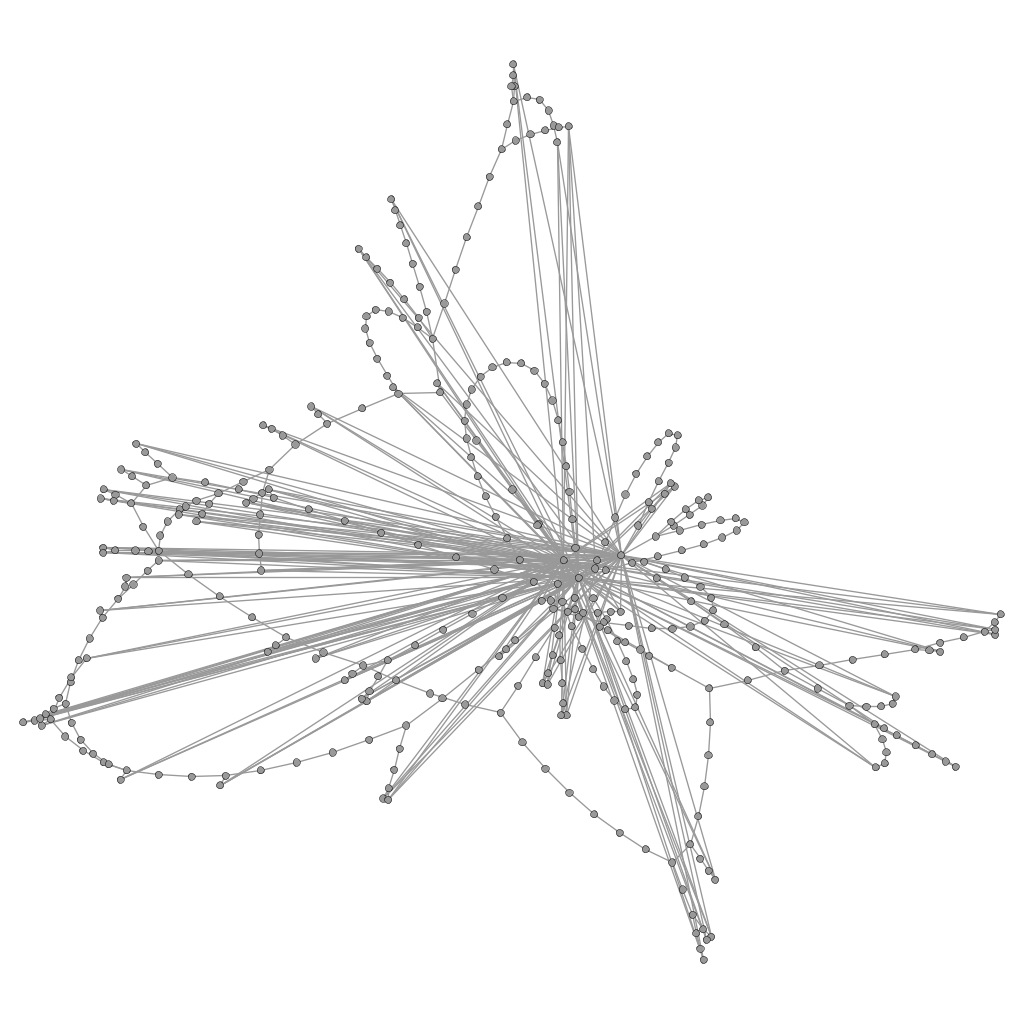}}  
      \subfloat[3$^{rd}$ stage of evolution (i.e., +75\% edges).]{\label{fig:diss75}\includegraphics[scale=.2]{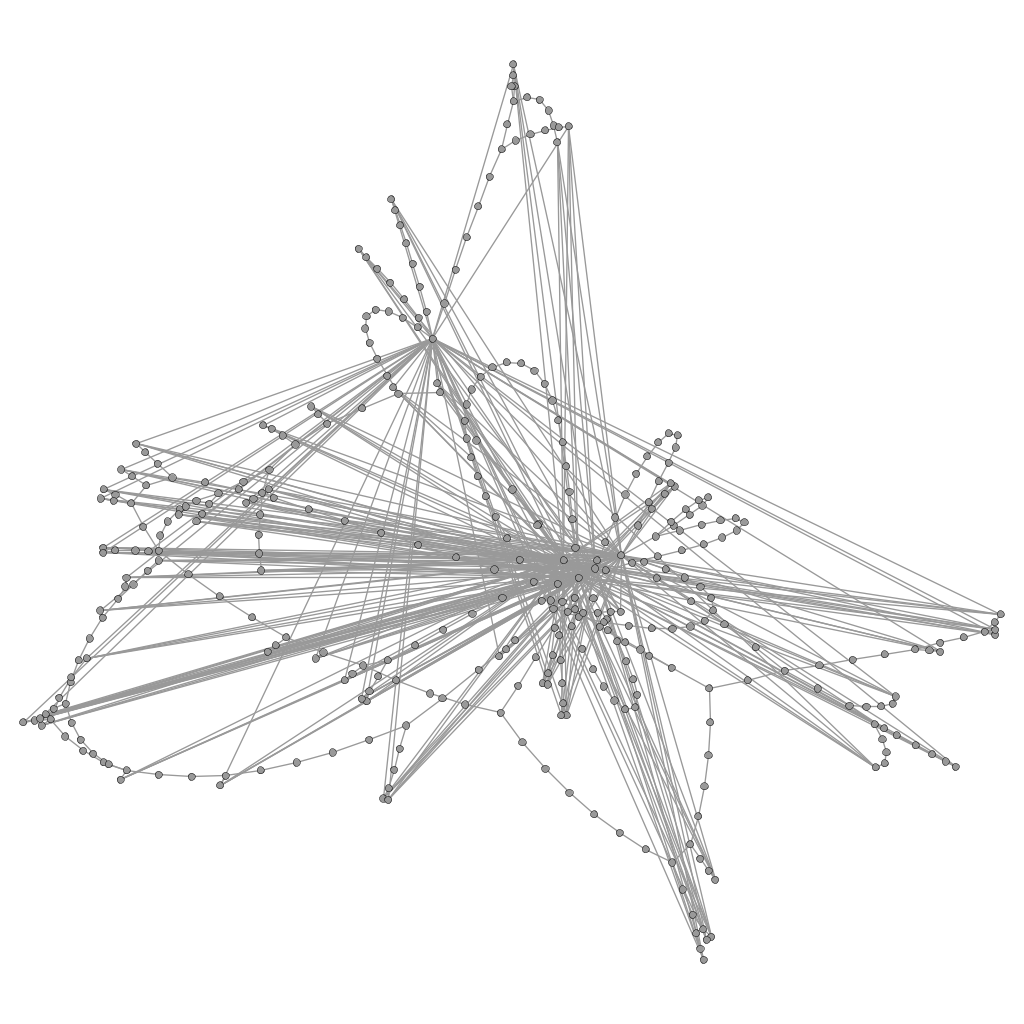}} \\
       \subfloat[4$^{th}$ stage of evolution (i.e., +100\% edges).]{\label{fig:diss100}\includegraphics[scale=.2]{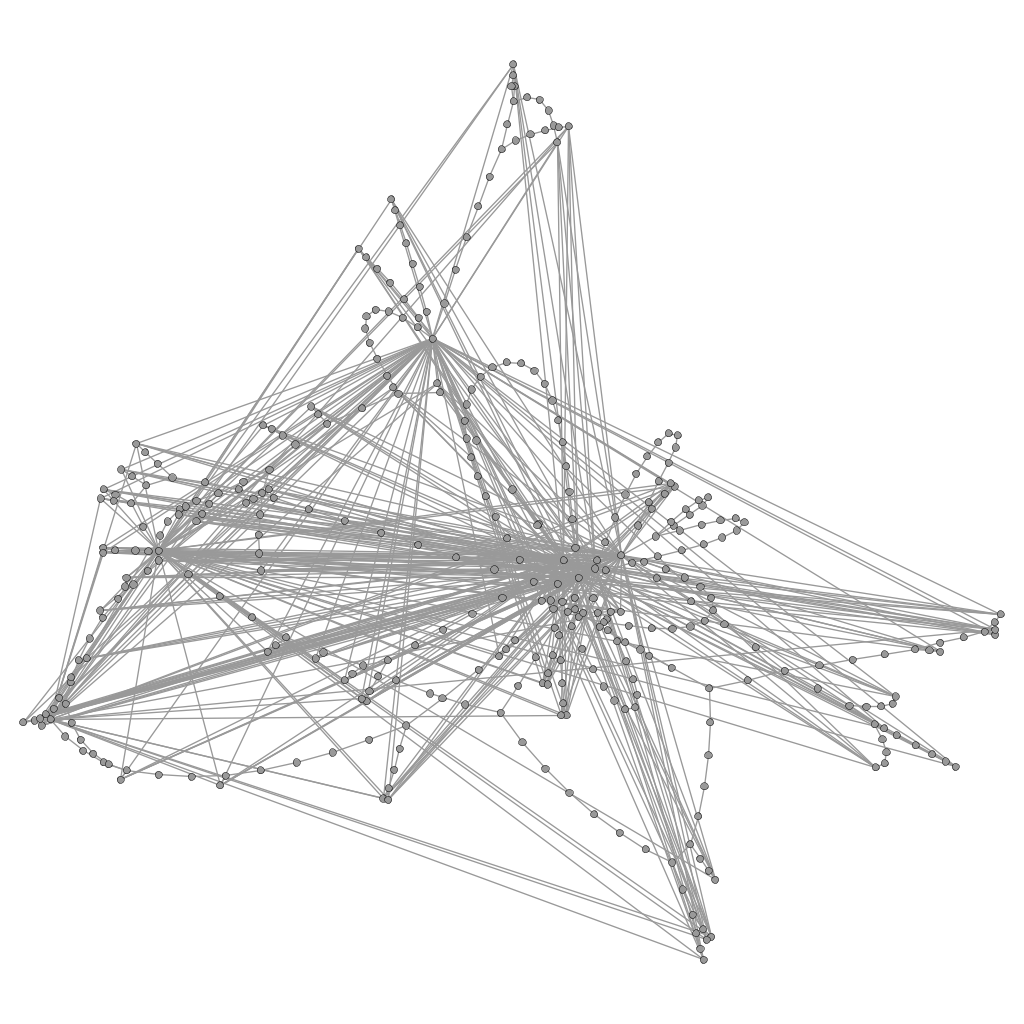}}  %\\
    \caption{Stages of evolution of \MV sample \#8 following  the dissortative strategy.}
    \label{fig:evolchartDiss}
\end{figure}

\subsubsection*{Least distance evolution}

Table~\ref{tab:mvDist} contains for each sample of the \MV \G (column one) the values for main topological quantities: \textit{order} and \textit{size} in columns three and four, respectively; average node degree in the fifth column; the \cpl is reported in column six; the clustering coefficient follows in column seven; robustness is shown in eight column; the cost in term of redundant path length closes the data series (column nine).

\begin{figure}
 \captionsetup{type=figure}
    \centering
    \subfloat[Random node removal robustness.]{\label{fig:rndlea}\includegraphics[scale=.4]{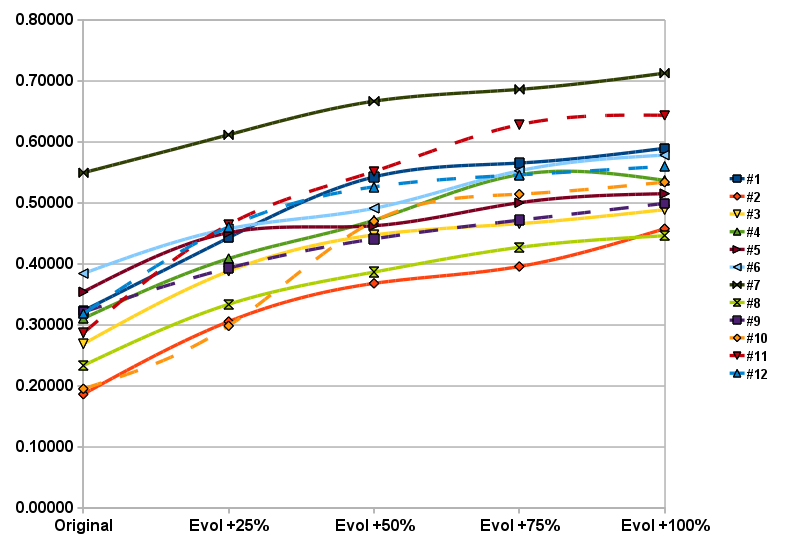}} \subfloat[Targeted node removal robustness.]{\label{fig:tarlea}\includegraphics[scale=.4]{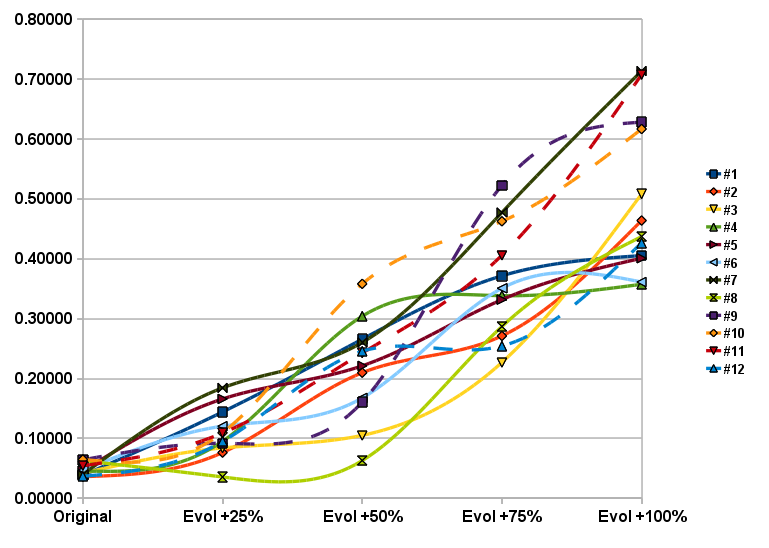}} %\\
    \caption{Evolution of robustness sub-metrics.}
    \label{fig:leaRob}
\end{figure}

The least distance  strategy aims at connecting the nodes (i.e., substations in the physical Distribution Grid) that are geographically closer to each other, therefore there is no specific topological strategy in such a type of evolution.  Considering the \cpl metric, the increase in the connectivity is beneficial. The samples on average reduce the \cpl by 33\% after the doubling of the \textit{size} of the graph. However, we note some variability: some samples improve more than 50\%, while for the biggest sample (i.e., sample \#7) the improvement is just 20\%. Even in this type of evolution the tendency is that of saturation with further steps in connectivity that are less beneficial. The clustering coefficient evolution for such  strategy provides benefits that are not more than the three orders of magnitude. The benefits obtained are comparable by those achieved with the triangle closure strategy at least in the first two stages of evolution; when more links are added the nodes that are connected are less and less belonging to the same neighborhood (i.e., nodes with longer distances are added), and this is the main difference at the final stage of evolution compared to the triangle closure strategy.  For the metric concerning robustness, the average over all samples gives a doubling of the metric compared to its original value. The greatest improvement in robustness is achieved after the first two steps of the evolution (about 75\% improvement), a lower increase is achieved in the following two steps (about 30\%). This strategy of evolution benefits both the random and targeted attacks towards nodes. In particular, in the forth stage of the evolution the components of the robustness metric (cf.~\ref{sec:metrics}) have almost the same value, actually for some samples it is more damaging the random attack than the attack targeting the most connected nodes. Such situation is represented in Figures~\ref{fig:rndlea} and~\ref{fig:tarlea}.
Considering the values for the redundant path robustness, the same considerations done for the \cpl apply. The overall reduction in the redundant \apl is about 50\% compared to the initial value of the samples. Just the first evolution step provides a reduction of more than 30\%. Even in this path-related metric we see a certain variability by samples that have consistent variations (e.g., sample \#3), and samples whose improvement are limited (e.g., sample \#7).

Table~\ref{tab:mvDistBet} contains for each sample of the \MV \G (column one) the values of metrics related to betweenness. In addition to \textit{order} and \textit{size} (columns three and four), average betweenness is provided in columns five, while a value of average betweenness normalized by the \textit{order} of the graph is computed in column six in order to compare the different samples. A measure of the statistical variation of betweenness is the coefficient of variation which is shown in the seventh column.

The strategy of adding edges between nodes that are geographically closest provides small benefits in terms of reduction of betweenness compared to the other strategies proposed. Actually, in the first step of the evolution there are four samples (samples \#2, \#3, \#9 and \#12) whose betweenness increases. In the last step of the evolution, over all the samples, the reduction is about 20\%, despite samples \#4, \#9 and \#12 that have an increment in the betweenness metric. The same trend is followed by the normalized value of average betweenness divided by the \textit{order} of the network; in the last step of evolution there are three sample which have a betweenness to \textit{order} ratio which is higher than 9. Considering the variability of betweenness, we see a general increase in the coefficient of variation. Only sample \#5 and sample \#8 show a decrease for such a metric. This strategy of adding the connections does not clearly have a strategy from the topology point of view. Therefore, it makes sense that there is not a clear trend in the results of betweenness since every sample has different characteristics in the physical distance between the nodes (substations in the network).

\begin{table}[htbp]
\begin{footnotesize}
\begin{center}
\begin{tabular}{|@p{1.7cm}|^p{1.7cm}|^p{0.9cm}|^p{0.6cm}|^p{1cm}|^p{1cm}|^p{1.2cm}|^p{1.5cm}|^p{2cm}|}
\hline
\rowstyle{\bfseries} Sample ID &Network type  & \textit{Order} & \textit{Size} & {Avg. deg.} & {CPL} & {CC} & {Removal robustness ($Rob_N$)} & {Redundancy cost ($APL_{10^{th}}$)} \\ \hline
\hline

\multicolumn{9}{|c|}{\textbf{original \textit{order} +25\%}} \\ \hline
1 & MV & 444 & 607 & 2.734 & 9.536 & 0.13338 & 0.293 & 12.753 \\ \hline
2 & MV & 472 & 632 & 2.678 & 14.726 & 0.14207 & 0.191 & 17.222 \\ \hline
3 & MV & 238 & 306 & 2.571 & 10.395 & 0.14795 & 0.236 & 13.316 \\ \hline
4 & MV & 263 & 360 & 2.738 & 10.431 & 0.15275 & 0.252 & 13.923 \\ \hline
5 & MV & 217 & 286 & 2.636 & 8.815 & 0.08661 & 0.308 & 11.384 \\ \hline
6 & MV & 191 & 258 & 2.702 & 8.132 & 0.11575 & 0.289 & 10.958 \\ \hline
7 & MV & 884 & 1323 & 2.993 & 8.527 & 0.12721 & 0.398 & 10.388 \\ \hline
8 & MV & 366 & 477 & 2.607 & 12.838 & 0.14166 & 0.184 & 16.860 \\ \hline
9 & MV & 218 & 290 & 2.661 & 9.606 & 0.13026 & 0.242 & 12.764 \\ \hline
10 & MV & 201 & 255 & 2.537 & 12.570 & 0.13361 & 0.204 & 15.755 \\ \hline
11 & MV & 202 & 266 & 2.634 & 9.284 & 0.12775 & 0.286 & 13.146 \\ \hline
%12 & MV & 25 & 30 & 2.400 & 4.583 & 0.27600 & 0.371 & 7.712 \\ \hline
12 & MV & 464 & 623 & 2.685 & 10.884 & 0.11793 & 0.278 & 13.828 \\ \hline \hline
\multicolumn{9}{|c|}{\textbf{original \textit{order} +50\%}} \\ \hline

1 & MV & 444 & 729 & 3.284 & 8.647 & 0.21344 & 0.404 & 10.725 \\ \hline
2 & MV & 472 & 759 & 3.216 & 13.190 & 0.19715 & 0.289 & 15.654 \\ \hline
3 & MV & 238 & 367 & 3.084 & 9.876 & 0.20415 & 0.276 & 11.840 \\ \hline
4 & MV & 263 & 432 & 3.285 & 9.565 & 0.22611 & 0.388 & 11.919 \\ \hline
5 & MV & 217 & 343 & 3.161 & 8.352 & 0.16083 & 0.342 & 10.662 \\ \hline
6 & MV & 191 & 310 & 3.246 & 7.768 & 0.15339 & 0.329 & 10.455 \\ \hline
7 & MV & 884 & 1588 & 3.593 & 8.097 & 0.18306 & 0.463 & 9.674 \\ \hline
8 & MV & 366 & 573 & 3.131 & 12.177 & 0.19587 & 0.224 & 15.782 \\ \hline
9 & MV & 218 & 348 & 3.193 & 9.184 & 0.20451 & 0.301 & 12.243 \\ \hline
10 & MV & 201 & 306 & 3.045 & 8.985 & 0.17791 & 0.414 & 11.746 \\ \hline
11 & MV & 202 & 319 & 3.158 & 8.296 & 0.18381 & 0.397 & 10.626 \\ \hline
%12 & MV & 25 & 36 & 2.880 & 3.708 & 0.47200 & 0.463 & 8.038 \\ \hline
12 & MV & 464 & 748 & 3.224 & 10.014 & 0.20459 & 0.386 & 12.681 \\ \hline \hline
\multicolumn{9}{|c|}{\textbf{original \textit{order} +75\%}} \\ \hline

1 & MV & 444 & 850 & 3.829 & 8.336 & 0.27250 & 0.468 & 10.525 \\ \hline
2 & MV & 472 & 885 & 3.750 & 12.644 & 0.25483 & 0.333 & 13.812 \\ \hline
3 & MV & 238 & 428 & 3.597 & 9.120 & 0.23636 & 0.346 & 11.277 \\ \hline
4 & MV & 263 & 504 & 3.833 & 8.912 & 0.28127 & 0.442 & 10.163 \\ \hline
5 & MV & 217 & 400 & 3.687 & 7.968 & 0.23789 & 0.416 & 10.177 \\ \hline
6 & MV & 191 & 362 & 3.791 & 6.979 & 0.22299 & 0.451 & 9.251 \\ \hline
7 & MV & 884 & 1853 & 4.192 & 7.849 & 0.24197 & 0.581 & 9.411 \\ \hline
8 & MV & 366 & 668 & 3.650 & 11.858 & 0.23799 & 0.357 & 14.784 \\ \hline
9 & MV & 218 & 406 & 3.725 & 8.760 & 0.25211 & 0.497 & 11.866 \\ \hline
10 & MV & 201 & 357 & 3.552 & 8.145 & 0.20495 & 0.488 & 10.736 \\ \hline
11 & MV & 202 & 372 & 3.683 & 7.326 & 0.21618 & 0.517 & 9.444 \\ \hline
%12 & MV & 25 & 42 & 3.360 & 3.417 & 0.48476 & 0.616 & 5.910 \\ \hline
12 & MV & 464 & 873 & 3.763 & 9.635 & 0.25206 & 0.400 & 11.813 \\ \hline \hline
\multicolumn{9}{|c|}{\textbf{original \textit{order} +100\%}} \\ \hline

1 & MV & 444 & 972 & 4.378 & 8.023 & 0.31095 & 0.497 & 9.485 \\ \hline
2 & MV & 472 & 1012 & 4.288 & 11.792 & 0.31725 & 0.461 & 13.000 \\ \hline
3 & MV & 238 & 490 & 4.118 & 8.842 & 0.26450 & 0.498 & 11.043 \\ \hline
4 & MV & 263 & 576 & 4.380 & 8.588 & 0.33434 & 0.447 & 9.674 \\ \hline
5 & MV & 217 & 458 & 4.221 & 7.505 & 0.29041 & 0.458 & 9.504 \\ \hline
6 & MV & 191 & 414 & 4.335 & 6.574 & 0.28652 & 0.470 & 8.522 \\ \hline
7 & MV & 884 & 2118 & 4.792 & 7.565 & 0.29363 & 0.713 & 9.048 \\ \hline
8 & MV & 366 & 764 & 4.175 & 11.340 & 0.28634 & 0.441 & 14.256 \\ \hline
9 & MV & 218 & 464 & 4.257 & 8.233 & 0.29032 & 0.564 & 10.460 \\ \hline
10 & MV & 201 & 408 & 4.060 & 7.605 & 0.23156 & 0.575 & 9.656 \\ \hline
11 & MV & 202 & 426 & 4.218 & 6.963 & 0.26151 & 0.676 & 9.056 \\ \hline
%12 & MV & 25 & 48 & 3.840 & 2.917 & 0.50476 & 0.505 & 5.051 \\ \hline
12 & MV & 464 & 998 & 4.302 & 9.472 & 0.28112 & 0.493 & 11.417 \\ \hline
\end{tabular}
\caption{Metrics for least distance strategy \MV samples evolution.}\label{tab:mvDist}
\end{center}
\end{footnotesize}
\end{table}

\begin{center}
\begin{table}[htbp]
\centering
\begin{footnotesize}
\begin{tabular}{|@p{1.7cm}|^p{1.5cm}|^p{0.8cm}|^p{0.8cm}|^p{1.5cm}|^p{1.5cm}|^p{1.2cm}|}
\hline
\rowstyle{\bfseries}
Sample  ID  &Network type &\textit{Order} & \textit{Size} & {Avg. betweenness} & {Avg. betw/order} & {Coeff. variation} \\ \hline
\hline
\multicolumn{7}{|c|}{\textbf{original \textit{order} +25\%}} \\ \hline
1 & MV & 444 & 607 & 3557.62 & 8.013 & 2.148 \\ \hline
2 & MV & 472 & 632 & 6274.899 & 13.294 & 1.704 \\ \hline
3 & MV & 238 & 306 & 2137.507 & 8.981 & 1.587 \\ \hline
4 & MV & 263 & 360 & 2297.157 & 8.734 & 1.419 \\ \hline
5 & MV & 217 & 286 & 1721.047 & 7.931 & 1.6 \\ \hline
6 & MV & 191 & 258 & 1416.741 & 7.417 & 1.923 \\ \hline
7 & MV & 884 & 1323 & 6810.632 & 7.704 & 2.915 \\ \hline
8 & MV & 366 & 477 & 4475.009 & 12.227 & 1.566 \\ \hline
9 & MV & 218 & 290 & 1840.431 & 8.442 & 1.547 \\ \hline
10 & MV & 201 & 255 & 2398.327 & 11.932 & 1.275 \\ \hline
11 & MV & 202 & 266 & 1741.949 & 8.624 & 1.368 \\ \hline
%12 & MV & 25 & 30 & 96.72 & 3.869 & 1.088 \\ \hline
12 & MV & 464 & 623 & 4814.58 & 10.376 & 2.003 \\ \hline \hline
\multicolumn{7}{|c|}{\textbf{original \textit{order} +50\%}} \\ \hline

1 & MV & 444 & 729 & 3221.546 & 7.256 & 2.191 \\ \hline
2 & MV & 472 & 759 & 5636.531 & 11.942 & 1.758 \\ \hline
3 & MV & 238 & 367 & 1999.372 & 8.401 & 1.577 \\ \hline
4 & MV & 263 & 432 & 2120.025 & 8.061 & 1.495 \\ \hline
5 & MV & 217 & 343 & 1627.028 & 7.498 & 1.638 \\ \hline
6 & MV & 191 & 310 & 1344.032 & 7.037 & 1.993 \\ \hline
7 & MV & 884 & 1588 & 6475.508 & 7.325 & 3.024 \\ \hline
8 & MV & 366 & 573 & 4287.187 & 11.714 & 1.552 \\ \hline
9 & MV & 218 & 348 & 1768.297 & 8.111 & 1.571 \\ \hline
10 & MV & 201 & 306 & 1655.439 & 8.236 & 1.305 \\ \hline
11 & MV & 202 & 319 & 1539.98 & 7.624 & 1.374 \\ \hline
%12 & MV & 25 & 36 & 67.28 & 2.691 & 1.117 \\ \hline
12 & MV & 464 & 748 & 4523.666 & 9.749 & 2.081 \\ \hline
 \hline
\multicolumn{7}{|c|}{\textbf{original \textit{order} +50\%}} \\ \hline
1 & MV & 444 & 850 & 3068.074 & 6.91 & 2.263 \\ \hline
2 & MV & 472 & 885 & 5388.36 & 11.416 & 1.783 \\ \hline
3 & MV & 238 & 428 & 1905.408 & 8.006 & 1.646 \\ \hline
4 & MV & 263 & 504 & 1956.884 & 7.441 & 1.41 \\ \hline
5 & MV & 217 & 400 & 1544.364 & 7.117 & 1.68 \\ \hline
6 & MV & 191 & 362 & 1239.704 & 6.491 & 2.067 \\ \hline
7 & MV & 884 & 1853 & 6210.444 & 7.025 & 3.074 \\ \hline
8 & MV & 366 & 668 & 4115.775 & 11.245 & 1.58 \\ \hline
9 & MV & 218 & 406 & 1687.799 & 7.742 & 1.625 \\ \hline
10 & MV & 201 & 357 & 1535.122 & 7.637 & 1.344 \\ \hline
11 & MV & 202 & 372 & 1358.508 & 6.725 & 1.423 \\ \hline
%12 & MV & 25 & 42 & 59.68 & 2.387 & 1.21 \\ \hline
12 & MV & 464 & 873 & 4329.609 & 9.331 & 2.14 \\ \hline
 \hline
\multicolumn{7}{|c|}{\textbf{original \textit{order} +50\%}} \\ \hline
1 & MV & 444 & 972 & 2937.585 & 6.616 & 2.309 \\ \hline
2 & MV & 472 & 1012 & 5016.908 & 10.629 & 1.896 \\ \hline
3 & MV & 238 & 490 & 1832.179 & 7.698 & 1.651 \\ \hline
4 & MV & 263 & 576 & 1852.488 & 7.044 & 1.484 \\ \hline
5 & MV & 217 & 458 & 1467.346 & 6.762 & 1.744 \\ \hline
6 & MV & 191 & 414 & 1168.878 & 6.12 & 2.098 \\ \hline
7 & MV & 884 & 2118 & 5956.625 & 6.738 & 3.085 \\ \hline
8 & MV & 366 & 764 & 3973.55 & 10.857 & 1.583 \\ \hline
9 & MV & 218 & 464 & 1602.22 & 7.35 & 1.689 \\ \hline
10 & MV & 201 & 408 & 1449.816 & 7.213 & 1.29 \\ \hline
11 & MV & 202 & 426 & 1296.051 & 6.416 & 1.408 \\ \hline
%12 & MV & 25 & 48 & 50.32 & 2.013 & 1.152 \\ \hline
12 & MV & 464 & 998 & 4229.495 & 9.115 & 2.099 \\ \hline
\end{tabular}
\caption{Betweenness for least distance strategy \MV samples evolution.}\label{tab:mvDistBet}
\end{footnotesize}
\end{table}
\end{center}

\begin{figure}
 \captionsetup{type=figure}
    \centering
    \subfloat[Physical sample.]{\label{fig:rndLea}\includegraphics[scale=.23]{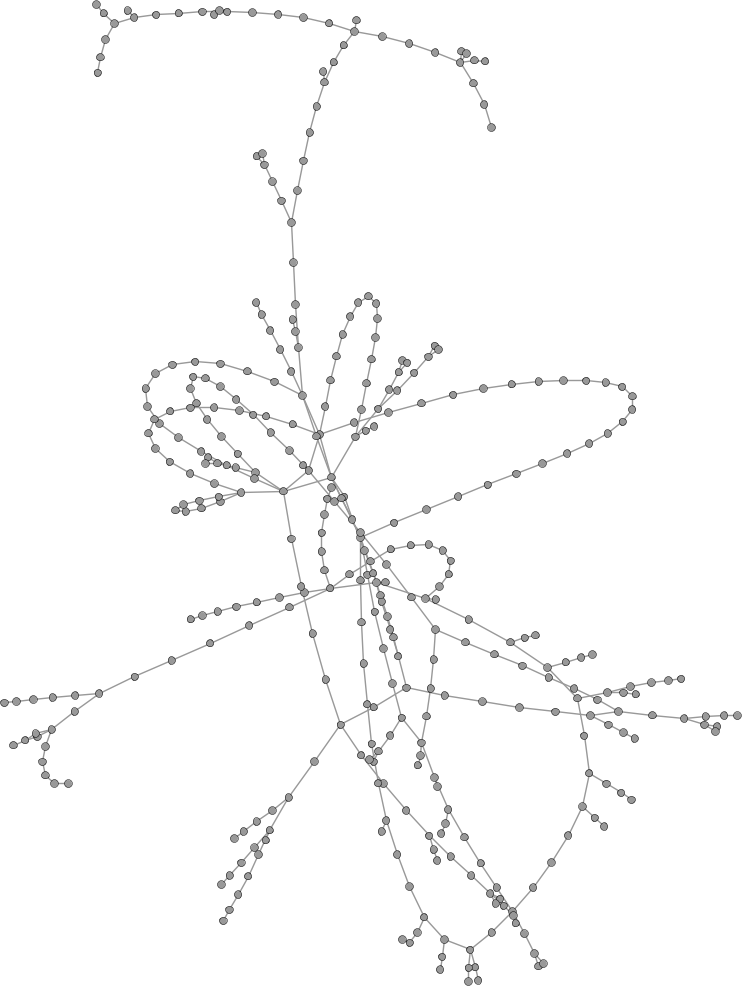}} 
    \subfloat[1$^{st}$ stage of evolution (i.e., +25\% edges).]{\label{fig:ecoRel251}\includegraphics[scale=.23]{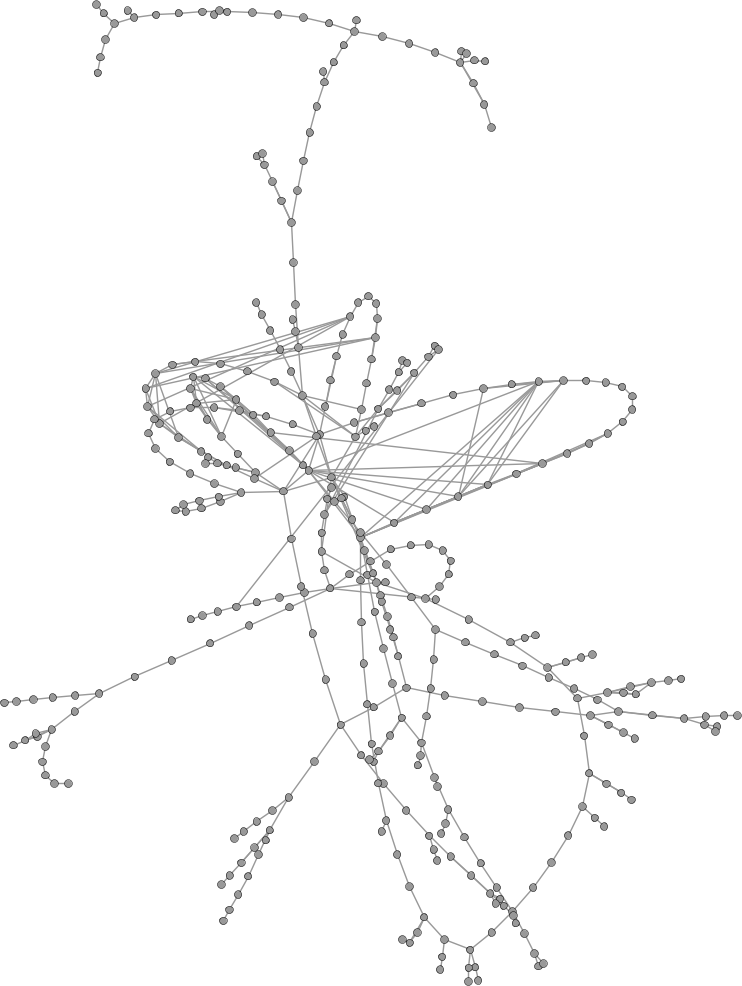}}\\
     \hspace*{-1.8cm}\subfloat[2$^{nd}$ stage of evolution (i.e., +50\% edges)]{\label{fig:ecoRel252}\includegraphics[scale=.23]{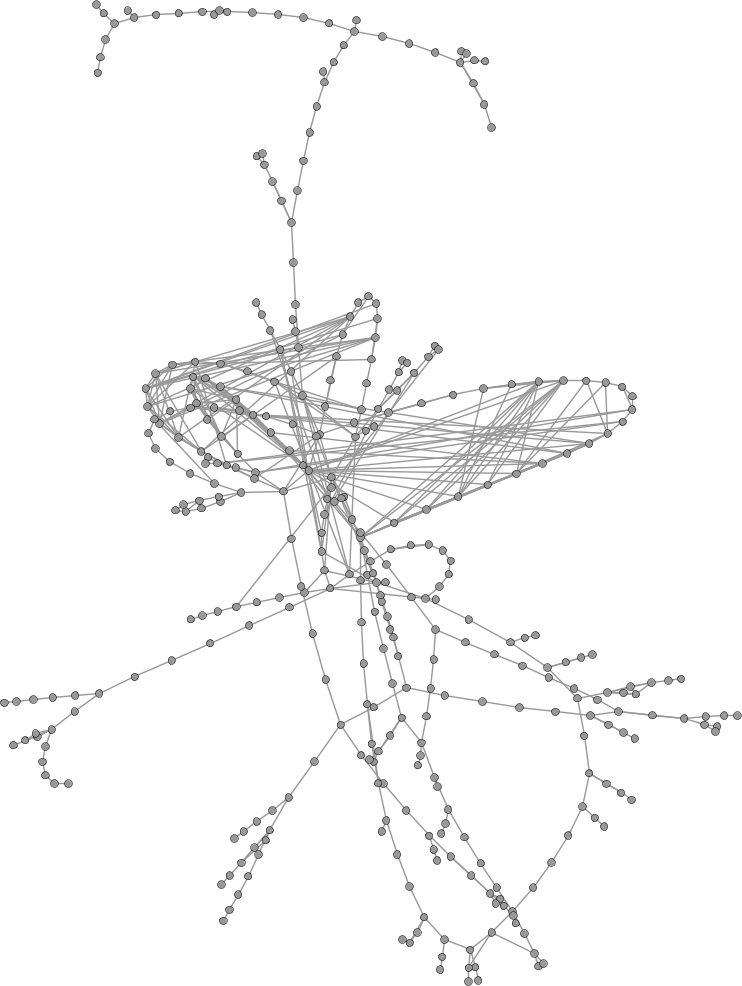}}  
      \subfloat[3$^{rd}$ stage of evolution (i.e., +75\% edges).]{\label{fig:ecoRel253}\includegraphics[scale=.23]{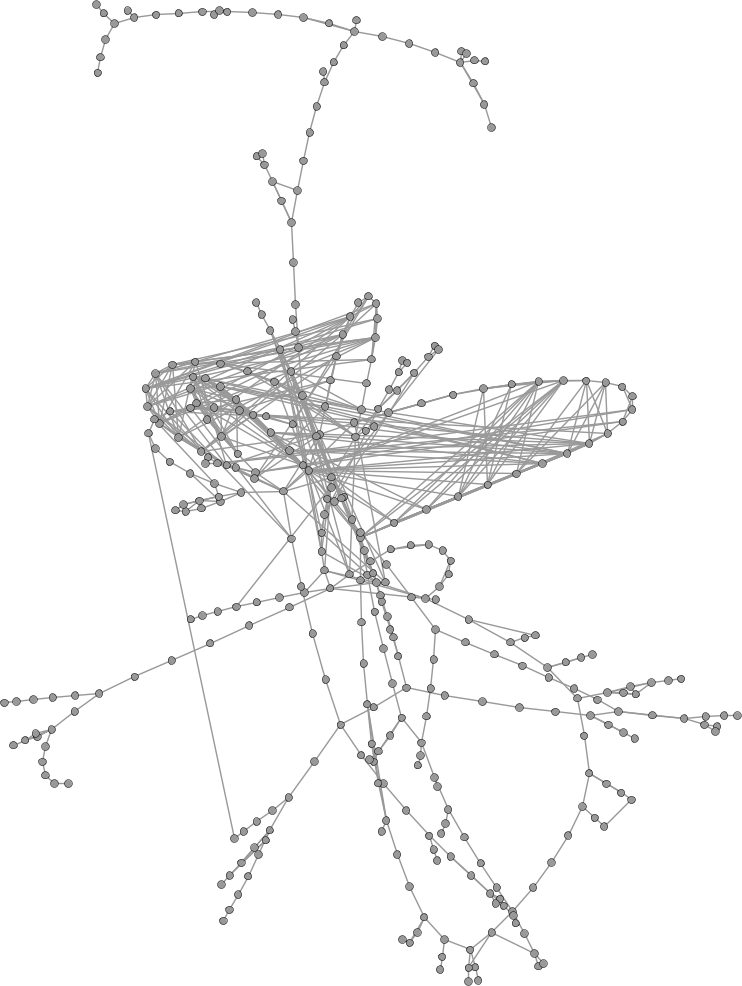}} 
       \subfloat[4$^{th}$ stage of evolution (i.e., +100\% edges).]{\label{fig:ecoRel254}\includegraphics[scale=.23]{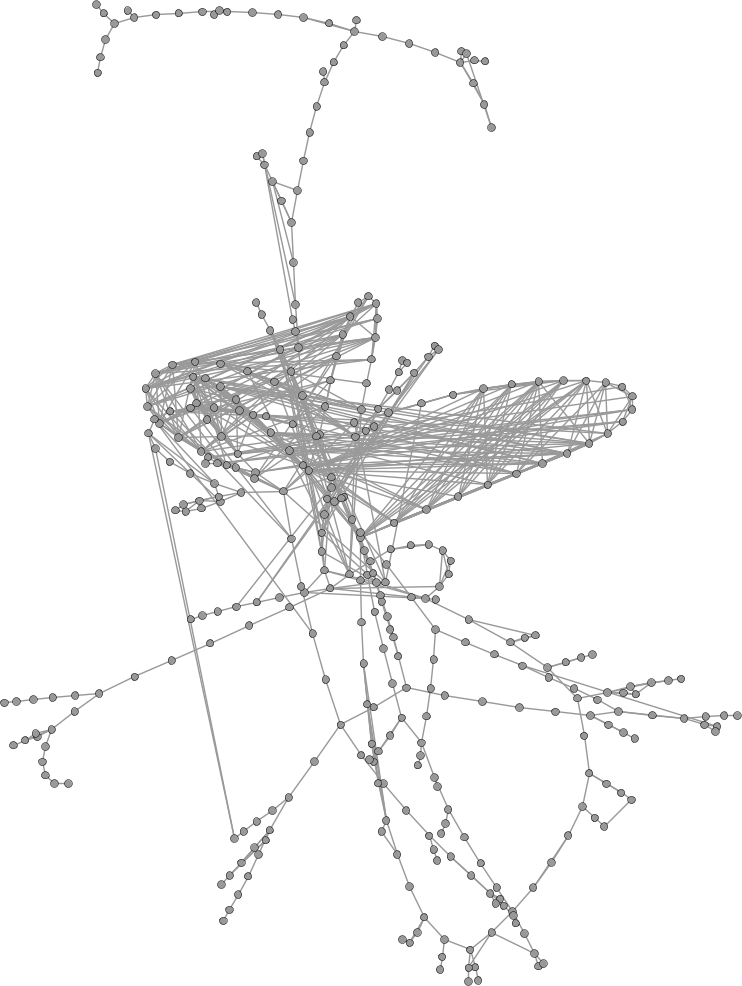}}  %\\
    \caption{Stages of evolution of \MV sample \#8 following  least distance strategy.}
    \label{fig:evolchartLea}
\end{figure}

\subsubsection*{Random evolution}

Table~\ref{tab:mvRnd} contains for each sample of the \MV \G (column one) the values for main topological quantities: \textit{order} and \textit{size} in columns three and four, respectively; average node degree in the fifth column; the \cpl is reported in column six; the clustering coefficient follows in column seven; robustness is shown in eight column; the cost in term of redundant path length closes the data series (column nine).

\begin{figure}[htbp]
 %\begin{minipage}[htbp]{7cm}
   \centering   
  \includegraphics[width=0.7\textwidth]{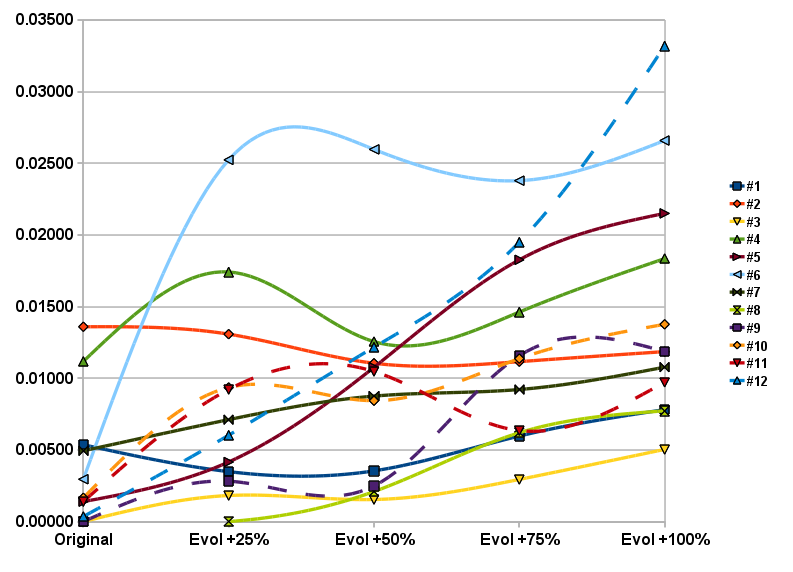}
   \caption{Evolution of the \cc metric.}
\label{fig:ccRndEvolMV}
\end{figure}

The random evolution strategy is mainly intended to consider the difference between evolutions driven by a specific target or property of nodes and an evolution driven by mere random picking of nodes.   Considering the \cpl metric the increase in the connectivity with such a strategy is extremely beneficial. The samples score for this metric below 5 after the third evolution step, with the exception of sample \#12 which scores 5.014. The last step of connectivity addition brings the total benefit for reduction in \cpl to a value higher than 65\%. The clustering coefficient evolution for such strategy provides limited benefits. In fact, to improve the clustering a special addition scheme needs to be followed targeting the neighbors of a node. The situation varies between the samples, but the majority of them improve in the \cc by one order of magnitude when the edges are double the initial number. Figure~\ref{fig:ccRndEvolMV} shows that, although a general improvement of the metric, there is not a common trend or tendency for all the samples: some reach a peak in intermediate stages of evolution (e.g., sample \#4), others a rapid increase (e.g., sample \#12), others have a slightly decreasing (e.g., sample \#2) or monotonously increasing (e.g., sample \#7) clustering coefficient. Robustness is the metric that improves the most. Values for this metric of 0.7 are obtained already after the third evolution step. Edge addition following this strategy provides an increase in both the random and targeted component of this metric. In particular, the random addition of nodes tends to make the whole network more robust such that the effects of the random and targeted attack towards nodes have the same effect on the network. All the samples reach a value close to 0.75 that seems an upper bound in such conditions of \textit{order} and \textit{size}; we already noted such an upper bound behavior for various types of synthetic topologies in our previous work~\cite{PaganiEvol2013}.
Considering the values for the redundant path robustness the same considerations done for the \cpl apply. Especially, the rates of reduction (in percentage) of the $10^{th}$ \apl at each evolution step are exactly the same as those shown by the \cpl. When the number of edges is doubled the redundant path length for all the samples stays below 6.5 which is just 2 hops higher than the \cpl of the network. 
%The same exception mentioned in the previous subsection still remains for sample \#12 which shows limited improvement from the evolution process, and actually the value for the redundant path increased in in the first and second step of the evolution compared to the initial value. Such increase is due to the problems that affect this metric when small networks (in terms of \textit{order} and \textit{size}) are considered since they do not have redundant path between nodes; such redundant path emerge when more connectivity is added to the network.

Table~\ref{tab:mvRndBet} contains for each sample of the \MV \G (column one) the values of metrics related to betweenness. In addition to \textit{order} and \textit{size} (columns three and four), average betweenness is provided in columns five, while a value of average betweenness normalized by the \textit{order} of the graph is computed in column six in order to compare the different samples. A measure of the statistical variation of betweenness is the coefficient of variation which is shown in the seventh column.

Random addition of edges in the networks provides the highest improvement in the betweenness metric. Average betweenness reduces to about 67\% in the final step of evolution. Already from the first step of the network evolution, the betweenness on average reduces to slightly more than 40\% of the original value. The only exception is sample \#4 that shows an increment in the first evolution stage. The same trend is followed by the normalized value of average betweenness divided by the \textit{order} of the network; already in the first step only two samples exceeds six, and in the final stage no sample exceeds 3.6. Considering the variability of betweenness, in each of the samples we note a decrease for this metric. In particular, in the last stage of the evolution, all samples for the coefficient of variation are below the unity with the exception of sample \#7 that scores just below 1.1. The random addition of edges tend to even split the betweenness among all the nodes of the network and this is shown by the sharp fall of standard deviation of betweenness which is responsible for a coefficient of variation that is below the unit in the last stage of the evolution. However, the highest fraction of nodes for all the samples has zero betweenness, the median in each step for each sample is zero.

\begin{table}[h]
\begin{footnotesize}
\begin{center}
\begin{tabular}{|@p{1.7cm}|^p{1.7cm}|^p{0.9cm}|^p{0.6cm}|^p{1cm}|^p{1cm}|^p{1.2cm}|^p{1.5cm}|^p{2cm}|}
\hline
\rowstyle{\bfseries} Sample ID &Network type  & \textit{Order} & \textit{Size} & {Avg. deg.} & {CPL} & {CC} & {Removal robustness ($Rob_N$)} & {Redundancy cost ($APL_{10^{th}}$)} \\ \hline
\hline

\multicolumn{9}{|c|}{\textbf{original \textit{order} +25\%}} \\ \hline
1 & MV & 444 & 607 & 2.734 & 6.672 & 0.00349 & 0.402 & 9.997 \\ \hline
2 & MV & 472 & 632 & 2.678 & 6.945 & 0.01308 & 0.375 & 10.496 \\ \hline
3 & MV & 238 & 306 & 2.571 & 6.679 & 0.00182 & 0.423 & 10.775 \\ \hline
4 & MV & 263 & 360 & 2.738 & 6.065 & 0.01742 & 0.418 & 9.945 \\ \hline
5 & MV & 217 & 286 & 2.636 & 6.361 & 0.00416 & 0.409 & 10.384 \\ \hline
6 & MV & 191 & 258 & 2.702 & 5.726 & 0.02524 & 0.387 & 9.448 \\ \hline
7 & MV & 884 & 1323 & 2.993 & 6.557 & 0.00711 & 0.572 & 9.231 \\ \hline
8 & MV & 366 & 477 & 2.607 & 6.919 & 0.00000 & 0.381 & 10.944 \\ \hline
9 & MV & 218 & 290 & 2.661 & 6.014 & 0.00282 & 0.387 & 9.237 \\ \hline
10 & MV & 201 & 255 & 2.537 & 6.595 & 0.00936 & 0.433 & 11.033 \\ \hline
11 & MV & 202 & 266 & 2.634 & 5.920 & 0.00922 & 0.408 & 9.705 \\ \hline
%12 & MV & 25 & 30 & 2.400 & 3.792 & 0.04667 & 0.526 & 8.147 \\ \hline
12 & MV & 464 & 623 & 2.685 & 6.996 & 0.00603 & 0.374 & 10.053 \\ \hline \hline
\multicolumn{9}{|c|}{\textbf{original \textit{order} +50\%}} \\ \hline
1 & MV & 444 & 729 & 3.284 & 5.367 & 0.00354 & 0.694 & 7.872 \\ \hline
2 & MV & 472 & 759 & 3.216 & 5.525 & 0.01105 & 0.684 & 8.243 \\ \hline
3 & MV & 238 & 367 & 3.084 & 5.135 & 0.00154 & 0.644 & 8.086 \\ \hline
4 & MV & 263 & 432 & 3.285 & 4.958 & 0.01256 & 0.704 & 7.587 \\ \hline
5 & MV & 217 & 343 & 3.161 & 4.944 & 0.01076 & 0.629 & 7.701 \\ \hline
6 & MV & 191 & 310 & 3.246 & 4.637 & 0.02597 & 0.645 & 7.277 \\ \hline
7 & MV & 884 & 1588 & 3.593 & 5.472 & 0.00876 & 0.728 & 7.716 \\ \hline
8 & MV & 366 & 573 & 3.131 & 5.519 & 0.00209 & 0.634 & 8.431 \\ \hline
9 & MV & 218 & 348 & 3.193 & 4.714 & 0.00249 & 0.509 & 7.299 \\ \hline
10 & MV & 201 & 306 & 3.045 & 5.120 & 0.00843 & 0.657 & 8.385 \\ \hline
11 & MV & 202 & 319 & 3.158 & 4.808 & 0.01047 & 0.607 & 7.328 \\ \hline
%12 & MV & 25 & 36 & 2.880 & 2.958 & 0.10133 & 0.491 & 6.577 \\ \hline
12 & MV & 464 & 748 & 3.224 & 5.711 & 0.01216 & 0.622 & 8.642 \\ \hline \hline
\multicolumn{9}{|c|}{\textbf{original \textit{order} +75\%}} \\ \hline
1 & MV & 444 & 850 & 3.829 & 4.729 & 0.00597 & 0.757 & 6.932 \\ \hline
2 & MV & 472 & 885 & 3.750 & 4.855 & 0.01116 & 0.736 & 7.102 \\ \hline
3 & MV & 238 & 428 & 3.597 & 4.498 & 0.00294 & 0.702 & 6.920 \\ \hline
4 & MV & 263 & 504 & 3.833 & 4.328 & 0.01462 & 0.749 & 6.704 \\ \hline
5 & MV & 217 & 400 & 3.687 & 4.366 & 0.01826 & 0.734 & 6.597 \\ \hline
6 & MV & 191 & 362 & 3.791 & 4.089 & 0.02379 & 0.722 & 6.595 \\ \hline
7 & MV & 884 & 1853 & 4.192 & 4.909 & 0.00921 & 0.777 & 6.788 \\ \hline
8 & MV & 366 & 668 & 3.650 & 4.768 & 0.00620 & 0.743 & 7.194 \\ \hline
9 & MV & 218 & 406 & 3.725 & 4.270 & 0.01157 & 0.710 & 6.463 \\ \hline
10 & MV & 201 & 357 & 3.552 & 4.480 & 0.01137 & 0.744 & 6.937 \\ \hline
11 & MV & 202 & 372 & 3.683 & 4.221 & 0.00635 & 0.698 & 6.672 \\ \hline
%12 & MV & 25 & 42 & 3.360 & 2.583 & 0.11067 & 0.760 & 5.455 \\ \hline
12 & MV & 464 & 873 & 3.763 & 5.014 & 0.01949 & 0.748 & 7.244 \\ \hline \hline
\multicolumn{9}{|c|}{\textbf{original \textit{order} +100\%}} \\ \hline
1 & MV & 444 & 972 & 4.378 & 4.341 & 0.00781 & 0.769 & 6.345 \\ \hline
2 & MV & 472 & 1012 & 4.288 & 4.468 & 0.01186 & 0.769 & 6.440 \\ \hline
3 & MV & 238 & 490 & 4.118 & 4.025 & 0.00504 & 0.761 & 6.230 \\ \hline
4 & MV & 263 & 576 & 4.380 & 3.966 & 0.01835 & 0.765 & 5.898 \\ \hline
5 & MV & 217 & 458 & 4.221 & 3.949 & 0.02150 & 0.767 & 5.890 \\ \hline
6 & MV & 191 & 414 & 4.335 & 3.726 & 0.02659 & 0.774 & 5.764 \\ \hline
7 & MV & 884 & 2118 & 4.792 & 4.542 & 0.01077 & 0.786 & 6.412 \\ \hline
8 & MV & 366 & 764 & 4.175 & 4.349 & 0.00773 & 0.753 & 6.424 \\ \hline
9 & MV & 218 & 464 & 4.257 & 3.864 & 0.01187 & 0.740 & 5.923 \\ \hline
10 & MV & 201 & 408 & 4.060 & 3.965 & 0.01376 & 0.764 & 6.254 \\ \hline
11 & MV & 202 & 426 & 4.218 & 3.866 & 0.00972 & 0.760 & 6.050 \\ \hline
%12 & MV & 25 & 48 & 3.840 & 2.375 & 0.17638 & 0.752 & 4.910 \\ \hline
12 & MV & 464 & 998 & 4.302 & 4.559 & 0.03318 & 0.773 & 6.495 \\ \hline
\end{tabular}
\caption{Metrics for random strategy \MV samples evolution.}\label{tab:mvRnd}
\end{center}
\end{footnotesize}
\end{table}

\begin{center}
\begin{table}[h]
\centering
\begin{footnotesize}
\begin{tabular}{|@p{1.7cm}|^p{1.5cm}|^p{0.8cm}|^p{0.8cm}|^p{1.5cm}|^p{1.5cm}|^p{1.2cm}|}
\hline
\rowstyle{\bfseries}
Sample  ID  &Network type &\textit{Order} & \textit{Size} & {Avg. betweenness} & {Avg. betw/order} & {Coeff. variation} \\ \hline
\hline
\multicolumn{7}{|c|}{\textbf{original \textit{order} +25\%}} \\ \hline
1 & MV & 444 & 607 & 2259.57 & 5.089 & 1.22 \\ \hline
2 & MV & 472 & 632 & 2866.64 & 6.073 & 1.187 \\ \hline
3 & MV & 238 & 306 & 1202.027 & 5.051 & 1.139 \\ \hline
4 & MV & 263 & 360 & 1361.074 & 5.175 & 1.067 \\ \hline
5 & MV & 217 & 286 & 1210.579 & 5.579 & 1.232 \\ \hline
6 & MV & 191 & 258 & 946.963 & 4.958 & 1.193 \\ \hline
7 & MV & 884 & 1323 & 4945.671 & 5.595 & 1.722 \\ \hline
8 & MV & 366 & 477 & 2153.055 & 5.883 & 1.306 \\ \hline
9 & MV & 218 & 290 & 1130.838 & 5.187 & 1.201 \\ \hline
10 & MV & 201 & 255 & 1190.609 & 5.923 & 1.016 \\ \hline
11 & MV & 202 & 266 & 981.574 & 4.859 & 1.145 \\ \hline
%12 & MV & 25 & 30 & 70.72 & 2.829 & 0.887 \\ \hline
12 & MV & 464 & 623 & 2897.982 & 6.246 & 1.204 \\ \hline \hline
\multicolumn{7}{|c|}{\textbf{original \textit{order} +50\%}} \\ \hline
1 & MV & 444 & 729 & 1825.762 & 4.112 & 1.029 \\ \hline
2 & MV & 472 & 759 & 2145.374 & 4.545 & 1.068 \\ \hline
3 & MV & 238 & 367 & 937.829 & 3.94 & 1.188 \\ \hline
4 & MV & 263 & 432 & 1026.759 & 3.904 & 0.956 \\ \hline
5 & MV & 217 & 343 & 877.673 & 4.045 & 1.185 \\ \hline
6 & MV & 191 & 310 & 696.794 & 3.648 & 1.094 \\ \hline
7 & MV & 884 & 1588 & 3985.532 & 4.509 & 1.455 \\ \hline
8 & MV & 366 & 573 & 1692.859 & 4.625 & 1.014 \\ \hline
9 & MV & 218 & 348 & 819.162 & 3.758 & 1.204 \\ \hline
10 & MV & 201 & 306 & 842.223 & 4.19 & 0.897 \\ \hline
11 & MV & 202 & 319 & 777.015 & 3.847 & 0.957 \\ \hline
%12 & MV & 25 & 36 & 50.08 & 2.003 & 0.858 \\ \hline
12 & MV & 464 & 748 & 2249.061 & 4.847 & 1.058 \\ \hline
\hline
\multicolumn{7}{|c|}{\textbf{original \textit{order} +75\%}} \\ \hline
1 & MV & 444 & 850 & 1582.139 & 3.563 & 1.021 \\ \hline
2 & MV & 472 & 885 & 1830.748 & 3.879 & 0.919 \\ \hline
3 & MV & 238 & 428 & 835.813 & 3.512 & 0.968 \\ \hline
4 & MV & 263 & 504 & 864.724 & 3.288 & 0.883 \\ \hline
5 & MV & 217 & 400 & 726.953 & 3.35 & 1.006 \\ \hline
6 & MV & 191 & 362 & 599.589 & 3.139 & 0.972 \\ \hline
7 & MV & 884 & 1853 & 3465.824 & 3.921 & 1.275 \\ \hline
8 & MV & 366 & 668 & 1393.805 & 3.808 & 0.929 \\ \hline
9 & MV & 218 & 406 & 719.314 & 3.3 & 1.082 \\ \hline
10 & MV & 201 & 357 & 698.497 & 3.475 & 0.856 \\ \hline
11 & MV & 202 & 372 & 647.624 & 3.206 & 0.839 \\ \hline
%12 & MV & 25 & 42 & 38.96 & 1.558 & 0.848 \\ \hline
12 & MV & 464 & 873 & 1914.675 & 4.126 & 0.939 \\ \hline
\hline
\multicolumn{7}{|c|}{\textbf{original \textit{order} +100\%}} \\ \hline
1 & MV & 444 & 972 & 1433.794 & 3.229 & 0.921 \\ \hline
2 & MV & 472 & 1012 & 1626.893 & 3.447 & 0.84 \\ \hline
3 & MV & 238 & 490 & 725.687 & 3.049 & 0.875 \\ \hline
4 & MV & 263 & 576 & 752.496 & 2.861 & 0.842 \\ \hline
5 & MV & 217 & 458 & 642.514 & 2.961 & 0.953 \\ \hline
6 & MV & 191 & 414 & 533.958 & 2.796 & 0.852 \\ \hline
7 & MV & 884 & 2118 & 3132.693 & 3.544 & 1.092 \\ \hline
8 & MV & 366 & 764 & 1221.086 & 3.336 & 0.846 \\ \hline
9 & MV & 218 & 464 & 626.181 & 2.872 & 0.94 \\ \hline
10 & MV & 201 & 408 & 594.832 & 2.959 & 0.817 \\ \hline
11 & MV & 202 & 426 & 573.313 & 2.838 & 0.781 \\ \hline
%12 & MV & 25 & 48 & 33.28 & 1.331 & 0.769 \\ \hline
12 & MV & 464 & 998 & 1678.895 & 3.618 & 0.931 \\ \hline
\end{tabular}
\caption{Betweenness for random strategy \MV samples evolution.}\label{tab:mvRndBet}
\end{footnotesize}
\end{table}
\end{center}

\begin{figure}
 \captionsetup{type=figure}
    \centering
    \subfloat[Physical sample.]{\label{fig:rndOrig}\includegraphics[scale=.2]{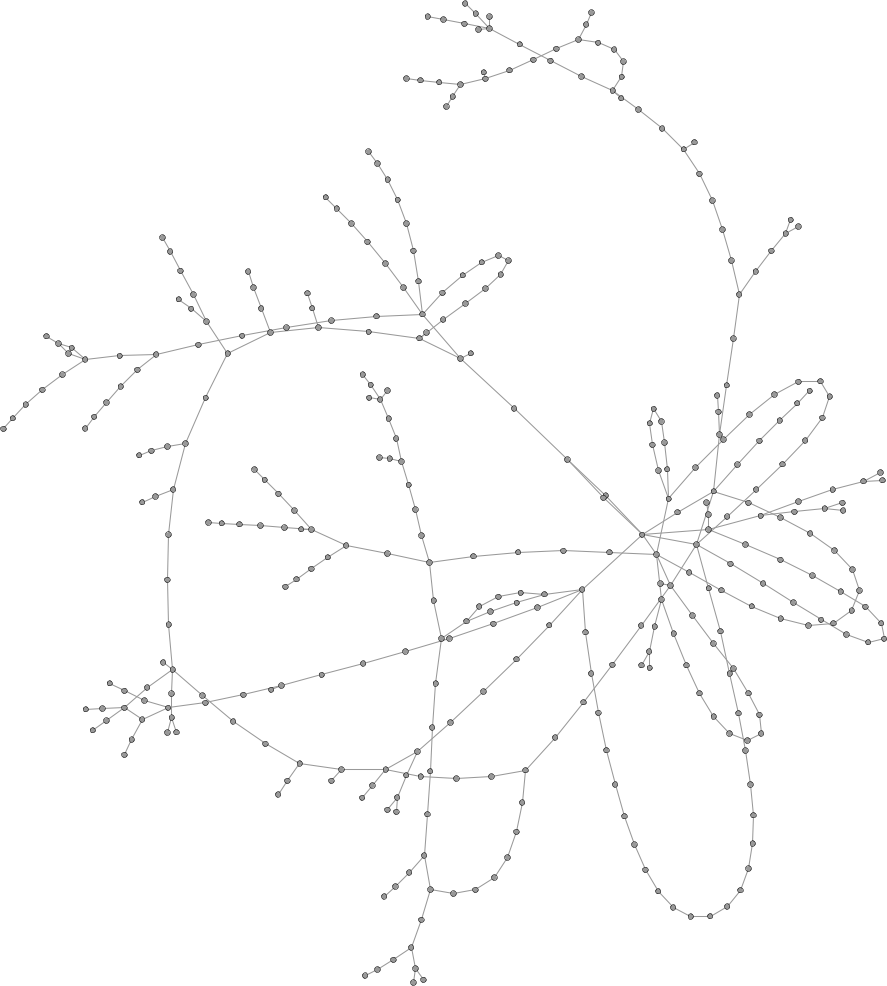}} 
    \subfloat[1$^{st}$ stage of evolution (i.e., +25\% edges).]{\label{fig:evol251}\includegraphics[scale=.2]{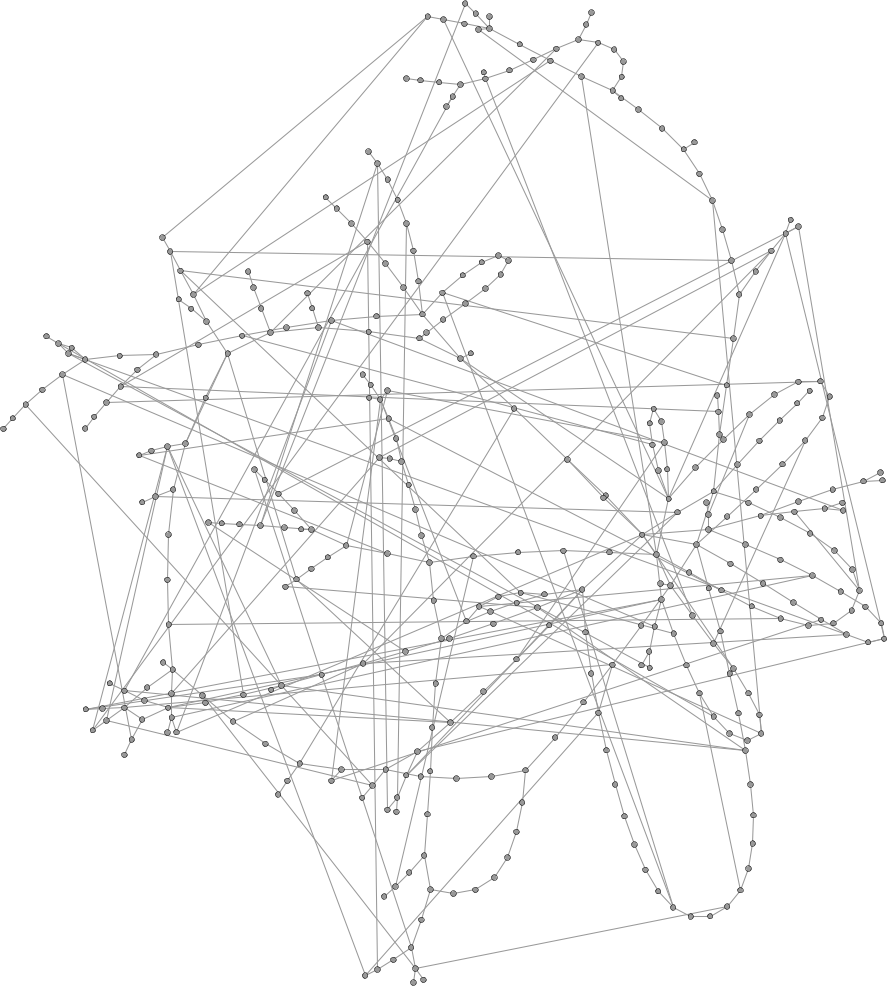}}\\
     \hspace*{-1.8cm}\subfloat[2$^{nd}$ stage of evolution (i.e., +50\% edges)]{\label{fig:evol252}\includegraphics[scale=.2]{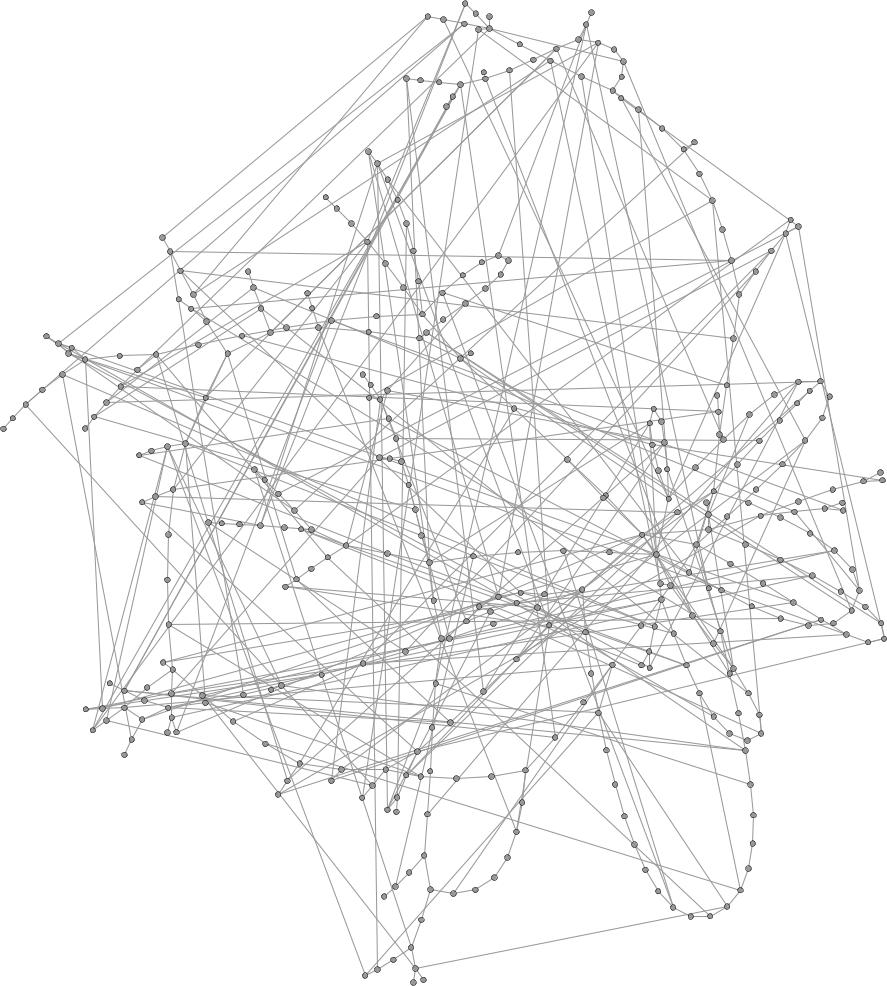}}  
      \subfloat[3$^{rd}$ stage of evolution (i.e., +75\% edges).]{\label{fig:evol253}\includegraphics[scale=.2]{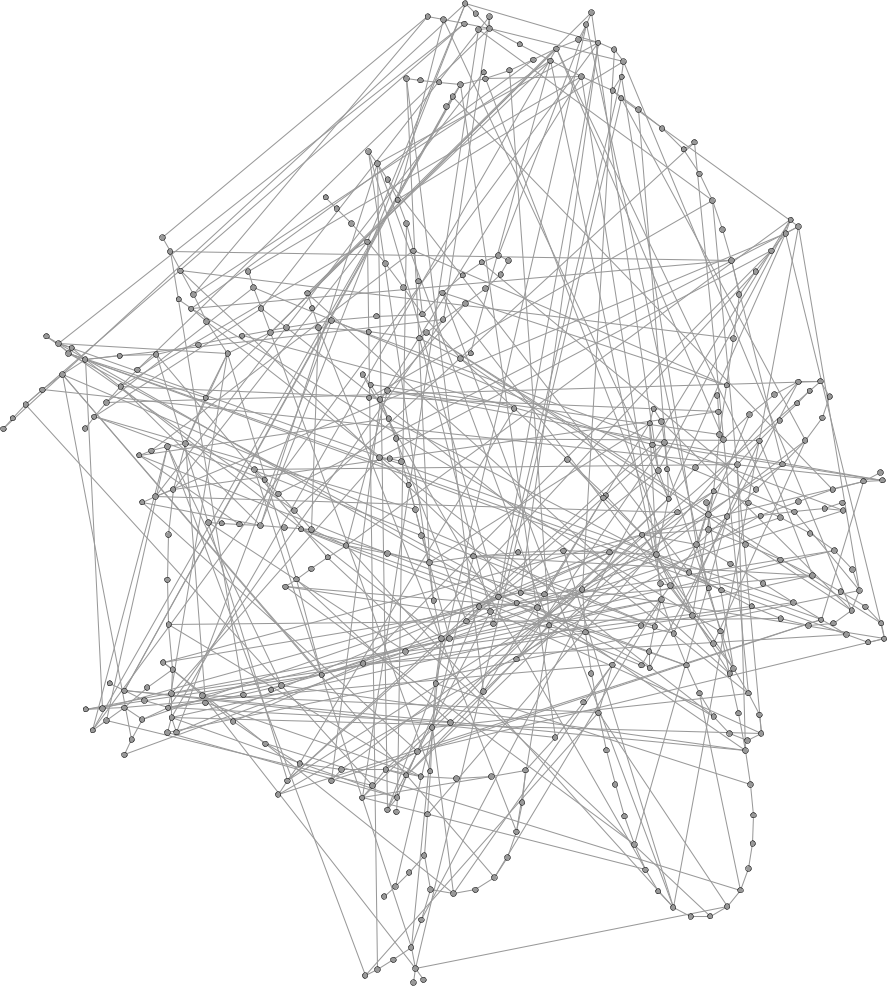}} 
       \subfloat[4$^{th}$ stage of evolution (i.e., +100\% edges).]{\label{fig:evol254}\includegraphics[scale=.2]{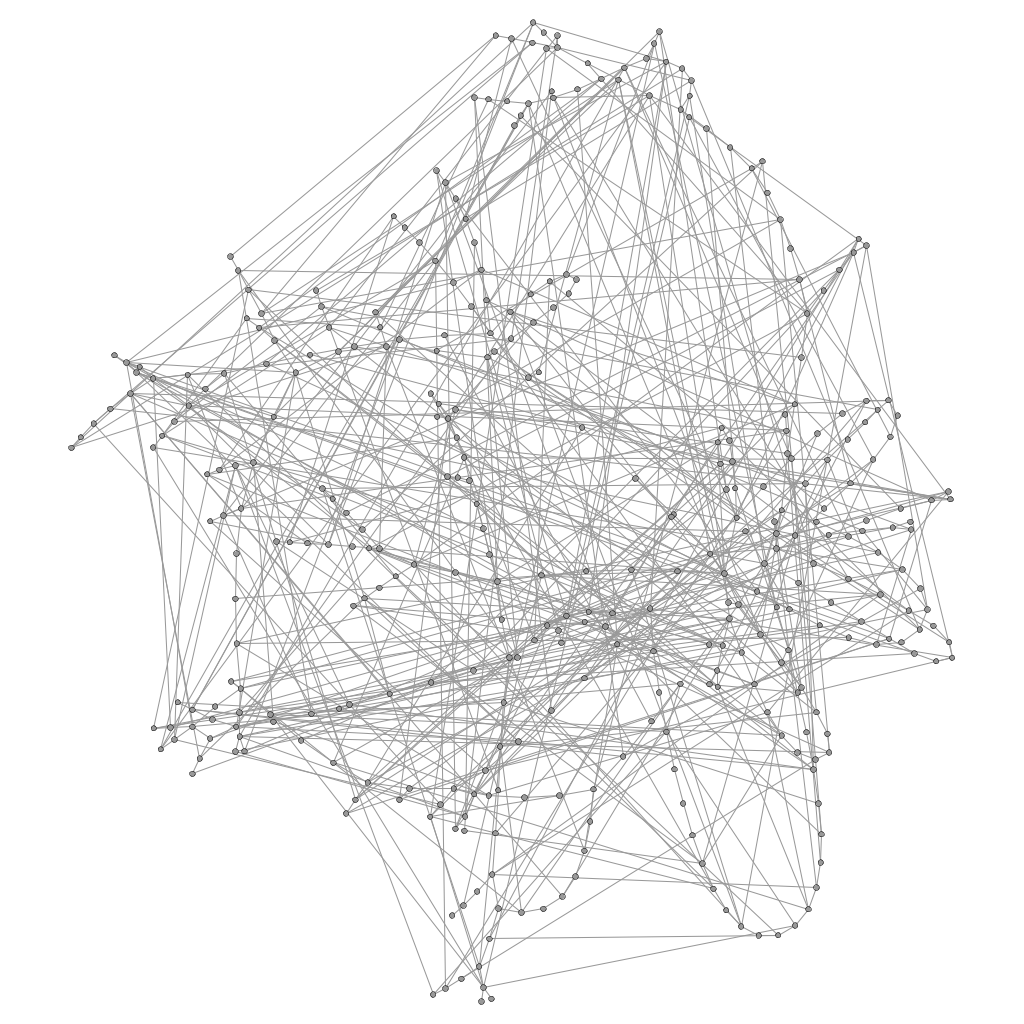}}  %\\
    \caption{Stages of evolution of \MV sample \#8 following the random strategy.}
    \label{fig:evolchartRnd}
\end{figure}

%****************************************************************************************************************************

\subsection{Evolution of the \LV Distribution \Gs}

The \LV Grid present some peculiarities caused by their practical use of being the terminals of the Grid. The majority of the networks (nine out of eleven samples) have null clustering coefficient, therefore it is difficult to analyze the improvement compared to the original as using percentage values as done for the \MV networks. Another peculiarity of this network is the limited  number of redundant paths (other than the shortest path) to connect any two nodes. In fact, paths in such networks especially the small samples are limited and mainly fixed by the radial topology of the network. Therefore, we consider for the evolution of such a metric only those samples for which there is a significant increase for the  $10^{th}$ \apl compared to the \cpl in the physical (not yet evolved) samples. In addition, we remark that for these samples there are no information available to us concerning the geographical position of the nodes and usually the samples cover a geographically very limited area with a small \textit{size} and \textit{order} of the corresponding graph. Therefore, we could not perform the distance based evolution of the \LV network samples.
%\footnote{not available to us, giusto? FATTO possiamo mettere altre motivazioni?HMMM DIREI DI NO}

\subsubsection*{Assortative high node degree evolution}

Table~\ref{tab:lvAss} contains for each sample of the \LV \G (column one) the values for main topological quantities: \textit{order} and \textit{size} in columns three and four, respectively; average node degree in the fifth column; the \cpl is reported in column six; the clustering coefficient follows in column seven; robustness is shown in eight column; the cost in term of redundant path length closes the data series (column nine).

The increase in the connectivity with the assortative high node degree strategy is extremely beneficial for the path length: just by adding 25\% of links it reduces, compared to the initial value of the original graphs, on average more than 30\%. The improvement is then reduced in the other steps, reaching 50\% when the number of link is doubled compared to the original \textit{size}. The two biggest samples benefit substantially from just the first evolution step: sample \#5 reduces the path from almost 19 to slightly more than 6, while sample \#9 from more than 11 to 4.75.
%[MAYBE IN THE DISCUSSION: This tendency of saturation in the reduction of the path length which reduces substantially just a few addition of nodes is explained by the small-world phenomenon that arises when a sufficient connectivity threshold is obtained, after this threshold is obtained the subsequent addition of nodes have an reduced effect on the improvement of the property. In this case, we see that this effect arises when the first two steps of the growth process are obtained since the \cpl then has no more significant improvement.] 
The improvement for the \cc is considerable: all the samples after the first step achieve a non null coefficient and the two samples that already have a significant \cc have a ten fold improvement. The improvement in robustness on average is about 70\% at the final step of the evolution. The addition of edges according to this strategy is mainly beneficial in the random attacks towards nodes, while the fraction of the metric that consider targeted attacks against the most connected nodes is marginally affected. This evolution strategy, indeed, leave the hierarchy of the nodes almost untouched reinforcing the node degree of those nodesthat had already an high degree at the beginning of the evolution process. For some small samples, the improvement in the targeted attack robustness is not monotonic, but it increases, then decreases in one intermediate step and then it increases again. Considering the values for the redundant path robustness the same considerations done for the \cpl apply. There is a reduction more than 35\% compared to the initial value of the samples already when the networks are evolved with a 25\% increase in the \textit{size} of the graph. The final improvement is about 63\%. However, an  exception can be noted: sample \#2 shows an increase in the value for the redundant path in the first step of the evolution compared to the initial value. Such increase is due to the problems that affect this metric when small networks (in terms of \textit{order} and \textit{size}) are considered since they do not have redundant path between nodes; such redundant path emerge when more connectivity is added to the network.

Table~\ref{tab:lvAssBet} contains for each sample of the \LV \G (column one) the values of the metrics related to betweenness. In addition to \textit{order} and \textit{size} (columns three and four), average betweenness is provided in column five, while a value of average betweenness normalized by the \textit{order} of the graph is computed in column six in order to compare the different samples. A measure of the statistical variation of betweenness is the coefficient of variation which is shown in the seventh column.

Considering betweenness, the assortative strategy involving the nodes with highest node degree is beneficial in reducing the average betweenness of all the \LV samples. The benefit that is provided after the four evolution steps is more than 50\%. Considering the two biggest samples (i.e., sample \#5 and \#9) the benefits reach the value of 70\% at the end of the evolution process, and a value closer to 65\% already after the first step.  The same trend is followed by the normalized value of average betweenness divided by the \textit{order} of the network.  Considering the variability of betweenness in each of the samples, we note a general increase for this metric. In particular, only one of the eleven samples presents a decrease when the connectivity in the network is double the original \textit{size} of the network. Such behavior is due to a slower decrease of the standard deviation of betweenness compared to its average. This strategy of adding the connections by providing more connections between the nodes that already have the highest connectivity and usually highest betweenness, reduces the number of ``bottleneck'' nodes. However, this strategy is not helpful for substantially reducing the variability of betweenness. In fact, the median of the betweenness  for the biggest samples is always zero. This aspect reinforces the knowledge that the majority of nodes are terminal nodes that are not involved in any path between other nodes. Only small samples have non zero values in the later stages of the evolution process.

\begin{table}[htbp]
\begin{footnotesize}
\begin{center}
\begin{tabular}{|@p{1.7cm}|^p{1.7cm}|^p{0.9cm}|^p{0.6cm}|^p{1cm}|^p{1cm}|^p{1.2cm}|^p{1.5cm}|^p{2cm}|}
\hline
\rowstyle{\bfseries} Sample ID &Network type  & \textit{Order} & \textit{Size} & {Avg. deg.} & {CPL} & {CC} & {Removal robustness ($Rob_N$)} & {Redundancy cost ($APL_{10^{th}}$)} \\ \hline
\hline

\multicolumn{9}{|c|}{\textbf{original \textit{order} +25\%}} \\ \hline
1 & LV & 17 & 22 & 2.588 & 2.563 & 0.08627 & 0.534 & 7.903 \\ \hline
2 & LV & 15 & 18 & 2.400 & 2.571 & 0.08667 & 0.379 & 8.107 \\ \hline
3 & LV & 21 & 27 & 2.571 & 3.200 & 0.06803 & 0.533 & 8.136 \\ \hline
4 & LV & 24 & 28 & 2.333 & 3.348 & 0.05000 & 0.426 & 9.493 \\ \hline
5 & LV & 186 & 236 & 2.538 & 6.286 & 0.03886 & 0.316 & 9.226 \\ \hline
6 & LV & 10 & 11 & 2.200 & 2.222 & 0.12000 & 0.370 & 4.040 \\ \hline
7 & LV & 63 & 77 & 2.444 & 3.839 & 0.04743 & 0.316 & 5.958 \\ \hline
8 & LV & 28 & 33 & 2.357 & 3.074 & 0.04464 & 0.413 & 7.036 \\ \hline
9 & LV & 133 & 175 & 2.632 & 4.750 & 0.04700 & 0.363 & 8.185 \\ \hline
10 & LV & 124 & 172 & 2.774 & 4.581 & 0.05247 & 0.349 & 6.835 \\ \hline
11 & LV & 31 & 37 & 2.387 & 3.167 & 0.04588 & 0.495 & 7.971 \\ \hline \hline
\multicolumn{9}{|c|}{\textbf{original \textit{order} +50\%}} \\ \hline

1 & LV & 17 & 27 & 3.176 & 2.188 & 0.29483 & 0.555 & 5.444 \\ \hline
2 & LV & 15 & 22 & 2.933 & 2.429 & 0.33926 & 0.452 & 5.589 \\ \hline
3 & LV & 21 & 33 & 3.143 & 2.200 & 0.22589 & 0.526 & 5.518 \\ \hline
4 & LV & 24 & 34 & 2.833 & 2.522 & 0.16982 & 0.453 & 5.931 \\ \hline
5 & LV & 186 & 283 & 3.043 & 6.065 & 0.06621 & 0.312 & 8.256 \\ \hline
6 & LV & 10 & 13 & 2.600 & 2.111 & 0.21333 & 0.479 & 5.360 \\ \hline
7 & LV & 63 & 93 & 2.952 & 3.516 & 0.05858 & 0.368 & 5.691 \\ \hline
8 & LV & 28 & 40 & 2.857 & 2.778 & 0.08413 & 0.441 & 4.709 \\ \hline
9 & LV & 133 & 210 & 3.158 & 4.417 & 0.08113 & 0.378 & 6.123 \\ \hline
10 & LV & 124 & 207 & 3.339 & 4.419 & 0.08386 & 0.351 & 6.188 \\ \hline
11 & LV & 31 & 45 & 2.903 & 2.833 & 0.11022 & 0.437 & 4.954 \\ \hline \hline
\multicolumn{9}{|c|}{\textbf{original \textit{order} +75\%}} \\ \hline 

1 & LV & 17 & 31 & 3.647 & 1.875 & 0.51569 & 0.540 & 4.347 \\ \hline
2 & LV & 15 & 26 & 3.467 & 1.929 & 0.41657 & 0.465 & 4.554 \\ \hline
3 & LV & 21 & 38 & 3.619 & 1.950 & 0.43312 & 0.552 & 4.500 \\ \hline
4 & LV & 24 & 40 & 3.333 & 2.370 & 0.26865 & 0.464 & 5.028 \\ \hline
5 & LV & 186 & 330 & 3.548 & 5.876 & 0.08132 & 0.311 & 8.240 \\ \hline
6 & LV & 10 & 15 & 3.000 & 1.889 & 0.30000 & 0.585 & 4.000 \\ \hline
7 & LV & 63 & 108 & 3.429 & 2.726 & 0.05858 & 0.398 & 4.828 \\ \hline
8 & LV & 28 & 47 & 3.357 & 2.778 & 0.13645 & 0.422 & 4.464 \\ \hline
9 & LV & 133 & 245 & 3.684 & 4.333 & 0.10548 & 0.399 & 5.377 \\ \hline
10 & LV & 124 & 241 & 3.887 & 4.346 & 0.10261 & 0.440 & 5.988 \\ \hline
11 & LV & 31 & 52 & 3.355 & 2.633 & 0.15801 & 0.454 & 4.325 \\ \hline \hline
\multicolumn{9}{|c|}{\textbf{original \textit{order} +100\%}} \\ \hline

1 & LV & 17 & 36 & 4.235 & 1.813 & 0.58796 & 0.493 & 3.931 \\ \hline
2 & LV & 15 & 30 & 4.000 & 1.857 & 0.51378 & 0.529 & 4.089 \\ \hline
3 & LV & 21 & 44 & 4.190 & 1.950 & 0.47279 & 0.532 & 4.027 \\ \hline
4 & LV & 24 & 46 & 3.833 & 2.304 & 0.29664 & 0.508 & 4.368 \\ \hline
5 & LV & 186 & 378 & 4.065 & 4.722 & 0.12554 & 0.343 & 6.808 \\ \hline
6 & LV & 10 & 18 & 3.600 & 1.556 & 0.30000 & 0.730 & 3.880 \\ \hline
7 & LV & 63 & 124 & 3.937 & 2.210 & 0.05858 & 0.408 & 4.149 \\ \hline
8 & LV & 28 & 54 & 3.857 & 2.685 & 0.23988 & 0.471 & 4.214 \\ \hline
9 & LV & 133 & 280 & 4.211 & 4.167 & 0.13065 & 0.397 & 5.543 \\ \hline
10 & LV & 124 & 276 & 4.452 & 4.191 & 0.12194 & 0.439 & 5.465 \\ \hline
11 & LV & 31 & 60 & 3.871 & 2.567 & 0.19894 & 0.438 & 4.217 \\ \hline
\end{tabular}
\caption{Metrics for assortative high node degree strategy \LV samples evolution.}\label{tab:lvAss}
\end{center}
\end{footnotesize}
\end{table}

\begin{center}
\begin{table}[htbp]
\centering
\begin{footnotesize}
\begin{tabular}{|@p{1.7cm}|^p{1.5cm}|^p{0.8cm}|^p{0.8cm}|^p{1.5cm}|^p{1.5cm}|^p{1.2cm}|}
\hline
\rowstyle{\bfseries}
Sample  ID  &Network type &\textit{Order} & \textit{Size} & {Avg. betweenness} & {Avg. betw/order} & {Coeff. variation} \\ \hline
\hline
\multicolumn{7}{|c|}{\textbf{original \textit{order} +25\%}} \\ \hline
1 & LV & 17 & 22 & 32.933 & 1.937 & 0.773 \\ \hline
2 & LV & 15 & 18 & 25.231 & 1.682 & 0.887 \\ \hline
3 & LV & 21 & 27 & 44.571 & 2.122 & 1.041 \\ \hline
4 & LV & 24 & 28 & 57 & 2.375 & 1.346 \\ \hline
5 & LV & 186 & 236 & 940.227 & 5.055 & 1.661 \\ \hline
6 & LV & 10 & 11 & 9 & 0.9 & 1.291 \\ \hline
7 & LV & 63 & 77 & 161.607 & 2.565 & 2.007 \\ \hline
8 & LV & 28 & 33 & 55.929 & 1.997 & 1.935 \\ \hline
9 & LV & 133 & 175 & 481.488 & 3.62 & 2.044 \\ \hline
10 & LV & 124 & 172 & 415.696 & 3.352 & 1.894 \\ \hline
11 & LV & 31 & 37 & 63.871 & 2.06 & 1.822 \\ \hline \hline
\multicolumn{7}{|c|}{\textbf{original \textit{order} +50\%}} \\ \hline
1 & LV & 17 & 27 & 32.933 & 1.937 & 0.773 \\ \hline
2 & LV & 15 & 22 & 25.231 & 1.682 & 0.887 \\ \hline
3 & LV & 21 & 33 & 27.524 & 1.311 & 2.194 \\ \hline
4 & LV & 24 & 34 & 38.667 & 1.611 & 1.691 \\ \hline
5 & LV & 186 & 283 & 904.648 & 4.864 & 1.513 \\ \hline
6 & LV & 10 & 13 & 8 & 0.8 & 1.173 \\ \hline
7 & LV & 63 & 93 & 151.065 & 2.398 & 2.158 \\ \hline
8 & LV & 28 & 40 & 50 & 1.786 & 1.589 \\ \hline
9 & LV & 133 & 210 & 423.442 & 3.184 & 1.842 \\ \hline
10 & LV & 124 & 207 & 396.786 & 3.2 & 1.645 \\ \hline
11 & LV & 31 & 45 & 53.419 & 1.723 & 1.734 \\ \hline \hline
\multicolumn{7}{|c|}{\textbf{original \textit{order} +75\%}} \\ \hline
1 & LV & 17 & 31 & 32.933 & 1.937 & 0.773 \\ \hline
2 & LV & 15 & 26 & 22.308 & 1.487 & 1.094 \\ \hline
3 & LV & 21 & 38 & 20.381 & 0.971 & 3.138 \\ \hline
4 & LV & 24 & 40 & 35.5 & 1.479 & 1.751 \\ \hline
5 & LV & 186 & 330 & 888.966 & 4.779 & 1.484 \\ \hline
6 & LV & 10 & 15 & 6 & 0.6 & 1 \\ \hline
7 & LV & 63 & 108 & 114.903 & 1.824 & 2.81 \\ \hline
8 & LV & 28 & 47 & 44.071 & 1.574 & 1.636 \\ \hline
9 & LV & 133 & 245 & 407.922 & 3.067 & 1.684 \\ \hline
10 & LV & 124 & 241 & 389.571 & 3.142 & 1.575 \\ \hline
11 & LV & 31 & 52 & 49.935 & 1.611 & 1.511 \\ \hline \hline
\multicolumn{7}{|c|}{\textbf{original \textit{order} +100\%}} \\ \hline
1 & LV & 17 & 36 & 20.533 & 1.208 & 0.83 \\ \hline
2 & LV & 15 & 30 & 18.923 & 1.262 & 1.434 \\ \hline
3 & LV & 21 & 44 & 19.619 & 0.934 & 2.66 \\ \hline
4 & LV & 24 & 46 & 33.5 & 1.396 & 1.409 \\ \hline
5 & LV & 186 & 378 & 882.58 & 4.745 & 1.482 \\ \hline
6 & LV & 10 & 18 & 3.75 & 0.375 & 0.702 \\ \hline
7 & LV & 63 & 124 & 76.548 & 1.215 & 4.604 \\ \hline
8 & LV & 28 & 54 & 40.643 & 1.452 & 1.692 \\ \hline
9 & LV & 133 & 280 & 395.039 & 2.97 & 1.66 \\ \hline
10 & LV & 124 & 276 & 383.45 & 3.092 & 1.588 \\ \hline
11 & LV & 31 & 60 & 45.935 & 1.482 & 1.436 \\ \hline

\end{tabular}
\caption{Betweenness for assortative high node degree strategy \LV samples evolution.}\label{tab:lvAssBet}
\end{footnotesize}
\end{table}
\end{center}

\subsubsection*{Assortative low node degree evolution}

Table~\ref{tab:lvLDAss} contains for each sample of the \LV \G (column one) the values for the main topological quantities: \textit{order} and \textit{size} in columns three and four, respectively; average node degree in the fifth column; the \cpl is reported in column six; the clustering coefficient follows in column seven; robustness is shown in eight column; the cost in term of redundant path length closes the data series (column nine).

The increase in the connectivity with the assortative low node degree strategy is beneficial for the path length. The first two evolution steps in particular, already provide for a reduction in the \cpl of more than 42\%, while the final step reaches a 50\% improvement. Even with this evolution strategy, the two biggest samples benefit substantially: sample \#5 reduces the path from almost 19 to less than 6, while sample \#9 from more than 11 to 4. Once again with this evolution strategy the saturation tendency in the reduction of the path length arises. This tendency is a consequence of the small-world phenomenon that arises when a sufficient connectivity threshold is reached. Once  this threshold is obtained, the subsequent addition of nodes has a reduced effect on the improvement of the property. In this case, we see that this phenomenon arises when the first two steps of the growth process are obtained since the \cpl then has no more significant improvement. 
The improvement for the \cc is considerable, but it takes three evolution steps before all the samples have non-zero clustering. In the final evolution step the clustering values are all higher than 0.1. Robustness almost triples, and eight out of the eleven samples are around or higher 0.7 for this metric. The addition of edges according to this strategy is beneficial almost only against the  random attacks towards nodes; while the component of the metric that consider targeted attacks against the most connected nodes is only partially affected. 
Considering the values for the redundant path robustness, the same considerations done for the \cpl apply. There is a reduction of more than 50\% compared to the initial value when two steps of evolution are completed. The final improvement is of about 63\%.

Table~\ref{tab:lvAssBetLD} contains for each sample of the \LV \G (column one) the values of metrics related to betweenness. In addition to \textit{order} and \textit{size} (columns three and four), average betweenness is provided in columns five, while a value of average betweenness normalized by the \textit{order} of the graph is computed in column six in order to compare the different samples. A measure of the statistical variation of betweenness is the coefficient of variation which is shown in the seventh column.

The assortative strategy involving the nodes with lowest node degree is beneficial in reducing the average betweenness of all the \LV samples.
%\footnote{\MV or \LV ? FATTO, SORRY} 
The benefit that is provided after the four evolution steps is more than 60\%. Considering the two biggest samples (i.e., samples \#5 and \#9) the benefits reach up to a value more than 70\% at the end of the evolution process.  The same trend is, of course, followed by the normalized value of average betweenness divided by the \textit{order} of the network.  Considering the coefficient of variation for betweenness, the general trend for the samples is an increase compared to the original values. Actually in the first step of evolution six of the eleven samples experience a decrease in the coefficient of variation, but at the end of the process only four keep the reduction trend.  This behavior is due to a slower decrease of the standard deviation of betweenness compared to its average. This strategy of adding the connectivity provides more connections between the nodes that have small connectivity that are usually the nodes at the periphery of the network and therefore that do not have paths traversing them in the initial sample. A proof is that the median of the betweenness for each sample in the first stages of evolution is zero, then in the last stage of evolution all the samples have a non-zero median that is a sign that all nodes are more evenly involved in path between other nodes.

\begin{table}[htbp]
\begin{footnotesize}
\begin{center}
\begin{tabular}{|@p{1.7cm}|^p{1.7cm}|^p{0.9cm}|^p{0.6cm}|^p{1cm}|^p{1cm}|^p{1.2cm}|^p{1.5cm}|^p{2cm}|}
\hline
\rowstyle{\bfseries} Sample ID &Network type  & \textit{Order} & \textit{Size} & {Avg. deg.} & {CPL} & {CC} & {Removal robustness ($Rob_N$)} & {Redundancy cost ($APL_{10^{th}}$)} \\ \hline
\hline

\multicolumn{9}{|c|}{\textbf{original \textit{order} +25\%}} \\ \hline
1 & LV & 17 & 22 & 2.588 & 2.813 & 0.15490 & 0.473 & 8.000 \\ \hline
2 & LV & 15 & 18 & 2.400 & 3.000 & 0.06667 & 0.528 & 10.179 \\ \hline
3 & LV & 21 & 27 & 2.571 & 3.950 & 0.19955 & 0.529 & 8.209 \\ \hline
4 & LV & 24 & 28 & 2.333 & 3.239 & 0.00000 & 0.438 & 10.868 \\ \hline
5 & LV & 186 & 236 & 2.538 & 5.638 & 0.00701 & 0.318 & 9.589 \\ \hline
6 & LV & 10 & 11 & 2.200 & 2.111 & 0.00000 & 0.391 & 5.760 \\ \hline
7 & LV & 63 & 77 & 2.444 & 4.790 & 0.13086 & 0.275 & 7.416 \\ \hline
8 & LV & 28 & 33 & 2.357 & 3.704 & 0.04932 & 0.362 & 9.301 \\ \hline
9 & LV & 133 & 175 & 2.632 & 4.864 & 0.01128 & 0.356 & 9.266 \\ \hline
10 & LV & 124 & 172 & 2.774 & 4.622 & 0.00899 & 0.374 & 8.052 \\ \hline
11 & LV & 31 & 37 & 2.387 & 3.800 & 0.00000 & 0.410 & 10.871 \\ \hline \hline
\multicolumn{9}{|c|}{\textbf{original \textit{order} +50\%}} \\ \hline
1 & LV & 17 & 27 & 3.176 & 2.125 & 0.36061 & 0.517 & 5.028 \\ \hline
2 & LV & 15 & 22 & 2.933 & 2.500 & 0.23111 & 0.551 & 5.982 \\ \hline
3 & LV & 21 & 33 & 3.143 & 2.200 & 0.25763 & 0.604 & 5.309 \\ \hline
4 & LV & 24 & 34 & 2.833 & 2.804 & 0.07262 & 0.453 & 6.417 \\ \hline
5 & LV & 186 & 283 & 3.043 & 5.622 & 0.08388 & 0.371 & 7.650 \\ \hline
6 & LV & 10 & 13 & 2.600 & 2.000 & 0.00000 & 0.410 & 6.640 \\ \hline
7 & LV & 63 & 93 & 2.952 & 3.339 & 0.13006 & 0.329 & 6.233 \\ \hline
8 & LV & 28 & 40 & 2.857 & 2.630 & 0.04801 & 0.391 & 6.760 \\ \hline
9 & LV & 133 & 210 & 3.158 & 4.083 & 0.08720 & 0.415 & 6.604 \\ \hline
10 & LV & 124 & 207 & 3.339 & 4.073 & 0.07965 & 0.404 & 6.253 \\ \hline
11 & LV & 31 & 45 & 2.903 & 2.433 & 0.04328 & 0.460 & 6.533 \\ \hline \hline
\multicolumn{9}{|c|}{\textbf{original \textit{order} +75\%}} \\ \hline
1 & LV & 17 & 31 & 3.647 & 1.875 & 0.58095 & 0.519 & 4.222 \\ \hline
2 & LV & 15 & 26 & 3.467 & 2.000 & 0.40296 & 0.575 & 4.732 \\ \hline
3 & LV & 21 & 38 & 3.619 & 1.950 & 0.42519 & 0.610 & 4.409 \\ \hline
4 & LV & 24 & 40 & 3.333 & 2.739 & 0.14484 & 0.522 & 5.326 \\ \hline
5 & LV & 186 & 330 & 3.548 & 5.616 & 0.11999 & 0.406 & 6.828 \\ \hline
6 & LV & 10 & 15 & 3.000 & 1.778 & 0.12000 & 0.629 & 5.360 \\ \hline
7 & LV & 63 & 108 & 3.429 & 2.226 & 0.12992 & 0.374 & 4.475 \\ \hline
8 & LV & 28 & 47 & 3.357 & 2.426 & 0.09464 & 0.501 & 4.959 \\ \hline
9 & LV & 133 & 245 & 3.684 & 4.045 & 0.14127 & 0.505 & 5.632 \\ \hline
10 & LV & 124 & 241 & 3.887 & 4.041 & 0.14584 & 0.420 & 5.958 \\ \hline
11 & LV & 31 & 52 & 3.355 & 2.400 & 0.14910 & 0.534 & 5.158 \\ \hline \hline
\multicolumn{9}{|c|}{\textbf{original \textit{order} +100\%}} \\ \hline
1 & LV & 17 & 36 & 4.235 & 1.875 & 0.59314 & 0.544 & 3.889 \\ \hline
2 & LV & 15 & 30 & 4.000 & 1.857 & 0.51091 & 0.591 & 4.179 \\ \hline
3 & LV & 21 & 44 & 4.190 & 1.950 & 0.44977 & 0.628 & 3.991 \\ \hline
4 & LV & 24 & 46 & 3.833 & 2.674 & 0.20079 & 0.769 & 5.090 \\ \hline
5 & LV & 186 & 378 & 4.065 & 5.597 & 0.14059 & 0.404 & 6.622 \\ \hline
6 & LV & 10 & 18 & 3.600 & 1.556 & 0.25000 & 0.688 & 4.320 \\ \hline
7 & LV & 63 & 124 & 3.937 & 2.194 & 0.16174 & 0.394 & 4.201 \\ \hline
8 & LV & 28 & 54 & 3.857 & 2.389 & 0.17619 & 0.514 & 4.577 \\ \hline
9 & LV & 133 & 280 & 4.211 & 4.008 & 0.17196 & 0.508 & 5.175 \\ \hline
10 & LV & 124 & 276 & 4.452 & 3.992 & 0.18500 & 0.512 & 5.460 \\ \hline
11 & LV & 31 & 60 & 3.871 & 2.333 & 0.20753 & 0.508 & 4.533 \\ \hline
\end{tabular}
\caption{Metrics for assortative low node degree strategy \LV samples evolution.}\label{tab:lvLDAss}
\end{center}
\end{footnotesize}
\end{table}

\begin{center}
\begin{table}[htbp]
\centering
\begin{footnotesize}
\begin{tabular}{|@p{1.7cm}|^p{1.5cm}|^p{0.8cm}|^p{0.8cm}|^p{1.5cm}|^p{1.5cm}|^p{1.2cm}|}
\hline
\rowstyle{\bfseries}
Sample  ID  &Network type &\textit{Order} & \textit{Size} & {Avg. betweenness} & {Avg. betw/order} & {Coeff. variation} \\ \hline
\hline
\multicolumn{7}{|c|}{\textbf{original \textit{order} +25\%}} \\ \hline
1 & LV & 17 & 22 & 24.933 & 1.467 & 0.978 \\ \hline
2 & LV & 15 & 18 & 27.429 & 1.829 & 0.832 \\ \hline
3 & LV & 21 & 27 & 61.048 & 2.907 & 0.854 \\ \hline
4 & LV & 24 & 28 & 53.5 & 2.229 & 0.947 \\ \hline
5 & LV & 186 & 236 & 888.486 & 4.777 & 2.442 \\ \hline
6 & LV & 10 & 11 & 10.667 & 1.067 & 1.021 \\ \hline
7 & LV & 63 & 77 & 228.984 & 3.635 & 2.02 \\ \hline
8 & LV & 28 & 33 & 79.143 & 2.827 & 1.102 \\ \hline
9 & LV & 133 & 175 & 562.931 & 4.233 & 2.497 \\ \hline
10 & LV & 124 & 172 & 465.847 & 3.757 & 1.985 \\ \hline
11 & LV & 31 & 37 & 87.613 & 2.826 & 1.113 \\ \hline \hline
\multicolumn{7}{|c|}{\textbf{original \textit{order} +50\%}} \\ \hline
1 & LV & 17 & 1 & 14.267 & 0.839 & 2.177 \\ \hline
2 & LV & 15 & 2 & 21.429 & 1.429 & 0.898 \\ \hline
3 & LV & 21 & 3 & 26.762 & 1.274 & 2.128 \\ \hline
4 & LV & 24 & 4 & 42.583 & 1.774 & 0.912 \\ \hline
5 & LV & 186 & 5 & 886.077 & 4.764 & 1.781 \\ \hline
6 & LV & 10 & 6 & 6.889 & 0.689 & 0.867 \\ \hline
7 & LV & 63 & 7 & 163.377 & 2.593 & 2.184 \\ \hline
8 & LV & 28 & 8 & 47.214 & 1.686 & 1.696 \\ \hline
9 & LV & 133 & 9 & 414.901 & 3.12 & 2.379 \\ \hline
10 & LV & 124 & 10 & 404.085 & 3.259 & 1.946 \\ \hline
11 & LV & 31 & 11 & 49.484 & 1.596 & 2.238 \\ \hline
\multicolumn{7}{|c|}{\textbf{original \textit{order} +75\%}} \\ \hline
1 & LV & 17 & 1 & 10.267 & 0.604 & 3.352 \\ \hline
2 & LV & 15 & 2 & 13 & 0.867 & 2.042 \\ \hline
3 & LV & 21 & 3 & 20.286 & 0.966 & 3.114 \\ \hline
4 & LV & 24 & 4 & 39.5 & 1.646 & 0.818 \\ \hline
5 & LV & 186 & 5 & 883.901 & 4.752 & 1.497 \\ \hline
6 & LV & 10 & 6 & 5.556 & 0.556 & 0.652 \\ \hline
7 & LV & 63 & 7 & 83.645 & 1.328 & 4.621 \\ \hline
8 & LV & 28 & 8 & 41.857 & 1.495 & 1.636 \\ \hline
9 & LV & 133 & 9 & 409.145 & 3.076 & 1.912 \\ \hline
10 & LV & 124 & 10 & 402.305 & 3.244 & 1.627 \\ \hline
11 & LV & 31 & 11 & 47.742 & 1.54 & 1.977 \\ \hline
\multicolumn{7}{|c|}{\textbf{original \textit{order} +100\%}} \\ \hline
1 & LV & 17 & 1 & 9.6 & 0.565 & 2.951 \\ \hline
2 & LV & 15 & 2 & 11.286 & 0.752 & 2.184 \\ \hline
3 & LV & 21 & 3 & 19.524 & 0.93 & 2.774 \\ \hline
4 & LV & 24 & 4 & 37.417 & 1.559 & 0.725 \\ \hline
5 & LV & 186 & 5 & 879.801 & 4.73 & 1.323 \\ \hline
6 & LV & 10 & 6 & 4.444 & 0.444 & 0.743 \\ \hline
7 & LV & 63 & 7 & 79.839 & 1.267 & 4.655 \\ \hline
8 & LV & 28 & 8 & 41.143 & 1.469 & 1.275 \\ \hline
9 & LV & 133 & 9 & 399.252 & 3.002 & 1.695 \\ \hline
10 & LV & 124 & 10 & 400.576 & 3.23 & 1.447 \\ \hline
11 & LV & 31 & 11 & 45.548 & 1.469 & 1.559 \\ \hline
\end{tabular}
\caption{Betweenness for assortative low node degree strategy \LV samples evolution.}\label{tab:lvAssBetLD}
\end{footnotesize}
\end{table}
\end{center}

\subsubsection*{Triangle closure evolution}

Table~\ref{tab:lvCC} contains for each sample of the \LV \G (column one) the values for the main topological quantities: \textit{order} and \textit{size} in columns three and four, respectively; average node degree in the fifth column; the \cpl is reported in column six; the clustering coefficient follows in column seven; robustness is shown in eight column; the cost in term of redundant path length closes the data series (column nine).

As we already noted in the analysis of the \MV networks, the triangle closure strategy
%\footnote{ricontrolla tutte le volte che introduci una strategy per MV e LV. Bisogna che nel primo paragrafo la citi esplicitamente e non dici solo ``this strategy'' per LV te lo sto modificando, ma per MV mi potrebbe essere sfuggito.OK FATTO} 
aims at incrementing the clustering coefficient. Therefore it is not surprising that the improvement in the path length are below 30\% at the final stage of evolution. The increase in the improvement is on average linear between the various steps. For instance, the two biggest samples improve only marginally in the \cpl with values at the end of the 4$^{th}$ evolution step of almost 15 and 8 for samples \#5 and \#8, respectively. The evolution of clustering coefficient is the primary goal of the evolution, the values that  are obtained in the first evolution step are already around 0.1, reaching for the majority of the samples a value near 0.5 in the last stage.
The improvement for the \cc is considerable: all the samples after the first step achieve a non null coefficient and the two samples that already have a significant \cc have a ten fold improvement. The improvement in robustness is below 70\% at the final step of the evolution. The addition of edges according to this strategy is beneficial both against random attacks towards nodes and also towards targeted attacks against the most connected nodes. Actually, the improvement are much higher in tolerating the targeted attacks than the improvement observed against the random failures. Considering the values for the redundant path availability, the same considerations done for the \cpl apply. There is a reduction more than 50\% compared to the initial value of the samples  when the last stage of edge addition is reached, even if the real improvement is already reached after the third stage. However, an  exception can be noted: sample \#2 shows an increase in the value for the redundant path in the first step of the evolution compared to the initial value. Such increase is due to the already mentioned problems that affect this metric when small networks (in terms of \textit{order} and \textit{size}) are considered since they do not have redundant path between nodes; such redundant path emerge when more connectivity is added to the network.

Table~\ref{tab:lvCCBet} contains for each sample of the \LV \G (column one) the values of metrics related to betweenness. In addition to \textit{order} and \textit{size} (columns three and four), average betweenness is provided in columns five, while a value of average betweenness normalized by the \textit{order} of the graph is computed in column six in order to compare the different samples. A measure of the statistical variation of betweenness is the coefficient of variation which is shown in the seventh column.

The triangle closure strategy provides limited benefits in the reduction of average betweenness and the average betweenness to \textit{order} ratio. The benefits are quantifiable in less than 40\% after the whole evolution process. The improvements during the several evolution steps progress almost linearly with the addition of edges. Looking at the variability of betweenness, only few samples experience a reduction in the coefficient of variation at the end of the evolution process. In the first stage, the reduction compared to the original values take place only in six out of the eleven samples. The reason is once again the slower decrease of the standard deviation of betweenness compared to its average. The standard deviation, in fact, slowly decreases, with a mode for the betweenness over the four evolution stages that is zero for all the samples; the only exception is sample \#1.

%\footnote{nella caption: \LV? forse bisogna fare un search e ricontrollare tutta la parte LV per vedere se ci sono refusi di MV? CONTROLLATO}

\begin{table}[htbp]
\begin{footnotesize}
\begin{center}
\begin{tabular}{|@p{1.7cm}|^p{1.7cm}|^p{0.9cm}|^p{0.6cm}|^p{1cm}|^p{1cm}|^p{1.2cm}|^p{1.5cm}|^p{2cm}|}
\hline
\rowstyle{\bfseries} Sample ID &Network type  & \textit{Order} & \textit{Size} & {Avg. deg.} & {CPL} & {CC} & {Removal robustness ($Rob_N$)} & {Redundancy cost ($APL_{10^{th}}$)} \\ \hline
\hline

\multicolumn{9}{|c|}{\textbf{original \textit{order} +25\%}} \\ \hline
1 & LV & 17 & 22 & 2.588 & 3.063 & 0.11765 & 0.489 & 7.722 \\ \hline
2 & LV & 15 & 18 & 2.400 & 2.857 & 0.10667 & 0.513 & 8.000 \\ \hline
3 & LV & 21 & 27 & 2.571 & 3.850 & 0.13492 & 0.484 & 7.709 \\ \hline
4 & LV & 24 & 28 & 2.333 & 4.087 & 0.09583 & 0.367 & 6.611 \\ \hline
5 & LV & 186 & 236 & 2.538 & 17.443 & 0.08734 & 0.142 & 24.535 \\ \hline
6 & LV & 10 & 11 & 2.200 & 2.056 & 0.21333 & 0.350 & 4.280 \\ \hline
7 & LV & 63 & 77 & 2.444 & 5.323 & 0.13171 & 0.249 & 6.188 \\ \hline
8 & LV & 28 & 33 & 2.357 & 4.222 & 0.21548 & 0.359 & 5.607 \\ \hline
9 & LV & 133 & 175 & 2.632 & 10.174 & 0.09291 & 0.232 & 12.059 \\ \hline
10 & LV & 124 & 172 & 2.774 & 6.943 & 0.08714 & 0.280 & 9.905 \\ \hline
11 & LV & 31 & 37 & 2.387 & 4.967 & 0.15484 & 0.340 & 6.688 \\ \hline \hline
\multicolumn{9}{|c|}{\textbf{original \textit{order} +50\%}} \\ \hline
1 & LV & 17 & 27 & 3.176 & 2.625 & 0.22941 & 0.500 & 5.194 \\ \hline
2 & LV & 15 & 22 & 2.933 & 2.643 & 0.20889 & 0.524 & 5.304 \\ \hline
3 & LV & 21 & 33 & 3.143 & 3.500 & 0.33492 & 0.500 & 6.445 \\ \hline
4 & LV & 24 & 34 & 2.833 & 3.630 & 0.17817 & 0.407 & 5.382 \\ \hline
5 & LV & 186 & 283 & 3.043 & 16.557 & 0.15282 & 0.161 & 20.545 \\ \hline
6 & LV & 10 & 13 & 2.600 & 2.000 & 0.35667 & 0.378 & 5.400 \\ \hline
7 & LV & 63 & 93 & 2.952 & 5.274 & 0.14160 & 0.282 & 5.962 \\ \hline
8 & LV & 28 & 40 & 2.857 & 3.815 & 0.44643 & 0.328 & 6.209 \\ \hline
9 & LV & 133 & 210 & 3.158 & 9.295 & 0.21776 & 0.345 & 12.848 \\ \hline
10 & LV & 124 & 207 & 3.339 & 6.683 & 0.18182 & 0.428 & 9.559 \\ \hline
11 & LV & 31 & 45 & 2.903 & 4.067 & 0.29724 & 0.355 & 5.917 \\ \hline \hline
\multicolumn{9}{|c|}{\textbf{original \textit{order} +75\%}} \\ \hline
1 & LV & 17 & 31 & 3.647 & 2.500 & 0.30644 & 0.581 & 4.625 \\ \hline
2 & LV & 15 & 26 & 3.467 & 2.357 & 0.34222 & 0.598 & 4.554 \\ \hline
3 & LV & 21 & 38 & 3.619 & 2.950 & 0.41882 & 0.549 & 5.200 \\ \hline
4 & LV & 24 & 40 & 3.333 & 3.348 & 0.37718 & 0.433 & 5.069 \\ \hline
5 & LV & 186 & 330 & 3.548 & 15.400 & 0.29473 & 0.236 & 17.494 \\ \hline
6 & LV & 10 & 15 & 3.000 & 1.889 & 0.42333 & 0.504 & 4.440 \\ \hline
7 & LV & 63 & 108 & 3.429 & 5.161 & 0.20174 & 0.487 & 5.800 \\ \hline
8 & LV & 28 & 47 & 3.357 & 3.648 & 0.52262 & 0.416 & 5.643 \\ \hline
9 & LV & 133 & 245 & 3.684 & 8.265 & 0.34422 & 0.375 & 10.834 \\ \hline
10 & LV & 124 & 241 & 3.887 & 5.972 & 0.28127 & 0.416 & 7.653 \\ \hline
11 & LV & 31 & 52 & 3.355 & 3.733 & 0.38771 & 0.424 & 5.388 \\ \hline \hline
\multicolumn{9}{|c|}{\textbf{original \textit{order} +100\%}} \\ \hline
1 & LV & 17 & 36 & 4.235 & 2.125 & 0.41167 & 0.596 & 4.097 \\ \hline
2 & LV & 15 & 30 & 4.000 & 2.143 & 0.48937 & 0.669 & 4.018 \\ \hline
3 & LV & 21 & 44 & 4.190 & 2.750 & 0.56485 & 0.565 & 4.882 \\ \hline
4 & LV & 24 & 46 & 3.833 & 3.043 & 0.48585 & 0.458 & 4.660 \\ \hline
5 & LV & 186 & 378 & 4.065 & 14.792 & 0.44891 & 0.250 & 16.530 \\ \hline
6 & LV & 10 & 18 & 3.600 & 1.667 & 0.48333 & 0.673 & 3.720 \\ \hline
7 & LV & 63 & 124 & 3.937 & 5.065 & 0.29045 & 0.550 & 6.058 \\ \hline
8 & LV & 28 & 54 & 3.857 & 3.426 & 0.64345 & 0.406 & 5.230 \\ \hline
9 & LV & 133 & 280 & 4.211 & 8.015 & 0.39996 & 0.383 & 11.030 \\ \hline
10 & LV & 124 & 276 & 4.452 & 5.224 & 0.33963 & 0.526 & 6.988 \\ \hline
11 & LV & 31 & 60 & 3.871 & 3.300 & 0.51951 & 0.391 & 5.092 \\ \hline
\end{tabular}
\caption{Metrics for triangle closure strategy \LV samples evolution.}\label{tab:lvCC}
\end{center}
\end{footnotesize}
\end{table}

\begin{center}
\begin{table}[htbp]
\centering
\begin{footnotesize}
\begin{tabular}{|@p{1.7cm}|^p{1.5cm}|^p{0.8cm}|^p{0.8cm}|^p{1.5cm}|^p{1.5cm}|^p{1.2cm}|}
\hline
\rowstyle{\bfseries}
Sample  ID  &Network type &\textit{Order} & \textit{Size} & {Avg. betweenness} & {Avg. betw/order} & {Coeff. variation} \\ \hline
\hline
\multicolumn{7}{|c|}{\textbf{original \textit{order} +25\%}} \\ \hline
1 & LV & 17 & 22 & 28.4 & 1.671 & 0.649 \\ \hline
2 & LV & 15 & 18 & 22.154 & 1.477 & 0.646 \\ \hline
3 & LV & 21 & 27 & 60.476 & 2.88 & 0.561 \\ \hline
4 & LV & 24 & 28 & 71.917 & 2.997 & 0.967 \\ \hline
5 & LV & 186 & 236 & 2821.193 & 15.168 & 1.214 \\ \hline
6 & LV & 10 & 11 & 7 & 0.7 & 1.429 \\ \hline
7 & LV & 63 & 77 & 252.885 & 4.014 & 2.074 \\ \hline
8 & LV & 28 & 33 & 88.786 & 3.171 & 1.475 \\ \hline
9 & LV & 133 & 175 & 1217.519 & 9.154 & 1.495 \\ \hline
10 & LV & 124 & 172 & 678.804 & 5.474 & 1.395 \\ \hline
11 & LV & 31 & 37 & 120.258 & 3.879 & 1.313 \\ \hline
\hline
\multicolumn{7}{|c|}{\textbf{original \textit{order} +50\%}} \\ \hline
1 & LV & 17 & 1 & 24.8 & 1.459 & 0.633 \\ \hline
2 & LV & 15 & 2 & 18.923 & 1.262 & 0.715 \\ \hline
3 & LV & 21 & 3 & 53.143 & 2.531 & 0.621 \\ \hline
4 & LV & 24 & 4 & 63.833 & 2.66 & 0.982 \\ \hline
5 & LV & 186 & 5 & 2747.807 & 14.773 & 1.25 \\ \hline
6 & LV & 10 & 6 & 5.5 & 0.55 & 1.226 \\ \hline
7 & LV & 63 & 7 & 247.443 & 3.928 & 2.111 \\ \hline
8 & LV & 28 & 8 & 75.357 & 2.691 & 1.704 \\ \hline
9 & LV & 133 & 9 & 1096.527 & 8.245 & 1.594 \\ \hline
10 & LV & 124 & 10 & 653.232 & 5.268 & 1.424 \\ \hline
11 & LV & 31 & 11 & 94.258 & 3.041 & 1.46 \\ \hline
\hline
\multicolumn{7}{|c|}{\textbf{original \textit{order} +75\%}} \\ \hline
1 & LV & 17 & 1 & 21.333 & 1.255 & 0.596 \\ \hline
2 & LV & 15 & 2 & 17.846 & 1.19 & 0.731 \\ \hline
3 & LV & 21 & 3 & 39.905 & 1.9 & 0.898 \\ \hline
4 & LV & 24 & 4 & 55.25 & 2.302 & 1.03 \\ \hline
5 & LV & 186 & 5 & 2513.67 & 13.514 & 1.278 \\ \hline
6 & LV & 10 & 6 & 5 & 0.5 & 1.153 \\ \hline
7 & LV & 63 & 7 & 238.754 & 3.79 & 2.201 \\ \hline
8 & LV & 28 & 8 & 70.143 & 2.505 & 1.801 \\ \hline
9 & LV & 133 & 9 & 992.558 & 7.463 & 1.703 \\ \hline
10 & LV & 124 & 10 & 571.214 & 4.607 & 1.541 \\ \hline
11 & LV & 31 & 11 & 84.581 & 2.728 & 1.483 \\ \hline
\hline
\multicolumn{7}{|c|}{\textbf{original \textit{order} +100\%}} \\ \hline
1 & LV & 17 & 1 & 16.933 & 0.996 & 0.972 \\ \hline
2 & LV & 15 & 2 & 13.077 & 0.872 & 0.85 \\ \hline
3 & LV & 21 & 3 & 35.429 & 1.687 & 0.874 \\ \hline
4 & LV & 24 & 4 & 44.833 & 1.868 & 1.373 \\ \hline
5 & LV & 186 & 5 & 2375.205 & 12.77 & 1.34 \\ \hline
6 & LV & 10 & 6 & 4.25 & 0.425 & 1.157 \\ \hline
7 & LV & 63 & 7 & 234.885 & 3.728 & 2.202 \\ \hline
8 & LV & 28 & 8 & 62.071 & 2.217 & 1.967 \\ \hline
9 & LV & 133 & 9 & 951.163 & 7.152 & 1.716 \\ \hline
10 & LV & 124 & 10 & 509.321 & 4.107 & 1.603 \\ \hline
11 & LV & 31 & 11 & 73.419 & 2.368 & 1.659 \\ \hline
\end{tabular}
\caption{Betweenness for triangle closure strategy \LV samples evolution.}\label{tab:lvCCBet}
\end{footnotesize}
\end{table}
\end{center}

\subsubsection*{Dissortative node degree evolution}

Table~\ref{tab:lvDiss} contains for each sample of the \LV \G (column one) the values for the main topological quantities: \textit{order} and \textit{size} in columns three and four, respectively; average node degree in the fifth column; the \cpl is reported in column six; the clustering coefficient follows in column seven; robustness is shown in eight column; the cost in term of redundant path length closes the data series (column nine).

Following the dissortative node degree strategy, the improvement considering the \cpl increases sub-linearly in the four steps of evolution. The first set of edge addition provides an average improvement around 35\%, while in the final step the improvement reaches around 54\% compared to the initial value. The two biggest samples (i.e., samples \#5 and \#8) reach values of path reduction that are higher than 65\%. The evolution of clustering coefficient is less uniform than other strategies: the smallest samples in term of \textit{order} reach an high level of improvement (i.e., samples \#1, \#2, and \#3), while for the biggest samples the dissortative evolution scheme is less beneficial in the emergence of triangle structures in the network.
 The improvement in robustness is about 77\% at the final step of the evolution. The trend in the four evolution steps is almost linear with an improvement at each step that is higher than 10\%. The addition of edges according to this strategy is beneficial mainly against random attacks. The effect towards targeted attacks against the most connected nodes improves usually only in the final stages of the evolution (i.e. fourth step). Considering the values for the redundant path robustness one notes a double in the path reduction between the first and the last step of evolution. For the biggest samples, the difference between the value of \cpl and the value of $10^{th}$ \apl is around or below three. However, an  exception can be noticed: sample \#2 shows an increase in the value for the redundant path  in the first step of the evolution compared to the initial value. This increase is due to the problems that affect this metric when small networks (in terms of \textit{order} and \textit{size}) are considered since they do not have redundant path between nodes; such redundant path emerge when more connectivity is added to the network.

Table~\ref{tab:lvDissBet} contains for each sample of the \LV \G (column one) the values of metrics related to betweenness. In addition to \textit{order} and \textit{size} (columns three and four), average betweenness is provided in columns five, while a value of average betweenness normalized by the \textit{order} of the graph is computed in column six in order to compare the different samples. A measure of the statistical variation of betweenness is the coefficient of variation which is shown in the seventh column.

The evolution of the betweenness metric realized by the dissortative strategy provides a quite good reduction. Overall the reduction is more than 65\%, and the major achievement is gained in the first two steps (54\% of improvement). This trend is followed both by the average betweenness and the normalized value of average betweenness divided by the \textit{order} of the network. Considering the coefficient of variation, all the samples present an increase in such a metric. As we have described the average betweenness experiences a consistent reduction, while the standard deviation does not follow this trend. For the biggest samples in term of \textit{order}, the standard deviation reduces, while for the smallest ones the standard deviation increases while more edges are added in the network.  This tendency is due to the very nature of this strategy that aims at connecting different classes of nodes in terms of node degree and therefore enabling the creation of new paths between the nodes at the edge of the network (having node degree one or two) and the nodes at the core of the network with high node degree. Such nodes, previously unaffected by managing the shortest paths, are now involved. This more even distribution of shortest paths is shown by the mode of betweenness that already from the second evolution step provides non-zero values for the majority of the samples. Later stages show all samples (except the small sample \#6) with a non-zero mode.

\begin{table}[htbp]
\begin{footnotesize}
\begin{center}
\begin{tabular}{|@p{1.7cm}|^p{1.7cm}|^p{0.9cm}|^p{0.6cm}|^p{1cm}|^p{1cm}|^p{1.2cm}|^p{1.5cm}|^p{2cm}|}
\hline
\rowstyle{\bfseries} Sample ID &Network type  & \textit{Order} & \textit{Size} & {Avg. deg.} & {CPL} & {CC} & {Removal robustness ($Rob_N$)} & {Redundancy cost ($APL_{10^{th}}$)} \\ \hline
\hline

\multicolumn{9}{|c|}{\textbf{original \textit{order} +25\%}} \\ \hline
1 & LV & 17 & 22 & 2.588 & 2.563 & 0.11373 & 0.539 & 8.111 \\ \hline
2 & LV & 15 & 18 & 2.400 & 2.571 & 0.27143 & 0.510 & 9.214 \\ \hline
3 & LV & 21 & 27 & 2.571 & 2.900 & 0.05873 & 0.440 & 8.427 \\ \hline
4 & LV & 24 & 28 & 2.333 & 3.326 & 0.08426 & 0.396 & 7.313 \\ \hline
5 & LV & 186 & 236 & 2.538 & 5.354 & 0.02151 & 0.356 & 9.082 \\ \hline
6 & LV & 10 & 11 & 2.200 & 2.000 & 0.40952 & 0.352 & 4.360 \\ \hline
7 & LV & 63 & 77 & 2.444 & 3.484 & 0.00000 & 0.319 & 7.449 \\ \hline
8 & LV & 28 & 33 & 2.357 & 3.241 & 0.04827 & 0.375 & 9.204 \\ \hline
9 & LV & 133 & 175 & 2.632 & 4.152 & 0.01128 & 0.372 & 8.367 \\ \hline
10 & LV & 124 & 172 & 2.774 & 4.195 & 0.00899 & 0.389 & 7.962 \\ \hline
11 & LV & 31 & 37 & 2.387 & 3.567 & 0.00000 & 0.390 & 7.663 \\ \hline \hline
\multicolumn{9}{|c|}{\textbf{original \textit{order} +50\%}} \\ \hline
1 & LV & 17 & 27 & 3.176 & 2.250 & 0.31634 & 0.519 & 5.417 \\ \hline
2 & LV & 15 & 22 & 2.933 & 2.500 & 0.25873 & 0.509 & 5.893 \\ \hline
3 & LV & 21 & 33 & 3.143 & 2.450 & 0.28903 & 0.541 & 5.936 \\ \hline
4 & LV & 24 & 34 & 2.833 & 2.326 & 0.14274 & 0.438 & 6.333 \\ \hline
5 & LV & 186 & 283 & 3.043 & 4.897 & 0.02869 & 0.362 & 7.066 \\ \hline
6 & LV & 10 & 13 & 2.600 & 1.778 & 0.81111 & 0.368 & 6.520 \\ \hline
7 & LV & 63 & 93 & 2.952 & 2.435 & 0.02118 & 0.365 & 5.843 \\ \hline
8 & LV & 28 & 40 & 2.857 & 2.426 & 0.05030 & 0.413 & 6.230 \\ \hline
9 & LV & 133 & 210 & 3.158 & 3.803 & 0.01128 & 0.413 & 5.593 \\ \hline
10 & LV & 124 & 207 & 3.339 & 3.951 & 0.01976 & 0.405 & 6.717 \\ \hline
11 & LV & 31 & 45 & 2.903 & 2.267 & 0.00000 & 0.448 & 6.304 \\ \hline \hline
\multicolumn{9}{|c|}{\textbf{original \textit{order} +75\%}} \\ \hline
1 & LV & 17 & 31 & 3.647 & 1.875 & 0.51036 & 0.574 & 4.333 \\ \hline
2 & LV & 15 & 26 & 3.467 & 2.214 & 0.19222 & 0.647 & 4.446 \\ \hline
3 & LV & 21 & 38 & 3.619 & 2.050 & 0.50913 & 0.489 & 4.627 \\ \hline
4 & LV & 24 & 40 & 3.333 & 2.304 & 0.11278 & 0.476 & 4.757 \\ \hline
5 & LV & 186 & 330 & 3.548 & 4.562 & 0.02511 & 0.409 & 6.651 \\ \hline
6 & LV & 10 & 15 & 3.000 & 1.778 & 0.83333 & 0.385 & 5.040 \\ \hline
7 & LV & 63 & 108 & 3.429 & 2.161 & 0.03188 & 0.388 & 4.682 \\ \hline
8 & LV & 28 & 47 & 3.357 & 2.315 & 0.04837 & 0.435 & 4.418 \\ \hline
9 & LV & 133 & 245 & 3.684 & 3.712 & 0.05067 & 0.443 & 5.044 \\ \hline
10 & LV & 124 & 241 & 3.887 & 3.569 & 0.01706 & 0.413 & 5.372 \\ \hline
11 & LV & 31 & 52 & 3.355 & 2.233 & 0.02222 & 0.457 & 4.796 \\ \hline \hline
\multicolumn{9}{|c|}{\textbf{original \textit{order} +100\%}} \\ \hline
1 & LV & 17 & 36 & 4.235 & 1.813 & 0.59804 & 0.580 & 3.903 \\ \hline
2 & LV & 15 & 30 & 4.000 & 1.857 & 0.25603 & 0.646 & 4.036 \\ \hline
3 & LV & 21 & 44 & 4.190 & 1.850 & 0.62824 & 0.490 & 3.964 \\ \hline
4 & LV & 24 & 46 & 3.833 & 2.283 & 0.15157 & 0.474 & 4.431 \\ \hline
5 & LV & 186 & 378 & 4.065 & 4.362 & 0.03534 & 0.409 & 7.533 \\ \hline
6 & LV & 10 & 18 & 3.600 & 1.667 & 0.69167 & 0.487 & 4.000 \\ \hline
7 & LV & 63 & 124 & 3.937 & 2.161 & 0.02121 & 0.402 & 4.147 \\ \hline
8 & LV & 28 & 54 & 3.857 & 2.130 & 0.03622 & 0.485 & 4.276 \\ \hline
9 & LV & 133 & 280 & 4.211 & 3.652 & 0.07177 & 0.477 & 5.164 \\ \hline
10 & LV & 124 & 276 & 4.452 & 3.301 & 0.03482 & 0.510 & 4.964 \\ \hline
11 & LV & 31 & 60 & 3.871 & 2.167 & 0.02174 & 0.552 & 4.383 \\ \hline
\end{tabular}
\caption{Metrics for dissortative node degree strategy \LV samples evolution.}\label{tab:lvDiss}
\end{center}
\end{footnotesize}
\end{table}

\begin{center}
\begin{table}[htbp]
\centering
\begin{footnotesize}
\begin{tabular}{|@p{1.7cm}|^p{1.5cm}|^p{0.8cm}|^p{0.8cm}|^p{1.5cm}|^p{1.5cm}|^p{1.2cm}|}
\hline
\rowstyle{\bfseries}
Sample  ID  &Network type &\textit{Order} & \textit{Size} & {Avg. betweenness} & {Avg. betw/order} & {Coeff. variation} \\ \hline
\hline
\multicolumn{7}{|c|}{\textbf{original \textit{order} +25\%}} \\ \hline
1 & LV & 17 & 22 & 26 & 1.529 & 0.99 \\ \hline
2 & LV & 15 & 18 & 22.143 & 1.476 & 1.315 \\ \hline
3 & LV & 21 & 27 & 40.571 & 1.932 & 1.204 \\ \hline
4 & LV & 24 & 28 & 55.25 & 2.302 & 1.617 \\ \hline
5 & LV & 186 & 236 & 843.381 & 4.534 & 2.559 \\ \hline
6 & LV & 10 & 11 & 6 & 0.6 & 2 \\ \hline
7 & LV & 63 & 77 & 180.129 & 2.859 & 2.143 \\ \hline
8 & LV & 28 & 33 & 65.643 & 2.344 & 1.31 \\ \hline
9 & LV & 133 & 175 & 431.169 & 3.242 & 3.003 \\ \hline
10 & LV & 124 & 172 & 393.897 & 3.177 & 2.39 \\ \hline
11 & LV & 31 & 37 & 79.032 & 2.549 & 1.808 \\ \hline
\hline
\multicolumn{7}{|c|}{\textbf{original \textit{order} +50\%}} \\ \hline
1 & LV & 17 & 1 & 19 & 1.118 & 1.57 \\ \hline
2 & LV & 15 & 2 & 20.429 & 1.362 & 1.11 \\ \hline
3 & LV & 21 & 3 & 31.714 & 1.51 & 1.788 \\ \hline
4 & LV & 24 & 4 & 36.583 & 1.524 & 2.225 \\ \hline
5 & LV & 186 & 5 & 800.796 & 4.305 & 1.92 \\ \hline
6 & LV & 10 & 6 & 5.556 & 0.556 & 2.828 \\ \hline
7 & LV & 63 & 7 & 102.871 & 1.633 & 4.007 \\ \hline
8 & LV & 28 & 8 & 40.571 & 1.449 & 2.307 \\ \hline
9 & LV & 133 & 9 & 362.397 & 2.725 & 2.76 \\ \hline
10 & LV & 124 & 10 & 377.458 & 3.044 & 2.045 \\ \hline
11 & LV & 31 & 11 & 43.355 & 1.399 & 2.934 \\ \hline
\hline
\multicolumn{7}{|c|}{\textbf{original \textit{order} +75\%}} \\ \hline
1 & LV & 17 & 1 & 13.375 & 0.787 & 2.611 \\ \hline
2 & LV & 15 & 2 & 15 & 1 & 1.021 \\ \hline
3 & LV & 21 & 3 & 22 & 1.048 & 2.876 \\ \hline
4 & LV & 24 & 4 & 33.833 & 1.41 & 1.751 \\ \hline
5 & LV & 186 & 5 & 737.348 & 3.964 & 1.751 \\ \hline
6 & LV & 10 & 6 & 5.556 & 0.556 & 2.828 \\ \hline
7 & LV & 63 & 7 & 77.226 & 1.226 & 4.985 \\ \hline
8 & LV & 28 & 8 & 36.857 & 1.316 & 2.114 \\ \hline
9 & LV & 133 & 9 & 349.13 & 2.625 & 2.351 \\ \hline
10 & LV & 124 & 10 & 328.119 & 2.646 & 2.069 \\ \hline
11 & LV & 31 & 11 & 41.613 & 1.342 & 2.69 \\ \hline
\hline
\multicolumn{7}{|c|}{\textbf{original \textit{order} +100\%}} \\ \hline
1 & LV & 17 & 1 & 11.125 & 0.654 & 3.411 \\ \hline
2 & LV & 15 & 2 & 12.429 & 0.829 & 1.127 \\ \hline
3 & LV & 21 & 3 & 15.81 & 0.753 & 3.915 \\ \hline
4 & LV & 24 & 4 & 32.167 & 1.34 & 1.523 \\ \hline
5 & LV & 186 & 5 & 705.503 & 3.793 & 1.659 \\ \hline
6 & LV & 10 & 6 & 5.333 & 0.533 & 2.763 \\ \hline
7 & LV & 63 & 7 & 74.871 & 1.188 & 4.318 \\ \hline
8 & LV & 28 & 8 & 33 & 1.179 & 2.256 \\ \hline
9 & LV & 133 & 9 & 338.931 & 2.548 & 2.14 \\ \hline
10 & LV & 124 & 10 & 274.627 & 2.215 & 2.247 \\ \hline
11 & LV & 31 & 11 & 39.226 & 1.265 & 2.162 \\ \hline
\end{tabular}
\caption{Betweenness for dissortative node degree strategy \LV samples evolution.}\label{tab:lvDissBet}
\end{footnotesize}
\end{table}
\end{center}

\subsubsection*{Random evolution}

Table~\ref{tab:lvRnd} contains for each sample of the \LV \G (column one) the values for the  main topological quantities: \textit{order} and \textit{size} in columns three and four, respectively; average node degree in the fifth column; the \cpl is reported in column six; the clustering coefficient follows in column seven; robustness is shown in eight column; the cost in term of redundant path length closes the data series (column nine).

The random addition of edges in the network reduces the \cpl in the samples up to 50\%. The first step in the evolution already provides a reduction of 30\% compared to the original non-evolved samples. The two biggest samples are those which benefits the most of the increase of connectivity with a reduction of 70\% or more.  
Considering the clustering coefficient, we observe a tendency that we already noted in the \MV network: triangle creation in the small samples is easier and high values for this metric appear, while the bigger samples have smaller values. The first evolution steps still show two samples with null clustering coefficient, but after this step all the networks have a significant value for this metric. Robustness almost doubles in the whole evolution process compared to the initial values. The addition of edges according to this strategy is beneficial both in facing random failures, and also against targeted attacks where actually there is most of the improvement. Eight of the eleven samples have values in the last step of the evolution that reach values around or higher than 70\%. 
For the significant samples the best improvement is obtained in the second step where on average the redundant path length drops of about 50\%. The last two steps add smaller benefits in order to reach a final reduction about 63\%.

Table~\ref{tab:lvRndBet} contains for each sample of the \MV \G (column one) the values of metrics related to betweenness. In addition to \textit{order} and \textit{size} (columns three and four), average betweenness is provided in columns five, while a value of average betweenness normalized by the \textit{order} of the graph is computed in column six in order to compare the different samples. A measure of the statistical variation of betweenness is the coefficient of variation which is shown in the seventh column.

The random evolution strategy provides a consistent reduction in betweenness that reaches 60\% of the initial value in the final stage of evolution and in the final step all the samples have a value of average betweenness to \textit{order} that is below 3. In addition eight out of the eleven samples have a value that is below two. The improvement through the stages of the evolution is sub-linear with already 35\% improvement in the first step.  The random addition of edges provides also a reduction in the variability of betweenness with a coefficient of variation that in the last stage of the evolution process is on average 15\% lower than in the original samples. In addition, eight out of the eleven samples show a coefficient of variation that is below one. Despite this benefits in the average betweenness and variation, the mode of betweenness stays zero for almost all samples through all the stages of evolution.

\begin{table}[htbp]
\begin{footnotesize}
\begin{center}
\begin{tabular}{|@p{1.7cm}|^p{1.7cm}|^p{0.9cm}|^p{0.6cm}|^p{1cm}|^p{1cm}|^p{1.2cm}|^p{1.5cm}|^p{2cm}|}
\hline
\rowstyle{\bfseries} Sample ID &Network type  & \textit{Order} & \textit{Size} & {Avg. deg.} & {CPL} & {CC} & {Removal robustness ($Rob_N$)} & {Redundancy cost ($APL_{10^{th}}$)} \\ \hline
\hline

\multicolumn{9}{|c|}{\textbf{original \textit{order} +25\%}} \\ \hline
1 & LV & 17 & 22 & 2.588 & 2.563 & 0.12157 & 0.541 & 7.556 \\ \hline
2 & LV & 15 & 18 & 2.400 & 2.786 & 0.00000 & 0.530 & 9.679 \\ \hline
3 & LV & 21 & 27 & 2.571 & 3.150 & 0.00000 & 0.422 & 9.036 \\ \hline
4 & LV & 24 & 28 & 2.333 & 3.413 & 0.06250 & 0.487 & 9.979 \\ \hline
5 & LV & 186 & 236 & 2.538 & 6.038 & 0.00269 & 0.402 & 9.931 \\ \hline
6 & LV & 10 & 11 & 2.200 & 2.222 & 0.06000 & 0.365 & 5.040 \\ \hline
7 & LV & 63 & 77 & 2.444 & 3.823 & 0.04321 & 0.352 & 8.029 \\ \hline
8 & LV & 28 & 33 & 2.357 & 3.759 & 0.08095 & 0.445 & 7.408 \\ \hline
9 & LV & 133 & 175 & 2.632 & 5.205 & 0.01078 & 0.435 & 8.984 \\ \hline
10 & LV & 124 & 172 & 2.774 & 4.805 & 0.03015 & 0.407 & 8.297 \\ \hline
11 & LV & 31 & 37 & 2.387 & 3.467 & 0.01183 & 0.460 & 8.508 \\ \hline \hline
\multicolumn{9}{|c|}{\textbf{original \textit{order} +50\%}} \\ \hline
1 & LV & 17 & 27 & 3.176 & 2.375 & 0.29412 & 0.677 & 5.181 \\ \hline
2 & LV & 15 & 22 & 2.933 & 2.357 & 0.05556 & 0.743 & 5.964 \\ \hline
3 & LV & 21 & 33 & 3.143 & 2.500 & 0.01587 & 0.598 & 5.609 \\ \hline
4 & LV & 24 & 34 & 2.833 & 2.652 & 0.05278 & 0.613 & 6.347 \\ \hline
5 & LV & 186 & 283 & 3.043 & 4.830 & 0.00520 & 0.546 & 8.064 \\ \hline
6 & LV & 10 & 13 & 2.600 & 2.056 & 0.34667 & 0.441 & 6.080 \\ \hline
7 & LV & 63 & 93 & 2.952 & 3.435 & 0.04780 & 0.403 & 6.180 \\ \hline
8 & LV & 28 & 40 & 2.857 & 3.481 & 0.16378 & 0.438 & 5.903 \\ \hline
9 & LV & 133 & 210 & 3.158 & 4.318 & 0.02016 & 0.683 & 7.078 \\ \hline
10 & LV & 124 & 207 & 3.339 & 4.045 & 0.02746 & 0.580 & 6.771 \\ \hline
11 & LV & 31 & 45 & 2.903 & 2.833 & 0.01183 & 0.690 & 6.379 \\ \hline \hline
\multicolumn{9}{|c|}{\textbf{original \textit{order} +75\%}} \\ \hline
1 & LV & 17 & 31 & 3.647 & 2.188 & 0.34146 & 0.745 & 4.528 \\ \hline
2 & LV & 15 & 26 & 3.467 & 2.071 & 0.17778 & 0.689 & 4.786 \\ \hline
3 & LV & 21 & 38 & 3.619 & 2.300 & 0.12748 & 0.633 & 4.564 \\ \hline
4 & LV & 24 & 40 & 3.333 & 2.457 & 0.15615 & 0.621 & 5.347 \\ \hline
5 & LV & 186 & 330 & 3.548 & 4.251 & 0.01116 & 0.700 & 6.807 \\ \hline
6 & LV & 10 & 15 & 3.000 & 1.889 & 0.28667 & 0.508 & 5.480 \\ \hline
7 & LV & 63 & 108 & 3.429 & 3.081 & 0.09340 & 0.438 & 5.438 \\ \hline
8 & LV & 28 & 47 & 3.357 & 2.870 & 0.22636 & 0.528 & 5.541 \\ \hline
9 & LV & 133 & 245 & 3.684 & 3.886 & 0.03401 & 0.715 & 6.126 \\ \hline
10 & LV & 124 & 241 & 3.887 & 3.589 & 0.03719 & 0.706 & 5.783 \\ \hline
11 & LV & 31 & 52 & 3.355 & 2.667 & 0.06237 & 0.712 & 5.554 \\ \hline \hline
\multicolumn{9}{|c|}{\textbf{original \textit{order} +100\%}} \\ \hline
1 & LV & 17 & 36 & 4.235 & 2.063 & 0.38277 & 0.750 & 4.083 \\ \hline
2 & LV & 15 & 30 & 4.000 & 1.929 & 0.14000 & 0.722 & 4.214 \\ \hline
3 & LV & 21 & 44 & 4.190 & 2.100 & 0.12611 & 0.635 & 4.255 \\ \hline
4 & LV & 24 & 46 & 3.833 & 2.283 & 0.15258 & 0.744 & 4.757 \\ \hline
5 & LV & 186 & 378 & 4.065 & 3.886 & 0.02847 & 0.754 & 6.114 \\ \hline
6 & LV & 10 & 18 & 3.600 & 1.722 & 0.31667 & 0.737 & 4.320 \\ \hline
7 & LV & 63 & 124 & 3.937 & 2.935 & 0.10115 & 0.451 & 5.012 \\ \hline
8 & LV & 28 & 54 & 3.857 & 2.556 & 0.25003 & 0.620 & 4.679 \\ \hline
9 & LV & 133 & 280 & 4.211 & 3.538 & 0.04299 & 0.731 & 5.646 \\ \hline
10 & LV & 124 & 276 & 4.452 & 3.289 & 0.04395 & 0.744 & 5.179 \\ \hline
11 & LV & 31 & 60 & 3.871 & 2.533 & 0.12734 & 0.696 & 4.888 \\ \hline
\end{tabular}
\caption{Metrics for random strategy \LV samples evolution.}\label{tab:lvRnd}
\end{center}
\end{footnotesize}
\end{table}

\begin{center}
\begin{table}[h]
\centering
\begin{footnotesize}
\begin{tabular}{|@p{1.7cm}|^p{1.5cm}|^p{0.8cm}|^p{0.8cm}|^p{1.5cm}|^p{1.5cm}|^p{1.2cm}|}
\hline
\rowstyle{\bfseries}
Sample  ID  &Network type &\textit{Order} & \textit{Size} & {Avg. betweenness} & {Avg. betw/order} & {Coeff. variation} \\ \hline
\hline
\multicolumn{7}{|c|}{\textbf{original \textit{order} +25\%}} \\ \hline
1 & LV & 17 & 22 & 20.667 & 1.216 & 0.78 \\ \hline
2 & LV & 15 & 18 & 23.429 & 1.562 & 0.854 \\ \hline
3 & LV & 21 & 27 & 42.19 & 2.009 & 1.256 \\ \hline
4 & LV & 24 & 28 & 59.833 & 2.493 & 0.841 \\ \hline
5 & LV & 186 & 236 & 898.716 & 4.832 & 1.213 \\ \hline
6 & LV & 10 & 11 & 8 & 0.8 & 1.256 \\ \hline
7 & LV & 63 & 77 & 191.129 & 3.034 & 1.724 \\ \hline
8 & LV & 28 & 33 & 82 & 2.929 & 1.263 \\ \hline
9 & LV & 133 & 175 & 558.279 & 4.198 & 1.195 \\ \hline
10 & LV & 124 & 172 & 452.348 & 3.648 & 1.146 \\ \hline
11 & LV & 31 & 37 & 75.161 & 2.425 & 1.226 \\ \hline \hline
\multicolumn{7}{|c|}{\textbf{original \textit{order} +50\%}} \\ \hline
1 & LV & 17 & 1 & 16.933 & 0.996 & 0.858 \\ \hline
2 & LV & 15 & 2 & 17.857 & 1.19 & 0.716 \\ \hline
3 & LV & 21 & 3 & 32 & 1.524 & 0.968 \\ \hline
4 & LV & 24 & 4 & 41.833 & 1.743 & 0.949 \\ \hline
5 & LV & 186 & 5 & 693.125 & 3.726 & 0.923 \\ \hline
6 & LV & 10 & 6 & 8.889 & 0.889 & 1.497 \\ \hline
7 & LV & 63 & 7 & 150.548 & 2.39 & 1.469 \\ \hline
8 & LV & 28 & 8 & 72.286 & 2.582 & 1.384 \\ \hline
9 & LV & 133 & 9 & 431.308 & 3.243 & 0.916 \\ \hline
10 & LV & 124 & 10 & 375.254 & 3.026 & 1.065 \\ \hline
11 & LV & 31 & 11 & 60 & 1.935 & 0.85 \\ \hline
\hline
\multicolumn{7}{|c|}{\textbf{original \textit{order} +75\%}} \\ \hline
1 & LV & 17 & 1 & 14.533 & 0.855 & 0.899 \\ \hline
2 & LV & 15 & 2 & 14.857 & 0.99 & 0.764 \\ \hline
3 & LV & 21 & 3 & 26.571 & 1.265 & 1.087 \\ \hline
4 & LV & 24 & 4 & 37 & 1.542 & 1.061 \\ \hline
5 & LV & 186 & 5 & 603.438 & 3.244 & 0.859 \\ \hline
6 & LV & 10 & 6 & 7.778 & 0.778 & 1.301 \\ \hline
7 & LV & 63 & 7 & 129.71 & 2.059 & 1.528 \\ \hline
8 & LV & 28 & 8 & 56.786 & 2.028 & 1.061 \\ \hline
9 & LV & 133 & 9 & 381.847 & 2.871 & 0.915 \\ \hline
10 & LV & 124 & 10 & 324.644 & 2.618 & 1.048 \\ \hline
11 & LV & 31 & 11 & 51.032 & 1.646 & 0.899 \\ \hline
\hline
\multicolumn{7}{|c|}{\textbf{original \textit{order} +100\%}} \\ \hline
1 & LV & 17 & 1 & 12.4 & 0.729 & 0.97 \\ \hline
2 & LV & 15 & 2 & 12.143 & 0.81 & 0.646 \\ \hline
3 & LV & 21 & 3 & 23.238 & 1.107 & 0.979 \\ \hline
4 & LV & 24 & 4 & 31.75 & 1.323 & 0.808 \\ \hline
5 & LV & 186 & 5 & 534.156 & 2.872 & 0.812 \\ \hline
6 & LV & 10 & 6 & 6.667 & 0.667 & 1.095 \\ \hline
7 & LV & 63 & 7 & 120.097 & 1.906 & 1.442 \\ \hline
8 & LV & 28 & 8 & 42.5 & 1.518 & 1.269 \\ \hline
9 & LV & 133 & 9 & 339.893 & 2.556 & 0.86 \\ \hline
10 & LV & 124 & 10 & 288.254 & 2.325 & 0.989 \\ \hline
11 & LV & 31 & 11 & 45.484 & 1.467 & 0.891 \\ \hline
\end{tabular}
\caption{Betweenness for random strategy \LV samples evolution.}\label{tab:lvRndBet}
\end{footnotesize}
\end{table}
\end{center}

\section{Discussion}\label{sec:discussion}

To determine which strategy, from a topological point of view, is more rewarding and provides the higher benefits one needs to compare the results presented in Section~\ref{sec:results}. If determining a ``best'' strategy is unfeasible, one should aim at identifying trade-offs between one strategy compared to another one. We evaluate all the strategies considering average values over all the samples and considering the satisfaction of the topological metrics required for the \SG explained in~\cite{PaganiEvol2013} and summarized in~\ref{sec:metrics}.  We consider, as we have done in the presentation of results, the \MV \G and the \LV \G separated and we compare each type of \G in a separate subsection.

\subsection{Comparison of the Evolution Strategies for the \MV Distribution Grid}

It is difficult and perhaps even wrong to establish ``the winner'' between the various evolution strategies proposed. Also because the improvement in the various phases of evolution are not linear, a strategy that scores best in one evolution step could result worse compared to another when more edges are added.

As a general remark we reinforce that the addition of links is beneficial to the topological metrics analyzed and one might speculate that the optimal solution in order to minimize (or maximize) the metrics is to realize a completely connected network (i.e., clique). With a real infrastructure such as the \PG this is impossible to achieve from the economical point of view, but also from a technical point of view (e.g., substations receiving thousand of connections).

The comparison of the averages over all samples in each evolution step for the various evolution strategies is shown in Table~\ref{tab:mvCompar}. Take the triangle closure strategy, for instance, it is the one that in every step of the evolution process leads to the maximal values of the clustering coefficient, but it is also the one that has the worst results together with the least distance strategy for the metrics related to path length (\cpl and $10^{th}$ redundant average path length). In addition, considering the robustness metric, the triangle closure offers the smallest improvements. Also considering the betweenness-related metrics, the triangle closure strategy results the worst compared to all the others with the second set of highest values for average betweenness and an increasing coefficient of variation. The same considerations apply to the least distance strategy that to a certain extent is very similar to the triangle closure. This strategy is the worst concerning the path length, having a \cpl more than double than the best evolution strategy (i.e., random); the same also applies for the $10^{th}$ redundant average path length. Concerning redundancy aspects at the end of the evolution process, the least distance strategy is the one (except for random) that have the best robustness results. In terms of metrics related to betweenness, the least distance strategy shows results that are even worse compared to the triangle closure one. Considering the high degree assortative evolution strategy the results show that, overall, it is the best evolution strategy in the first two steps of the evolution; in the first step it is even better than the random one (with the exception of robustness). Considering betweenness metrics, this strategy is second only to the random strategy, with the exception of coefficient of variation which is worse than all other metrics. In general, this is a good strategy of evolving the network especially concerning the path-related aspects, with just robustness that lacks compared to other evolution strategies. The assortative low degree strategy is not particularly appealing and it is outrun by the assortative high degree strategy in every metric except robustness and  coefficient of variation of betweenness, therefore we do not find it particularly interesting for evolution purposes of the \MV networks. The dissortative strategy does not excel particularly in any of the metrics considered, however its values are quite fair, especially in the initial stage of the evolution. In particular, since the values of the metric do not improve particularly in the following steps of the evolution process, such evolution strategy could be used as a slightly more robust alternative compared to the assortative high degree in those scenario where the number of edges to be added is minimal and there is no need of special excellence in one topological parameter. When enough connectivity is added, the evolution strategy that scores best compared to all the others for all the metrics, with the exception of the \cc metric, is the random one. It has already been noted by Casals \etal~\cite{Rosas-Casals2009} that some randomness in the network is beneficial especially for aspects related to robustness. In our analysis we have the same general results, with \cpl about 4 and a $10^{th}$ redundant \apl of just 6. In addition, the networks evolved according to this strategy are the most robust with a value higher than 0.7 when the maximal connectivity is reached. The same considerations apply for betweenness metrics that obtain the best results when the connectivity is enhanced reaching a betweenness to \textit{order} ratio smaller than three and a coefficient of variation below one at the last evolution step. However, it is difficult to propose for a Distribution provider to improve his \G in a random fashion, even if part of the weaknesses and inefficiencies come from the lack of such randomness. Considering rational and evolution strategies that come with a motivation we consider the strategy assortative high degree as the one that, by evaluating the various topological metrics, scores best in the evolution tests that we have performed on the Dutch \MV Grids.

\begin{table}[htbp]
\centering
\begin{footnotesize}
\begin{tabular}{|l||l|r|r|r|r|r|r|r|}
\hline
 & Evolution step & \multicolumn{1}{l|}{CPL} & \multicolumn{1}{l|}{CC} & \multicolumn{1}{l|}{rob} & \multicolumn{1}{l|}{$10^{th}$ red. path} & \multicolumn{1}{l|}{avg. bet.} & \multicolumn{1}{l|}{bet./\textit{order}} & \multicolumn{1}{l|}{coeff. var.} \\ \hline \hline
\multirow{4}{*}{Assortative high degree} & +25\% & 5.757 & 0.037 & 0.326 & 8.550 & 1682.845 & 4.744 & 2.258 \\ \cline{2-9}
 & +50\% & 5.121 & 0.067 & 0.344 & 7.095 & 1369.399 & 4.100 & 2.259 \\\cline{2-9}
 & +75\% & 4.724 & 0.098 & 0.361 & 6.493 & 1245.963 & 3.799 & 2.147 \\ \cline{2-9}
 & +100\% & 4.558 & 0.122 & 0.367 & 6.072 & 1226.062 & 3.717 & 1.904 \\ \hline
\multirow{4}{*}{Assortative low degree} & +25\% & 6.404 & 0.045 & 0.358 & 9.503 & 1955.047 & 5.598 & 2.039 \\ \cline{2-9}
 & +50\% & 6.320 & 0.090 & 0.408 & 8.492 & 1931.053 & 5.508 & 1.651 \\ \cline{2-9}
 & +75\% & 6.261 & 0.109 & 0.482 & 8.149 & 1913.409 & 5.440 & 1.504 \\ \cline{2-9}
 & +100\% & 6.172 & 0.121 & 0.496 & 8.014 & 1888.415 & 5.357 & 1.441 \\ \hline
\multirow{4}{*}{Triangle closure} & +25\% & 10.761 & 0.165 & 0.257 & 14.559 & 2698.910 & 8.130 & 1.786 \\ \cline{2-9}
 & +50\% & 9.504 & 0.288 & 0.321 & 11.680 & 2377.892 & 7.123 & 1.872 \\ \cline{2-9}
 & +75\% & 8.482 & 0.388 & 0.341 & 10.053 & 2104.554 & 6.289 & 1.927 \\ \cline{2-9}
 & +100\% & 7.618 & 0.483 & 0.384 & 8.997 & 1884.527 & 5.563 & 2.034 \\ \hline
\multirow{4}{*}{Dissortative}  & +25\% & 5.803 & 0.019 & 0.349 & 8.891 & 1727.708 & 4.949 & 2.433 \\ \cline{2-9}
 & +50\% & 5.591 & 0.028 & 0.408 & 7.904 & 1645.824 & 4.710 & 1.953 \\ \cline{2-9}
 & +75\% & 5.402 & 0.034 & 0.463 & 7.421 & 1594.361 & 4.532 & 1.739 \\ \cline{2-9}
 & +100\% & 5.217 & 0.033 & 0.468 & 7.053 & 1559.650 & 4.418 & 1.613 \\ \hline
\multirow{4}{*}{Least distance} & +25\% & 10.025 & 0.141 & 0.272 & 13.078 & 3044.817 & 9.042 & 1.703 \\ \cline{2-9}
 & +50\% & 9.066 & 0.214 & 0.360 & 11.696 & 2789.684 & 8.127 & 1.744 \\ \cline{2-9}
 & +75\% & 8.535 & 0.261 & 0.455 & 10.705 & 2646.133 & 7.652 & 1.788 \\ \cline{2-9}
 & +100\% & 8.109 & 0.304 & 0.523 & 10.013 & 2525.651 & 7.275 & 1.807 \\ \hline
\multirow{4}{*}{Random} & +25\% & 6.249 & 0.011 & 0.423 & 9.954 & 1785.946 & 5.265 & 1.194 \\ \cline{2-9}
 & +50\% & 4.990 & 0.016 & 0.634 & 7.780 & 1378.933 & 3.998 & 1.074 \\ \cline{2-9}
 & +75\% & 4.393 & 0.019 & 0.737 & 6.739 & 1178.359 & 3.394 & 0.964 \\ \cline{2-9}
 & +100\% & 4.000 & 0.027 & 0.764 & 6.080 & 1044.279 & 2.988 & 0.881 \\ \hline
\end{tabular}
\end{footnotesize}
\caption{Comparison of evolution strategies for \MV network.}\label{tab:mvCompar}
\end{table}

\subsection{Comparison of the Evolution Strategies for the \LV Distribution Grid}

As for the \MV evolution strategies, one cannot declare a winner. Also in this case some improvements between the various phases of evolution are not linear, therefore a strategy that scores best in one evolution step could result worse compared to another one when more edges are added.  We point out once again that the addition of links is beneficial to the topological metrics analyzed and one might speculate that the optimal solution to minimize (or maximize) the metrics is to realize a completely connected network (i.e., clique). With a real infrastructure such as the \PG this is impossible to achieve due to economical and techncial obvious considerations.

The comparison of the averages over all samples in each evolution step for the various evolution strategies is shown in Table~\ref{tab:lvCompar}. Once again, the strategy that scores best in comparison with the others is the random one for robustness and for metrics related to betweenness. For robustness in the last stage of evolution the value reaches on average 0.7. For betweenness-related metrics the two final stages of evolution are the best ones for average betweenness and average betweenness to \textit{order} ratio. This strategy is the only one for which the coefficient of variation results below one (on average for all the samples) in the final stage of evolution. However, the random strategy has its weak point in the \cc metric that score below all the others evolution strategies. Considering the assortative strategies, one sees that the two strategies have quite similar scores for the metrics. In the very first stage, the high degree assortative strategy is better, but later in the evolution the assortative low degree slightly outperforms the assortative high degree one for \cc and characteristic path length, while for the robustness the difference is limited. The results between these two metrics for betweenness are quite similar and here the only interesting difference is in the coefficient of variation which scores best for the high node degree strategy. These two strategies are comparable to the random one for \cpl matters, while scoring worse for robustness, but better for clustering. Considering the \cc metric alone, the strategy that outperforms the others is the triangle closure strategy with values that are double compared to the others at the end of the evolution process. However, this strategy is worse than all the others for the characteristic path length. Although it has the highest values for average betweenness, the coefficient of variation ranks second compared to the other strategies considered. The strategy that slightly outperforms the others (except the values for robustness of the random one) is the dissortative strategy. In fact, the path length is just smaller than 2.5 even better than the random edge addition. The clustering coefficient is in line with the assortative strategies and scores around 0.5. Concerning betweenness, this strategy scores best considering the average betweenness, but for the coefficient of variation this strategy scores worst.
In the \LV Distribution \G it is even more difficult to propose a strategy to follow for the evolution given the similarity for the values of the evolution strategies considered. Considering non-random evolution strategies we consider the dissortative strategy   that scores best in the evolution tests that we have performed on the Dutch \LV Grids.

\begin{table}[htbp]
\centering
\begin{footnotesize}
\begin{tabular}{|l||l|r|r|r|r|r|r|r|}
\hline
 & Evolution step & \multicolumn{1}{l|}{CPL} & \multicolumn{1}{l|}{CC} & \multicolumn{1}{l|}{rob} & \multicolumn{1}{l|}{$10^{th}$ red. path} & \multicolumn{1}{l|}{avg. bet.} & \multicolumn{1}{l|}{bet./\textit{order}} & \multicolumn{1}{l|}{coeff. var.} \\ \hline \hline
\multirow{4}{*}{Assortative high degree} & +25\% & 3.600 & 0.06248 & 0.408 & 7.535 & 207.959 & 2.515 & 1.518 \\ \cline{2-9}
 & +50\% & 3.225 & 0.15702 & 0.432 & 5.797 & 191.974 & 2.227 & 1.564 \\\cline{2-9}
 & +75\% & 2.973 & 0.23422 & 0.457 & 5.059 & 182.954 & 2.043 & 1.678\\ \cline{2-9}
 & +100\% & 2.729 & 0.27697 & 0.481 & 4.977 & 174.593 & 1.830 & 1.772 \\ \hline
\multirow{4}{*}{Assortative low degree} & +25\% & 3.866 & 0.05714 & 0.405 & 8.865 & 226.416 & 2.869 & 1.436 \\ \cline{2-9}
 & +50\% & 3.074 & 0.12673 & 0.446 & 6.310 & 188.824 & 2.093 & 1.746 \\ \cline{2-9}
 & +75\% & 2.827 & 0.22316 & 0.509 & 5.187 & 177.928 & 1.825 & 2.113 \\ \cline{2-9}
 & +100\% & 2.766 & 0.27706 & 0.551 & 4.730 & 175.312 & 1.765 & 1.939 \\ \hline
\multirow{4}{*}{Triangle closure} & +25\% & 5.908 & 0.13071 & 0.346 & 9.028 & 488.127 & 4.599 & 1.202 \\ \cline{2-9}
 & +50\% & 5.463 & 0.24961 & 0.383 & 8.070 & 461.893 & 2.044 & 1.247 \\ \cline{2-9}
 & +75\% & 5.020 & 0.35457 & 0.456 & 6.973 & 419.114 & 1.850 & 1.310 \\ \cline{2-9}
 & +100\% & 4.686 & 0.46154 & 0.497 & 6.573 & 392.781 & 1.732 & 1.428\\ \hline
\multirow{4}{*}{Dissortative}  & +25\% & 3.396 & 0.09343 & 0.403 & 7.923 & 194.838 & 2.413 & 1.849 \\ \cline{2-9}
 & +50\% & 2.826 & 0.17720 & 0.435 & 6.168 & 167.339 & 1.875 & 2.318 \\ \cline{2-9}
 & +75\% & 2.616 & 0.21392 & 0.465 & 4.924 & 150.914 & 0.667 & 2.459 \\ \cline{2-9}
 & +100\% & 2.476 & 0.23151 & 0.501 & 4.618 & 140.275 & 0.619 & 2.502 \\ \hline
\multirow{4}{*}{Random} & +25\% & 3.748 & 0.03852 & 0.440 & 8.404 & 219.250 & 2.650 & 1.159\\ \cline{2-9}
 & +50\% & 3.171 & 0.09465 & 0.583 & 6.323 & 172.730 & 2.113 & 1.054 \\ \cline{2-9}
 & +75\% & 2.841 & 0.14127 & 0.636 & 5.450 & 149.836 & 0.642 & 1.038 \\ \cline{2-9}
 & +100\% & 2.621 & 0.15564 & 0.689 & 4.831 & 132.417 & 0.567 & 0.978 \\ \hline
\end{tabular}
\end{footnotesize}
\caption{Comparison of evolution strategies for \LV network.}\label{tab:lvCompar}
\end{table}

%[MARCO: IS A COMPARISON BETWEEN THE MV AND LV NEEDED?] non direi

\subsection{Desiderata parameters satisfaction for \MV networks}\label{sec:desMV}

We now consider the satisfaction of the quantitative metrics that we have proposed in our previous work~\cite{pag:preprint12,PaganiEvol2013} to analyze the appropriateness of synthetic topologies to improve the Grid in reducing losses, facilitate local energy distribution, and increasing network robustness. The metrics  that we require to be satisfied in the future Smart Grid can be can be categorized into three macro categories with respect to how they affect a Power Grid: efficiency in the transfer of energy, resilience in providing alternative path if part of the network is compromised/congested, and robustness to failures for network connectivity. The satisfaction of these metrics, provides topologies where the cost of electricity should be reduced since the considered metrics directly influence the cost of distributing electricity~\cite{PaganiAielloTSG2011}. For a through insight of the metrics, we refer to the~\ref{sec:metrics} or to~\cite{PaganiAielloTSG2011} where a detailed explanation is provided.%\footnote{make this sentence more clear and explain more. Which metrics? why are these the one to satisfy? DONE}
The results are shown graphically in Tables~\ref{tab:desiderataSatisfMV1} and~\ref{tab:desiderataSatisfMV2} where for each evolution strategy and step of evolution the satisfaction of the parameter is marked with a ``tick'' sign (\tick), while the dissatisfaction is marked with a cross sign (\cross), and a value closer to satisfaction by an approximation sign ($\approx$).

\paragraph*{Assortative high degree}
The assortative high degree evolution strategy satisfies the desired values for the metrics concerning the \cpl already from the second step of evolution for ten out of twelve samples. All samples satisfy the requirements over the redundant path. Considering the clustering coefficient, the majority of samples satisfies the requirement at the third step of evolution where seven samples have a $CC \geq 5\times CC_{RG}$; in the final stage of evolution this metric is satisfied by all the samples except two. The real drawback of this metric, as shown in the table, is represented by the robustness that never reaches the goal of 0.45, but stops around 0.37. The unsatisfaction of the metric is also present for betweenness ones: all samples are one unit larger even in the last step of evolution. In addition, the coefficient of variation never reaches the target for the samples at any stage of evolution. 

\paragraph*{Assortative low degree}
The assortative low degree evolution strategy satisfies the desired values for the metrics concerning the \cc already from the second step of evolution with eleven out of twelve samples. All the samples satisfy the requirements over the redundant path already from the first step. Considering the clustering coefficient, the majority of samples satisfies the requirement at the second step of evolution where seven samples have a $CC \geq 5\times CC_{RG}$; in the final stage of evolution this metric is satisfied by all the samples except one. This metric almost provides the satisfaction of the robustness requirement from the third step on by having six samples fully satisfying it and three very close to the 0.45 threshold. The metrics concerning betweenness are not satisfied both for betweenness to \textit{order} ratio and for the coefficient of variation which never reach the target for the samples at any stage of evolution. 

\paragraph*{Triangle closure}
The triangle closure evolution strategy focuses on the clustering coefficient, therefore reaching the target already in the first step of the evolution. The evolved graphs do not only satisfy the requirement posed of having $CC \geq 5\times CC_{RG}$, but also a more restrictive requirement of $CC \geq 10\times CC_{RG}$ which can be considered the condition for satisfying the \sw requirement for this property cf.~\cite{Watts98}. Concerning the path length properties, the \cpl is never lower than the logarithm of the \textit{order} of the graph. The requirement over the $10^{th}$ \apl is satisfied already from the first addition of edges. On all other requirements this strategy is weak. For robustness just few samples reach values around 0.4, while the majority is about 0.3 when the most of the edges are added. For metrics that involve betweenness, this evolution strategy scores poorly and never even close for any single sample to the target for both average betweenness to \textit{order} ratio and for the coefficient of variation.

\paragraph*{Dissortative node degree}
The dissortative node degree strategy scores quite poorly. As mentioned before, this strategy does not excel in one specific metric, but all the metrics are slightly improved. Such aspects result in a limited crossing of the threshold for the desiderata parameters imposed. Only after the third evolution step the \cpl is almost satisfied: nine of the twelve samples reach the target, while one more reaches the target when even more edges are added. A similar condition is true for the robustness metric. In fact, six samples satisfy the target after the third step and other three are above the 0.4 value. Clustering coefficient is always well below the threshold, having values that often are just slightly higher than the clustering coefficient of a random graph with same \textit{order} and \textit{size}. The only metric fully satisfied is the one related to redundant paths in the network. While the metrics related to betweenness are not satisfied even at later stages of evolution.  

\paragraph*{Least distance}
The results of the least distance strategy are similar to those of triangle closure. The clustering coefficient desiderata is reached already in the first step of the evolution. From the second step on, the evolved graphs do not only satisfy the requirement posed of having $CC \geq 5\times CC_{RG}$, but also a more restrictive requirement of $CC \geq 10\times CC_{RG}$ which can be considered the condition for satisfying the \sw requirement for this property cf.~\cite{Watts98}. Concerning the path length properties, the \cpl is never lower than the logarithm of the \textit{order} of the graph. However, the easy requirement related to the $10^{th}$ \apl is satisfied already from the first addition of edges. Robustness is another positive note of this evolution strategy. From the third step on the seven samples reach the 0.45 target and other three are close to 0.4; in the last step basically all the \MV evolved samples satisfy the desiderata. The situation is not so positive for betweenness related metrics since they score worst compared to the others strategies for both average betweenness to \textit{order} ratio and for the coefficient of variation, therefore having two crosses for these cells in the table. 

\paragraph*{Random}
The random evolution strategy is the one that satisfies most of the desiderata parameters not only after the first step of evolution, but already after the second one. The metrics concerning the \cpl are satisfied basically already from the second step of evolution. The same applies to the redundant path whose goal is met already in the first evolution step. Robustness is really impressively achieved after the addition of 50\% of more edges, with all the samples above the 0.45 threshold and the majority of them scoring even higher, above 0.6. The metrics related to betweenness that usually fail for the other samples are here met completely on the coefficient of variation side after the third evolution stage (eight out the twelve samples meet the target). In addition, this evolution strategy is the one that also goes closer to the satisfaction of the  betweenness to \textit{order} ratio. The only problem of such a strategy is on the \cc side. With such a strategy the formation or closure of topological triangle structures is really difficult, therefore the cross sign for this parameter. 

\begin{sidewaystable}[htbp]
\centering
\begin{footnotesize}
\begin{tabular}{|l|c|c|c|c|c|c|c|c|c|c|c|c|}
  \hline
  \textbf{Desiderata}&\multicolumn{4}{|c|}{\textbf{Assortative HD}}&\multicolumn{4}{|c|}{\textbf{Assortative LD}}&\multicolumn{4}{|c|}{\textbf{Triangle closure}} \\
  \hline
  & \textit{order}+25\% & \textit{order}+50\%&\textit{order}+75\%&\textit{order}+100\%& \textit{order}+25\% & \textit{order}+50\%&\textit{order}+75\%&\textit{order}+100\%& \textit{order}+25\% & \textit{order}+50\%&\textit{order}+75\%&\textit{order}+100\%\\ \hline
  %Modularity & $\approx$ &\cross &\cross &\tick & $\approx$ &\cross &\cross &\tick & $\approx$ &\cross &\cross &\tick \\  \hline
  $CPL\leq ln(N)$  & \cross &$\approx$ &\tick &\tick & \cross &\cross &\cross &\cross & \cross &\cross &\cross & \cross \\  \hline
  $CC\geq 5\times CC_{RG}$ & \cross &\cross & $\approx$ &\tick & \cross &\tick &\tick &\tick & \tick &\tick &\tick &\tick \\  \hline
  $\overline{\upsilon}=\frac{\overline{\sigma}}{N} \approx 2.5$ & \cross &\cross &\cross & \cross & \cross &\cross &\cross &\cross & \cross &\cross &\cross &\cross \\ \hline
  $c_v \leq 1$ &\cross &\cross &\cross & \cross & \cross &\cross &\cross &\cross & \cross &\cross &\cross &\cross \\ \hline
  $Rob_N \geq 0.45$ & \cross &\cross &\cross & \cross & \cross &\cross &$\approx$ &$\approx$ & \cross &\cross &\cross &\cross \\ \hline
  $APL_{10^{th}} \leq 2 \times CPL$ & \tick &\tick &\tick &\tick & \tick &\tick &\tick &\tick & \tick &\tick &\tick &\tick \\ \hline
\end{tabular}
\end{footnotesize}
\caption{Desiderata parameter compliance fro \MV networks evolved according the assortative high degree, assortative low degree, and triangle closure strategy.}\label{tab:desiderataSatisfMV1}
\end{sidewaystable}

\begin{sidewaystable}[htbp]
\centering
\begin{footnotesize}
\begin{tabular}{|l|c|c|c|c|c|c|c|c|c|c|c|c|}
  \hline
  \textbf{Desiderata}&\multicolumn{4}{|c|}{\textbf{Dissortative}} &\multicolumn{4}{|c|}{\textbf{Least dist.}} &\multicolumn{4}{|c|}{\textbf{Random}} \\
  \hline
  & \textit{order}+25\% & \textit{order}+50\%&\textit{order}+75\%&\textit{order}+100\%& \textit{order}+25\% & \textit{order}+50\%&\textit{order}+75\%&\textit{order}+100\%& \textit{order}+25\% & \textit{order}+50\%&\textit{order}+75\%&\textit{order}+100\%\\ \hline
  %Modularity & $\approx$ &\cross &\cross &\tick & $\approx$ &\cross &\cross &\tick & $\approx$ &\cross &\cross &\tick \\  \hline
  $CPL\leq ln(N)$  & \cross &\cross &$\approx$ &$\approx$ & \cross &\cross &\cross &\cross &\cross &\tick  &\tick &\tick \\  \hline
  $CC\geq 5\times CC_{RG}$ & \cross &\cross &\cross &\cross & \tick  &\tick &\tick &\tick & \cross &\cross &\cross &\cross \\  \hline
  $\overline{\upsilon}=\frac{\overline{\sigma}}{N} \approx 2.5$ & \cross &\cross &\cross & \cross & \cross &\cross &\cross &\cross & \cross &\cross &\cross &\cross \\ \hline
  $c_v \leq 1$ &\cross &\cross &\cross & \cross & \cross &\cross &\cross &\cross & \cross &$\approx$ &\tick &\tick \\ \hline
  $Rob_N \geq 0.45$ & \cross &\cross &$\approx$ &$\approx$ & \cross &\cross & $\approx$&\tick &\cross  &\tick &\tick &\tick\\ \hline
  $APL_{10^{th}} \leq 2 \times CPL$ & \tick &\tick &\tick &\tick & \tick &\tick &\tick &\tick & \tick &\tick &\tick &\tick \\ \hline
\end{tabular}
\end{footnotesize}
\caption{Desiderata parameter compliance for \MV networks evolved according the dissortative, least distance, and random strategy.}\label{tab:desiderataSatisfMV2}
\end{sidewaystable}

\bigskip

As a general remark about the satisfaction of the desiderata parameters is what we have in part already noted in the comparison of the result for topological quantities. The random evolution strategy is the one that satisfies most of the desiderata requirements  (three parameters fully satisfied and one almost satisfied) already at the second stage of evolution and from the third step on, four requirements are fully satisfied. Concerning the other strategies that have an evolution with a specific goal, those that satisfy most of the requirements are the assortative high degree and the least distance one. Both strategies satisfy three out of the six parameters and the difference being that the assortative one has good performance for the \cpl parameters, while the least distance strategy reaches the target for the robustness aspects. Therefore, these sub optimal strategies could be used where such different requirements  are most needed.

%\footnote{da qualche parte dobbiamo trarre delle conclusioni e dire cosa ne pensiamo rispetto a quanti passi fare, e quale strategia utilizzare per quale evoluzione, un po' come abbiamo fatto per TSG2, dicendo che small worlds con K=4 era la scelta piu' promettente. FATTO}

From the comparison of the different evolution strategies and their steps in adding new edges as shown in Tables~\ref{tab:desiderataSatisfMV1} and~\ref{tab:desiderataSatisfMV2}, there are evolution strategies that tend to satisfy the majority of the metrics for the \SG that we have defined (c.f.~\ref{sec:metrics} or~\cite{pag:preprint12}). The evolution strategy that already in the second step fully satisfies three requirements is the random evolution. The requirements satisfied become four when 75\% of additional edges are added. This strategy is in line with the finding of our previous work~\cite{PaganiEvol2013} where a \sw model was the best solution for modeling a \SG topology. The addition of random edges goes into that direction: the network with the rational structure planned by the power engineers 
%(that we can consider the regular lattice in the \sw model) 
is modified by the addition of random links, thus in something between the rational topology and the fully random network. The other strategies that are quite successful are the assortative  high degree and the least distance but only when the edges are doubled. This is the situation considering a pure topological decision. When also the costs of the evolution are taken into the picture the optimal evolution strategy might change due to economical constraints. The extended analysis that consider economic aspects is performed in Section~\ref{sec:economics}.

\subsection{Desiderata Parameters Satisfaction for \LV Networks}\label{sec:desLV}

We now consider the satisfaction of the quantitative metrics that we have applied for the \MV networks for the \LV evolved samples. 
The aspects that these metrics assess deal with losses, local energy distribution, and increasing network robustness. These are the aspects that we require to be satisfied in the future Smart Grid. 
%The metrics can be can be categorized into three macro categories with respect to how they affect a Power Grid: efficiency in the transfer of energy, resilience in providing alternative path if part of the network is compromised/congested and robustness to failures for network connectivity.
%require to be satisfied in the future Smart Grid.\footnote{stesso problema, FATTO VERSIONE RIDOTTA DI SOPRA}
The results are shown graphically in Tables~\ref{tab:desiderataSatisfLV1} and~\ref{tab:desiderataSatisfLV2} where for each evolution strategy and step the satisfaction of the parameter is marked with a ``tick'' sign (\tick), while the dissatisfaction is marked with a cross sign (\cross), and a value closer to satisfaction by an approximation sign ($\approx$). In Tables~\ref{tab:desiderataSatisfLV1} and~\ref{tab:desiderataSatisfLV2}, we consider the satisfaction of the \cc metric, therefore assigning a ``tick'' sign, if the \cc desiderata requirement is satisfied by the three biggest samples in the network (i.e., sample \#5, \#9, and \#10). This is done in consideration of the extreme high values of \cc that the desiderata parameter impose when samples have a small \textit{order}.

\paragraph*{Assortative high degree}
The assortative high degree evolution strategy almost satisfies the desired values for the metrics concerning the \cpl  from the second step of evolution with eight out of eleven samples; while fully satisfaction is in the last evolution step. However, the requirement for the redundant path are never fully satisfied. Considering the biggest samples, they never satisfy the condition $CC \geq 5\times CC_{RG}$. Also robustness requirement are not satisfied, only six samples in the last two evolution stages reaches the goal of 0.45. The betweenness requirement is only partially satisfied: eight out of the eleven samples have a betwenness to \textit{order} ratio around the target of 2.5. Considering the coefficient of variation, only the two smallest samples satisfy this property. 

\paragraph*{Assortative low degree}
The assortative low degree evolution strategy satisfies or is close to satisfying the metrics in the last stage of the evolution process. For characteristic path length, already from the second step, ten of the eleven samples reach the target; on the contrary, the target is never reached for the redundant path requirement. The three biggest samples score sufficiently concerning the \cc requirement, while the others have high values, but never satisfy the high-demanding requirements. Robustness is close to satisfaction in the last stage of evolution where all samples reach the target except two that stop to a value around 0.4. The metrics concerning betweenness show an almost satisfaction of the average betweenness to \textit{order} ratio with eight of the samples well below the target; however the coefficient of variation never reaches the target for the samples at any stage of evolution.

\paragraph*{Triangle closure}
The triangle closure evolution strategy focuses on the clustering coefficient, therefore reaching the target already after the second step of the evolution. Two of the three biggest samples satisfy also a more strict condition than the desiderata requirement such as $CC \geq 10\times CC_{RG}$, that can be considered the condition for satisfying the \sw requirment for this property cf.~\cite{Watts98}. Concerning the path length properties, the \cpl is never lower than the logarithm of the \textit{order} of the graph. However, the requirement over the $10^{th}$ \apl is satisfied already after the second step of evolution. In fact, the redundant paths are smaller than twice the \cpl given the quite high \cpl that the networks have. For robustness, seven out of the eleven samples reach the target in the last step of evolution. For metrics that involve betweenness, only seven samples are below the target for the betweenness to \textit{order} metric, actually the smallest in \textit{order}, while the three biggest samples are far from the target. The coefficient of variation increases in the evolution, therefore not allowing for this metric to gain a ``tick'' sign.

\paragraph*{Dissortative node degree}
The dissortative node degree strategy scores quite well for evolving \LV networks. Especially in the last stage of evolution half of the metrics are satisfied or very close to satisfaction. Already in the second step of evolution, the \cpl is satisfied; actually already in the first step of evolution eight samples satisfy the condition. Robustness is satisfied with nine out of the eleven samples that fully comply with the desiderata and the other two that are above 0.4. The real weak point is the \cc which even for the biggest samples is never close to the target. Betweenness-related metrics have a dissimilar behavior: the betweenness to \textit{order} ratio satisfies the desiderata for nine samples, while the coefficient of variation has an increasing trend and in the final step does not comply with the requirements for any sample.

\paragraph*{Random}
The random evolution strategy proves to be one of the strategies that satisfies most of the parameters. The metrics concerning the \cpl are satisfied basically already from the second step of evolution with ten of the eleven samples compliant. A different behavior applies to the redundant paths whose goal is not met, but in the final step eight of the eleven samples reach the target. Robustness is almost achieved already after the second step: all the samples score above 0.4 and just three  do not reach the target in that step, but later in the evolution. In the last step, the satisfaction of the redundant path is almost satisfied with eight of the samples (containing also the three biggest ones). The metrics related to betweenness that usually fail for the other types of evolution are here almost entirely met concerning the coefficient of variation and the betweenness to \textit{order} ratio in the last stage of evolution. The main drawback is on the \cc side. With such a strategy the formation or closure of topological triangle structures is really difficult, therefore the cross sign for this parameter.  

\begin{sidewaystable}[htbp]
\centering
\begin{footnotesize}
\begin{tabular}{|l|c|c|c|c|c|c|c|c|c|c|c|c|}
  \hline
  \textbf{Desiderata}&\multicolumn{4}{|c|}{\textbf{Assortative HD}}&\multicolumn{4}{|c|}{\textbf{Assortative LD}}&\multicolumn{4}{|c|}{\textbf{Triangle closure}} \\
  \hline
  & \textit{order}+25\% & \textit{order}+50\%&\textit{order}+75\%&\textit{order}+100\%& \textit{order}+25\% & \textit{order}+50\%&\textit{order}+75\%&\textit{order}+100\%& \textit{order}+25\% & \textit{order}+50\%&\textit{order}+75\%&\textit{order}+100\%\\ \hline
  %Modularity & $\approx$ &\cross &\cross &\tick & $\approx$ &\cross &\cross &\tick & $\approx$ &\cross &\cross &\tick \\  \hline
  $CPL\leq ln(N)$  &  \cross &\tick &\tick & \tick & \cross &\tick &\tick &\tick & \cross &\cross &\cross & \cross \\  \hline
  $CC\geq 5\times CC_{RG}$ & \cross &\cross & \cross &\cross & \cross &\cross &\cross &\cross & \cross &\tick &\tick &\tick \\  \hline
  $\overline{\upsilon}=\frac{\overline{\sigma}}{N} \approx 2.5$ & \cross &\cross &$\approx$ & $\approx$ & \cross &$\approx$ &$\approx$ &$\approx$ & \cross &\cross &\cross &\cross \\ \hline
  $c_v \leq 1$ &\cross &\cross &\cross & \cross & \cross &\cross &\cross &\cross & \cross &\cross &\cross &\cross \\ \hline
  $Rob_N \geq 0.45$ & \cross &\cross &$\approx$ & $\approx$ & \cross &\cross  &\cross &\tick & \cross &\cross &\cross &$\approx$ \\ \hline
  $APL_{10^{th}} \leq 2 \times CPL$ & \cross &$\approx$ &$\approx$ & \cross & \cross &\cross &\cross & $\approx$ & \cross &$\approx$ &\tick &\tick \\ \hline
\end{tabular}
\end{footnotesize}
\caption{Desiderata parameter compliance for \LV networks evolved according assortative high degree, assortative low degree, and triangle closure strategy.}\label{tab:desiderataSatisfLV1}
\end{sidewaystable}

\begin{sidewaystable}[htbp]
\centering
\begin{footnotesize}
\begin{tabular}{|l|c|c|c|c|c|c|c|c|}
  \hline
  \textbf{Desiderata}&\multicolumn{4}{|c|}{\textbf{Dissortative}} &\multicolumn{4}{|c|}{\textbf{Random}} \\
  \hline
  & \textit{order}+25\% & \textit{order}+50\%&\textit{order}+75\%&\textit{order}+100\%& \textit{order}+25\% & \textit{order}+50\%&\textit{order}+75\%&\textit{order}+100\%\\ \hline
  %Modularity & $\approx$ &\cross &\cross &\tick & $\approx$ &\cross &\cross &\tick & $\approx$ &\cross &\cross &\tick \\  \hline
  $CPL\leq ln(N)$  & $\approx$ &\tick  &\tick  & \tick  &\cross &\tick  &\tick &\tick \\  \hline
  $CC\geq 5\times CC_{RG}$ & \cross &\cross &\cross &\cross & \cross &\cross &\cross &\cross \\  \hline
  $\overline{\upsilon}=\frac{\overline{\sigma}}{N} \approx 2.5$ & \cross &$\approx$ &$\approx$ &\tick & \cross &$\approx$ &$\approx$ &\tick \\ \hline
  $c_v \leq 1$ &\cross &\cross &\cross & \cross  & \cross &\cross &\cross &$\approx$ \\ \hline
  $Rob_N \geq 0.45$ & \cross &\cross &\cross &\tick & \cross  &$\approx$ &\tick &\tick\\ \hline
  $APL_{10^{th}} \leq 2 \times CPL$ &\cross &\cross &\cross &\cross & \cross &\cross &$\approx$ &\tick \\ \hline
\end{tabular}
\end{footnotesize}
\caption{Desiderata parameter compliance for \LV networks evolved according dissortative and random strategy.}\label{tab:desiderataSatisfLV2}
\end{sidewaystable}

\bigskip

The random evolution strategy is the one that satisfies most of the desiderata (two parameters fully satisfied and three almost satisfied) requirements at the final stage of evolution. Concerning the other strategies that have an evolution with a specific goal those that satisfy most of the requirements are the assortative low degree followed by the dissortative one which almost satisfy  four and three requirements respectively. Therefore these sub optimal strategies could be used too. In particular, the assortative low degree satisfies the requirements regarding the clustering coefficient, that are not satisfied neither by the random, nor by the dissortative strategy.

\bigskip

A commonality between the evolution for \MV and \LV is the best result achieved by the random evolution of network that scores best among the strategies considered in this study. An interesting difference lies in the sub-optimal strategies that score high in for the \MV network and \LV network evolution. For the \MV network, it is best to provide more connectivity between those nodes that already hold a high node degree, thus reinforcing their role as key components of the network. On the other hand, for the \LV network, it seems that adding more connectivity between the nodes that have a small connectivity provides better performances for evolution. The assortative low degree aims at creating more connections and (therefore hubs) where is less connectivity in the current samples. Also the other strategy that is sub-optimal for \LV networks, the dissortative, aims at giving a more important role to the nodes in the periphery of the network by connecting them to the more connected nodes. 
A synthesis of the topological performance of the different strategies is provided in Table~\ref{tab:steps}. For  the layer of the \PG considered (column one), each strategy (column two) is assessed with an optimality level (column three) based on the satisfaction of the topological metrics. The step during the evolution process in which the most of the metrics are satisfied is also provided in column four. The achievement of the good results depends on the network layer and on the strategy used. The optimality level is assigned based on the full satisfaction of the topological metrics described earlier in this section and more thoroughly in~\ref{sec:metrics}. One can see a distinction between the \MV and the \LV networks: the \LV achieve less topological optimality (less metrics are satisfied) although requiring on average more evolution steps. Another aspect to note is that some evolution strategies (i.e., assortative high degree and triangle closure) do not have an improvement between the various steps despite the addition of more connectivity. On the other hand, other strategies (i.e., dissortative and random) benefit more from additional connectivity by improving the satisfaction of the metrics.
%\footnote{ottimo questo testo, possiamo anche dire qualcosa sul numero dei step? possiamo fare una super sintetica tabellina di riassunto?DONE IN BLUE}

\begin{center}
\begin{table}[htbp]
\centering
\begin{footnotesize}
\begin{tabular}{|@p{1.3cm}|^p{2.8cm}|^p{1.5cm}|^p{2cm}|}
\hline
\rowstyle{\bfseries}
Network level  & Strategy & Optimality & Optimality reached at step  \\ \hline
\hline

MV    & Assortative HD & $\star\star\star$ & 4 \\ \hline
MV    & Assortative LD & $\star\star$ & 2 \\ \hline
MV    & Triangle closure & $\star\star$ &1 \\ \hline
MV    & Dissortative & $\star$ & 1 \\ \hline
MV    & Least distance& $\star\star\star$ &4\\ \hline
MV    & Random & $\star\star\star\star$ & 3 \\ \hline \hline
LV    & Assortative HD & $\star\star$ & 3 \\ \hline
LV    & Assortative LD & $\star\star$& 4 \\ \hline
LV    & Triangle closure  & $\star\star$ & 3 \\ \hline
LV    & Dissortative & $\star\star\star$ & 4 \\ \hline
LV    & Random & $\star\star\star\star$ & 4 \\ \hline
\end{tabular}
\caption{Parameter optimal satisfaction.}\label{tab:steps}
\end{footnotesize}
\end{table}
\end{center}

\section{Economic Considerations}~\label{sec:economics}

Traditionally the problem of evaluating the expansion of an electrical system is a complex task that involves both the use of modeling, usually based on operation research optimization techniques and linear programming~\cite{garver70,lee74,belagari75}, and the experience and vision of experts in the field aided by computer technologies~\cite{grigsby07}.
%\footnote{add a citation about a book on power systems? graduate level?DONE} 
In this latter case, computers acquire knowledge based on previous expert decision and, based on the electrical physical constraints of the domain, are then able to support \PG evolution decision finding the most suitable technical and economical solution~\cite{teive98}. With more distributed generating facilities at local scale, traditional methods have limits and need to be modified or updated to take into account the new scenarios of the Smart Grid. The strategies that we have so far analyzed as being candidates for evolving the current \G in the future \SG need also to be evaluated from the economic point of view. How much will it cost to deploy electrical infrastructures according to these models? What is the actual cost of adding an edge to the topology of the Grid, physically?

To answer these questions, we need to characterize the cables used for the upgrading process in terms of their physical properties and costs. The information of length for each cable in the existing sample networks is available. For the majority of the nodes in the samples belonging to the \MV we know the geographical coordinate information, so with some approximation it is possible to compute the length of the new cable connecting nodes that do not have a line connecting them yet. Therefore, each new added line is assigned a length. In order to define the properties characterizing the new cables, we adopt a k-nearest neighbor (KNN) classification based on the length of the cable~\cite{cover67,Dasarathy91}. The classification is performed using the Java Machine Learning Library (Java-ML) v0.1.5 (\protect\url{http://java-ml.sourceforge.net/}). For each new cable, we consider the k cables (we choose $k=5$) in the original network used as the training set for the classifier that have the length closer to the added one and then the type of cable is available. We choose this method to identify the cable  since it is simple and it has proven successful in several practical contexts~\cite{citeulike:2827911}. Once the properties of the new cables are identified, all the information we need for the economic analysis are then available: resistance per unit of length, cost per unit of length, maximal supported current. We remark that our proposed analysis does not aim at being a comprehensive investment analysis, for which other techniques are well established~\cite{stermole09}, simply an economic evaluation of the proposed evolution strategies to confirm their feasibility or unfeasibility. 

Once again, as in our previous work~\cite{pag:preprint12}, to assess these costs in the \MV infrastructure, we consider a simple relation where the cost of cabling and cost of substations are added:
\begin{equation}
C_{impl}=\sum_{j=1}^{N} Ssc_{j}+\sum_{i=1}^{M}Cc_i
\end{equation}
where $C_{impl}$ stands for implementation cost, $Ssc_j$ is the adaptation cost for the substation $j$ and $Cc_i$ is the cost for the cable $i$.  The cost of the cable can be expressed as a linear function of the distance the cable $i$ covers: $Cc_i= C_{uc_i}\cdot l_i$, where $C_{uc_i}$ is the cable cost per unit of length and $l_i$ is the length of the cable.  There are several types of cables used for power transmission and distribution with varying physical characteristics and costs. In addition, the cost for installation can vary significantly~\cite{natGrid}. In the present work, to provide an initial estimate, we simply consider cabling costs and ignore substation ones. While the former are directly tied to the topology and length of the links, the latter pricing is too dependent on other factors to be useful in the present analysis (e.g., different equipment in the substation). 
%As a source of data for cable type and pricing, we have been provided (courtesy of Enexis B.V. the Netherlands) with cable characteristics and prices, together with topological information, for 11 network samples belonging to the \LV network and 13 samples belonging to the \MV network of the Northern Netherlands. 

For the cost analysis, we limit our investigation to the three strategies that have scored best in the pure topological analysis (i.e., random, assortative high node degree and least distance) described in Section~\ref{sec:evolMec}. In addition, the economic analysis can be applied only to the \MV network samples since there is no geographical information about the location of the nodes. The results of the cost for evolution of the Dutch samples are shown in Tables~\ref{tab:costAssHD}, ~\ref{tab:costRnd}, and~\ref{tab:costLeast}. The first column of each table contains the sample ID, while the second provides the information about the evolution step considered, the third column has the information concerning the cost of the evolution of the network according to the specified strategy. The fourth column contains the information on the fraction of the cost that the evolution of the infrastructure impacts on the whole cost of the infrastructure. From the tables the most interesting result is the difference in the cost that the three methods of evolution require. On average the costs of network improvement is similar when considering the assortative and the random strategy: the cost of adding more edges is about 75\% of the cost for cabling of the whole \G infrastructure. The situation is radically different for the least distance strategy whose development impacts only marginally in the total cost of the infrastructure, by adding just less than 13\% of the cost of the infrastructure.
Purely from the point of view of cabling costs, the most promising evolution strategy in economic and topological terms for the \MV Grid is the strategy that connects the nodes that are geographically closer (i.e., least distance strategy).

\begin{table}[htbp]
\centering
\begin{footnotesize}
\begin{tabular}{|l||r|r|r|}
\hline
 & Evolution step & \multicolumn{1}{l|}{Evolution cost (euro)} & \multicolumn{1}{l|}{Fraction of evolution } \\ 
&  & &\multicolumn{1}{l|}{cost on the whole infrastructure} \\ \hline \hline
\multirow{4}{*}{Sample \#1}
&+25\% &  59334032 & 0.73 \\ \cline{2-4}
&+50\% &  59334032 & 0.73 \\ \cline{2-4}
&+75\% &  59334032 & 0.73 \\ \cline{2-4}
&+100\% &  119505699 & 0.85 \\ \hline
\multirow{4}{*}{Sample \#2}
&+25\% & 107698129 & 0.73 \\ \cline{2-4}
&+50\%  & 225049864 & 0.85 \\ \cline{2-4}
&+75\% & 255361286 & 0.87 \\ \cline{2-4}
&+100\%  & 298792505 & 0.88 \\ \hline
\multirow{4}{*}{Sample \#3}
&+25\%  & 23007544 & 0.72 \\ \cline{2-4}
&+50\%  & 44739934 & 0.83 \\ \cline{2-4}
&+75\%  & 72515143 & 0.89 \\ \cline{2-4}
&+100\%  & 105567690 & 0.92 \\ \hline
\multirow{4}{*}{Sample \#4}
&+25\% &  45762304 & 0.65 \\ \cline{2-4}
&+50\% & 55159897 & 0.69 \\ \cline{2-4}
&+75\% &  55159897 & 0.69 \\ \cline{2-4}
&+100\% &  55159897 & 0.69 \\ \hline
\multirow{4}{*}{Sample \#5}
&+25\% &  65393631 & 0.74 \\ \cline{2-4}
&+50\% &  117984090 & 0.84 \\ \cline{2-4}
&+75\% &  157876214 & 0.87 \\ \cline{2-4}
&+100\% &  194838593 & 0.90 \\ \hline
\multirow{4}{*}{Sample \#6}
&+25\% &  31063470 & 0.67 \\ \cline{2-4}
&+50\% &62971658 & 0.81 \\ \cline{2-4}
&+75\% &  104711687 & 0.87 \\ \cline{2-4}
&+100\% &  165938342 & 0.92 \\ \hline
\multirow{4}{*}{Sample \#7}
&+25\% &  79148152 & 0.61 \\ \cline{2-4}
&+50\% & 199600692 & 0.80 \\ \cline{2-4}
&+75\% & 229365781 & 0.82 \\ \cline{2-4}
&+100\% &  350782953 & 0.87 \\ \hline
\multirow{4}{*}{Sample \#8}
&+25\% &  32826377 & 0.67 \\ \cline{2-4}
&+50\% & 36976365 & 0.70 \\ \cline{2-4}
&+75\% &  83305090 & 0.84 \\ \cline{2-4}
&+100\% &  123241632 & 0.88 \\ \hline
\multirow{4}{*}{Sample \#9}
&+25\% &  11739091 & 0.46 \\ \cline{2-4}
&+50\% &  42894090 & 0.76 \\ \cline{2-4}
&+75\% &  72834999 & 0.84 \\ \cline{2-4}
&+100\% &  103147797 & 0.88 \\ \hline
\multirow{4}{*}{Sample \#10}
&+25\% &  14899834 & 0.60 \\ \cline{2-4}
&+50\% &  32136672 & 0.77 \\ \cline{2-4}
&+75\% & 45646927 & 0.82 \\ \cline{2-4}
&+100\% & 60546613 & 0.86 \\ \hline
\multirow{4}{*}{Sample \#11}
&+25\% &  13094364 & 0.50 \\ \cline{2-4}
&+50\% &  26377193 & 0.67 \\ \cline{2-4}
&+75\% &  38709007 & 0.75 \\ \cline{2-4}
&+100\% &  40774034 & 0.76 \\ \hline
\multirow{4}{*}{Sample \#12}
&+25\% &  93369553 & 0.70 \\ \cline{2-4}
&+50\% &  93369553 & 0.70 \\ \cline{2-4}
&+75\% &  138091611 & 0.77 \\ \cline{2-4}
&+100\% &  253056714 & 0.86 \\ \hline
\end{tabular}
\end{footnotesize}
\caption{Cost of \MV network evolution for assortative high degree strategy.}\label{tab:costAssHD}
\end{table}

\begin{table}[htbp]
\centering
\begin{footnotesize}
\begin{tabular}{|l||r|r|r|}
\hline
 & Evolution step & \multicolumn{1}{l|}{Evolution cost (euro)} & \multicolumn{1}{l|}{Fraction of evolution } \\ 
&  & &\multicolumn{1}{l|}{cost on the whole infrastructure} \\ \hline \hline
\multirow{4}{*}{Sample \#1}
&+25\%  & 49585634 & 0.70 \\ \cline{2-4}
&+50\%  & 99550100 & 0.82 \\ \cline{2-4}
&+75\%  & 147237698 & 0.87 \\ \cline{2-4}
&+100\%  & 193060613 & 0.90 \\ \hline
\multirow{4}{*}{Sample \#2}

&+25\%  & 81443232 & 0.67 \\ \cline{2-4}
&+50\%  & 173835139 & 0.81 \\ \cline{2-4}
&+75\%  & 270738228 & 0.87 \\ \cline{2-4}
&+100\%  & 352561226 & 0.90 \\ \hline
\multirow{4}{*}{Sample \#3}

&+25\% & 20401026 & 0.70 \\ \cline{2-4}
&+50\%  & 39401799 & 0.82 \\ \cline{2-4}
&+75\%  & 58698139 & 0.87 \\ \cline{2-4}
&+100\%  & 80038042 & 0.90 \\ \hline
\multirow{4}{*}{Sample \#4}

&+25\%  & 38550574 & 0.61 \\ \cline{2-4}
&+50\%  & 74405969 & 0.75 \\ \cline{2-4}
&+75\%  & 112507292 & 0.82 \\ \cline{2-4}
&+100\%  & 144092688 & 0.85 \\ \hline
\multirow{4}{*}{Sample \#5}

&+25\%  & 34665986 & 0.60 \\ \cline{2-4}
&+50\%  & 63522559 & 0.74 \\ \cline{2-4}
&+75\%  & 90171964 & 0.80 \\ \cline{2-4}
&+100\%  & 118746816 & 0.84 \\ \hline
\multirow{4}{*}{Sample \#6}

&+25\%  & 32756917 & 0.68 \\ \cline{2-4}
&+50\%  & 67956839 & 0.82 \\ \cline{2-4}
&+75\%  & 102189138 & 0.87 \\ \cline{2-4}
&+100\% & 140632238 & 0.90 \\ \hline
\multirow{4}{*}{Sample \#7}

&+25\%  & 85124154 & 0.63 \\ \cline{2-4}
&+50\%  & 181095750 & 0.78 \\ \cline{2-4}
&+75\%  & 262020773 & 0.84 \\ \cline{2-4}
&+100\% & 352245673 & 0.87 \\ \hline
\multirow{4}{*}{Sample \#8}

&+25\%  & 40434688 & 0.72 \\ \cline{2-4}
&+50\%  & 79435946 & 0.83 \\ \cline{2-4}
&+75\% & 107843650 & 0.87 \\ \cline{2-4}
&+100\%  & 139828749 & 0.90 \\ \hline
\multirow{4}{*}{Sample \#9}

&+25\%  & 31096020 & 0.70 \\ \cline{2-4}
&+50\%  & 57036433 & 0.81 \\ \cline{2-4}
&+75\% & 85190089 & 0.86 \\ \cline{2-4}
&+100\%  & 115442294 & 0.89 \\ \hline
\multirow{4}{*}{Sample \#10}

&+25\%  & 12677249 & 0.56 \\ \cline{2-4}
&+50\%  & 26126828 & 0.73 \\ \cline{2-4}
&+75\%  & 42325971 & 0.81 \\ \cline{2-4}
&+100\%  & 53357492 & 0.84 \\ \hline
\multirow{4}{*}{Sample \#11}

&+25\%  & 1214608 & 0.08 \\ \cline{2-4}
&+50\%  & 2553383 & 0.16 \\ \cline{2-4}
&+75\%  & 3117824 & 0.19 \\ \cline{2-4}
&+100\%  & 4494600 & 0.25 \\ \hline

\multirow{4}{*}{Sample \#12}

&+25\% & 105240341 & 0.72 \\ \cline{2-4}
&+50\%  & 223059810 & 0.85 \\ \cline{2-4}
&+75\%  & 317337397 & 0.89 \\ \cline{2-4}
&+100\%  & 417627345 & 0.91 \\ \hline
\end{tabular}
\end{footnotesize}
\caption{Cost of \MV network evolution for random strategy.}\label{tab:costRnd}
\end{table}

\begin{table}[htbp]
\centering
\begin{footnotesize}
\begin{tabular}{|l||r|r|r|}
\hline
 & Evolution step & \multicolumn{1}{l|}{Evolution cost (euro)} & \multicolumn{1}{l|}{Fraction of evolution } \\ 
&  & &\multicolumn{1}{l|}{cost on the whole infrastructure} \\ \hline \hline
\multirow{4}{*}{Sample \#1}

&+25\%  & 635014 & 0.03 \\ \cline{2-4}
&+50\% & 1944353 & 0.08 \\ \cline{2-4}
&+75\%  & 3536060 & 0.14 \\ \cline{2-4}
&+100\%  & 5539077 & 0.20 \\ \hline
\multirow{4}{*}{Sample \#2}

&+25\%  & 1832132 & 0.04 \\ \cline{2-4}
&+50\%  & 5362070 & 0.12 \\ \cline{2-4}
&+75\%  & 9454081 & 0.19 \\ \cline{2-4}
&+100\%  & 15488444 & 0.28 \\ \hline
\multirow{4}{*}{Sample \#3}

&+25\%  & 357093 & 0.04 \\ \cline{2-4}
&+50\%  & 1143678 & 0.11 \\ \cline{2-4}
&+75\%  & 2075073 & 0.19 \\ \cline{2-4}
&+100\% & 3206160 & 0.26 \\ \hline
\multirow{4}{*}{Sample \#4}

&+25\%  & 831466 & 0.03 \\ \cline{2-4}
&+50\%  & 2936596 & 0.11 \\ \cline{2-4}
&+75\%  & 5026907 & 0.17 \\ \cline{2-4}
&+100\%  & 7688212 & 0.24 \\ \hline
\multirow{4}{*}{Sample \#5}

&+25\%  & 801764 & 0.03 \\ \cline{2-4}
&+50\%  & 2128363 & 0.09 \\ \cline{2-4}
&+75\% & 3412290 & 0.13 \\ \cline{2-4}
&+100\% & 5187261 & 0.19 \\ \hline
\multirow{4}{*}{Sample \#6}

&+25\%  & 543105 & 0.03 \\ \cline{2-4}
&+50\%  & 1433992 & 0.09 \\ \cline{2-4}
&+75\%  & 2588723 & 0.15 \\ \cline{2-4}
&+100\% & 3822329 & 0.20 \\ \hline
\multirow{4}{*}{Sample \#7}

&+25\%  & 1104605 & 0.02 \\ \cline{2-4}
&+50\%  & 3521507 & 0.07 \\ \cline{2-4}
&+75\%  & 6534753 & 0.11 \\ \cline{2-4}
&+100\%  & 10042275 & 0.17 \\ \hline
\multirow{4}{*}{Sample \#8}

&+25\%  & 584039 & 0.03 \\ \cline{2-4}
&+50\%  & 2127943 & 0.12 \\ \cline{2-4}
&+75\%  & 4016394 & 0.20 \\ \cline{2-4}
&+100\%  & 5840352 & 0.27 \\ \hline
\multirow{4}{*}{Sample \#9}

&+25\%  & 661698 & 0.05 \\ \cline{2-4}
&+50\%  & 1861427 & 0.12 \\ \cline{2-4}
&+75\%  & 3476464 & 0.20 \\ \cline{2-4}
&+100\% & 5524643 & 0.29 \\ \hline
\multirow{4}{*}{Sample \#10}

&+25\%  & 314451 & 0.03 \\ \cline{2-4}
&+50\%  & 987993 & 0.09 \\ \cline{2-4}
&+75\%  & 1971402 & 0.17 \\ \cline{2-4}
&+100\%  & 2912718 & 0.23 \\ \hline
\multirow{4}{*}{Sample \#11}

&+25\%  & 478534 & 0.03 \\ \cline{2-4}
&+50\%  & 1230921 & 0.09 \\ \cline{2-4}
&+75\%  & 2224697 & 0.14 \\ \cline{2-4}
&+100\%  & 3262358 & 0.20 \\ \hline
\multirow{4}{*}{Sample \#12}

&+25\%  & 1069466 & 0.03 \\ \cline{2-4}
&+50\%  & 3143898 & 0.07 \\ \cline{2-4}
&+75\%  & 5723602 & 0.12 \\ \cline{2-4}
&+100\%  & 8686081 & 0.18 \\ \hline

\end{tabular}
\end{footnotesize}
\caption{Cost of \MV network evolution for least distance strategy.}\label{tab:costLeast}
\end{table}

One may then wonder if such investments are beneficial for the end users and the distribution companies in reducing the cost of electricity flows. We resort to a set of metrics that we have developed and already applied to the Northern Netherlands Distribution Grid~\cite{PaganiAielloTSG2011} and to synthetically generated networks~\cite{pag:preprint12,PaganiEvol2013}. 
The goal is to consider, from a topological perspective, those measures that are critical in contributing to the cost of electricity as elements in the Transmission and Distribution Networks as described in economic studies such as the one of Harris and Munasinghe~\cite{Harris06,Munasinghe84}:
\begin{itemize}
\item losses both in line and at transformer stations,
\item security and capacity factors,
\item line redundancy, and 
\item power transfer limits.
\end{itemize}
For each of these elements we associate topological quantities that are representative, so that we can assess their values and therefore compare them in the original networks and in the evolutions that we propose in this paper.

Here we assess the 
\begin{itemize}
\item Losses on the transmission/distribution line can be expressed by the quotient of the weighted
  characteristic path length :
\begin{align}WCPL_N\end{align}
\item The average resistance of a line (a weighted edge in the graph):
\begin{align}\overline{w}=\frac{1}{M}\sum_{i=1}^{M} w_i \end{align}
\item Losses at substation level are expressed as the number of nodes (on average) that are traversed when computing the weighted shortest path between all the nodes in the network:
\begin{align}L_{substation_N} = \overline{Nodes_{WCPL_N}} \label{eq:stationLoss}\end{align}
\item Robustness is evaluated with random removal strategy and the weighted-node-degree-based removal by computing the average of the order of maximal connected component between the two situations when the 20\% of the nodes of the original graph are removed. It can be written as:
  \begin{align}Rob_N = \frac{|MCC_{Random20\%}|+|MCC_{NodeDegree20\%}|}{2} \label{eq:robust}\end{align} 
\item Redundancy is evaluated by covering a random sample of the nodes
  in the network (40\% of the nodes whose half represents source nodes
  and the other half represents destination nodes) and computing for
  each source and destination pair the first ten shortest paths of
  increasing length. If there are less than ten paths available, the
  worst case path between the two nodes is considered. To have a
  measure of how these resilient paths have an increment in
  transportation cost, a normalization with the weighted
  characteristic path length is performed. We formalized it as:
\begin{align}Red_{N} = \frac{\sum_{i \in Sources, j \in Sinks}{SP_{w_{ij}}}}{WCPL} \label{eq:redund}\end{align}
\item Network capacity is considered as the value of the weighted characteristic path length, whose weights are the maximal operating current supported: 
\begin{align}Cap_{N} = WCPL_{current N}\end{align}
\item The average supported current of a line (a weighted edge in the graph):
\begin{align} \overline{w_{current}}=\frac{1}{M}\sum_{i=1}^{M} w_{icurrent}\end{align}
\end{itemize}
The aspects considered are just some of the factors (the ones closely coupled to topology) that influence the overall price of electricity.  Naturally, there are other factors that influence the final price, e.g., fuel prices, government strategies and taxation, etc., as illustrated for instance in the economic studies of Harris and Munasinghe~\cite{Harris06,Munasinghe84}. 

\begin{figure}
 \captionsetup{type=figure}
    \centering
    \subfloat[Losses-related metrics.]{\label{fig:ecoLos25}\includegraphics[scale=.5]{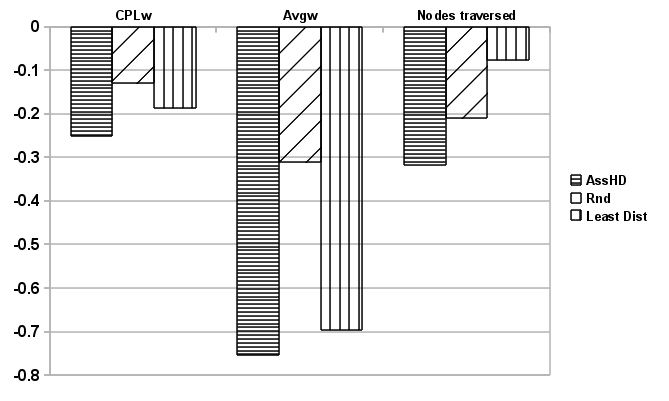}} 
    \subfloat[Reliability-related metrics.]{\label{fig:ecoRel25}\includegraphics[scale=.5]{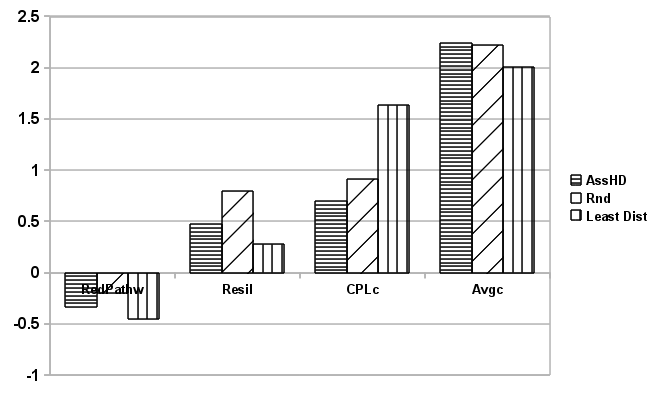}} %\\
    \caption{Comparison to the original networks of properties influencing electricity cost at 1$^{st}$ stage of evolution (i.e., +25\% edges).}
    \label{fig:ecoEvo25}
\end{figure}

\begin{figure}
 \captionsetup{type=figure}
    \centering
    \subfloat[Losses-related metrics.]{\label{fig:ecoLos50}\includegraphics[scale=.5]{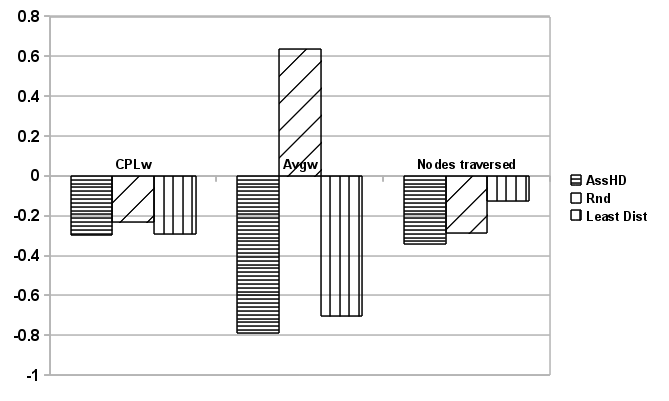}} 
    \subfloat[Reliability-related metrics.]{\label{fig:ecoRel50}\includegraphics[scale=.5]{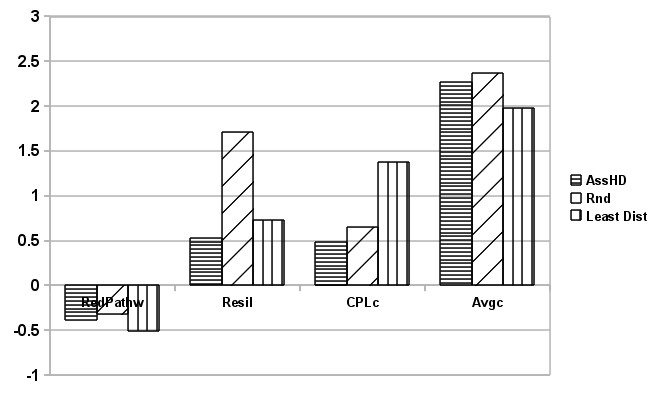}} %\\
    \caption{Comparison to the original networks of properties influencing electricity cost at 2$^{nd}$ stage of evolution (i.e., +50\% edges).}
    \label{fig:ecoEvo50}
\end{figure}

\begin{figure}
 \captionsetup{type=figure}
    \centering
    \subfloat[Losses-related metrics.]{\label{fig:ecoLos75}\includegraphics[scale=.5]{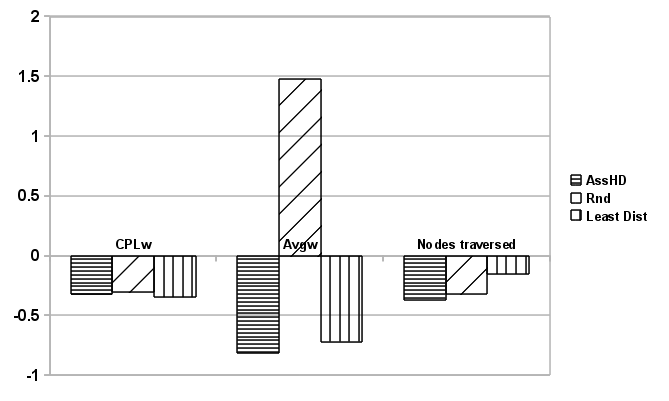}} 
    \subfloat[Reliability-related metrics.]{\label{fig:ecoRel75}\includegraphics[scale=.5]{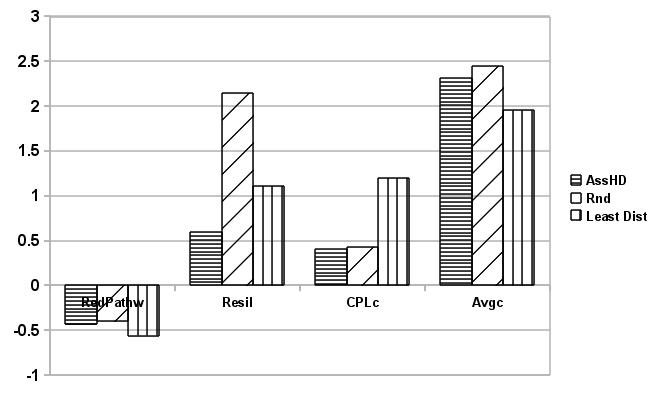}} %\\
    \caption{Comparison to the original networks of properties influencing electricity cost at 3$^{rd}$ stage of evolution (i.e., +75\% edges).}
    \label{fig:ecoEvo75}
\end{figure}

\begin{figure}
 \captionsetup{type=figure}
    \centering
    \subfloat[Losses-related metrics.]{\label{fig:ecoLos100}\includegraphics[scale=.5]{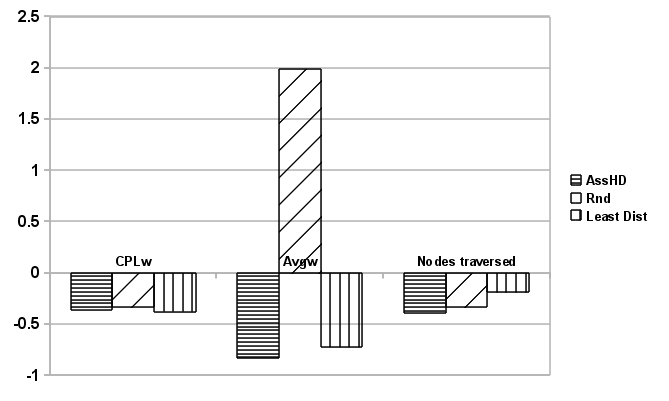}} 
    \subfloat[Reliability-related metrics.]{\label{fig:ecoRel100}\includegraphics[scale=.5]{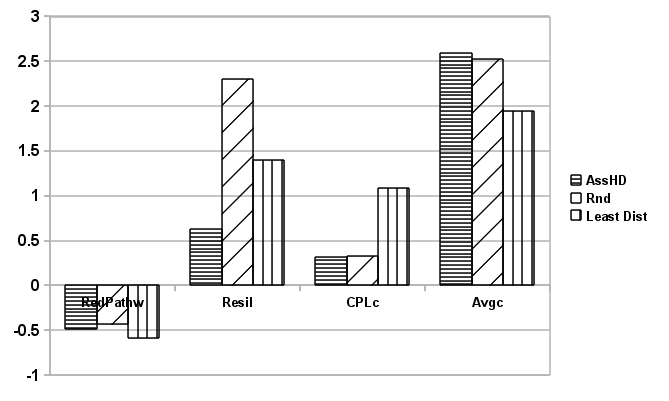}} %\\
    \caption{Comparison to the original networks of properties influencing electricity cost at 4$^{th}$ stage of evolution (i.e., +100\% edges).}
    \label{fig:ecoEvo100}
\end{figure}

Figures~\ref{fig:ecoEvo25},~\ref{fig:ecoEvo50},~\ref{fig:ecoEvo75}, and~\ref{fig:ecoEvo100} show the improvement as a fraction of the original values in the metrics related to the cost of electricity distribution respectively in the four step of evolution we consider (+25\%, +50\%, +75\%, and +100\% of edges) compared to the original samples. To give a general idea of the improvement, we use the average results over the 12 \MV networks samples. For each strategy, we look at the improvement, in percentage, for the same metrics compared to the initial samples. Already from a small increase in connectivity (addition of 25\% of edges) the resistance of a path decreases with every strategy with the best result obtained by the assortative high degree strategy. Adding more lines in general promotes a reduction of the average resistance of edges compared to the initial situation. This is even more true in the assortative and least distance strategy where short distance connections (and therefore less resistive) tend to dominate or more efficient cables are chosen in the KNN procedure selection. In addition, a considerable reduction in the losses that are experienced in traversing substation can be avoided with more dense networks: almost a 30\% less station traversed in the graph enhanced with assortative connections; of course less benefits take place with the least distance strategy that reduces the average number of traversed stations in a path by 7\%. This higher connectivity provides benefits to the robustness aspects: resilient paths that are less lossy (about 30\% reduction for the least distance strategy). An improvement is also experienced in the resilience of the network to node disruptions: about 80\% more resilience with the random strategy. Also the amount of current that is supported in a path (the weighted \CPL with maximal supported current as weight of the edges) is definitely higher (more than 2.5 times for the least distance strategy) compared to the initial samples.
Similar considerations can be done for the second evolution step (i.e., +50\% cables). Considering the loss aspect the benefits are for the three strategies on average about 30\% in the reduction of the weighted path. For the average weight (i.e., resistance) of a cable the random addition creates networks with much more cables with an high resistance. This is an indication that the random strategy creates long distance connections that act as shortcuts in the network from a topological point of view. However, in a physical system there are no significant benefits since the resistance grows with the distance. These limits of the benefits are shown by the weighted path analysis.  
%{\color{red}but that in the physical system with cables that have resistance that increase with distance do not provide much benefits as one sees by the weighted characteristic path length.}\footnote{sta frase in rosso non si capisce. Accorcia e semplifica.FATTO} 
Considering robustness oriented metrics, the most of the benefits take place in resilience where the networks on average stay twice as much as connected than the initial evolution step, except for the assortative strategy that lags behind. We see a small decrease for all the strategies in the characteristic path when the weight is the maximal current that can flow in the cables; this is not difficult to understand since the characteristic path (cf. Definition~\ref{def:cpl}) uses the most efficient (i.e., with smallest weight paths). 
When even more connectivity (+75\% and +100\%) is added one sees even better results and improvements but smaller in their magnitude. In these last two stages, the benefits for the losses reach values of reduction above 30\%, the same for the nodes to be traversed on average on a path. Only the least distance strategy has a small reduction, but this is due to the very nature of such evolution strategy that avoids long distance (topological shortcuts) cables. Considering network robustness, the main improvements are for resilience to node failures that for the least distance strategy allow the network to be twice as much as robust then the initial samples. Concerning the capacity of the network we assist to a slightly decrease for the same reasons that we have explained above. However, the networks with the least distance evolution are able on average to transport twice the amount of current than the initial samples.

Given the cost analysis for cable addition and the benefits that such an higher connectivity brings to the networks considering the different strategies in the topological aspects related to price of electricity, we provide a suggestion on how to evolve the analyzed samples. 
%We remark of course that this method involves a topological/\CNA based analysis and design of the network. We know that power distribution companies and power engineers have other more complex methods to assess and design the evolution of a network that have proved successful and reliable so far. 
Using our topological-based method, we consider that the least distance strategy with the addition of 75\% or 100\% of edges is a very good way of making the network more connected. Such an evolution strategy provides benefits from a topological point of view, and it keeps the costs low.  For the samples examined it requires an investment in cabling costs about 25\% of the value of the cables already on the ground. Such investment can provide benefits in the loss reduction of the network as well as in its robustness. The energy economic studies show~\cite{Harris06,Munasinghe84} that losses and robustness  are  directly related to the cost of electricity. These factors are tight to topological parameters as shown above, therefore such evolution would provide less cost in electricity distribution. It is difficult to translate in exact monetary terms for the end user the savings in the energy bill that an improvement in the topology parameters that influence the cost of energy distribution might bring.  For the realization of a \SG less costs in the distribution of electricity and a local reliable and robust network are the essential ingredients to enable a paradigm where energy is produced and distributed locally such as neighborhood or city level.
%\footnote{questo in rosso non \'e pi\'u appropriato per una discussion section? DIREI DI NO VISTO CHE INIZIO QUI A PARLARE DI ASPETTI ECONOMICI CHE NELLA PARTE PRECEDENTE NON SONO MENZIONATI, IO SAREI DEL PARERE DI LASCIARLA QUI}

\section{Related Work}\label{sec:relWks}

To the best of our knowledge there are no works that use the principles of \CNA to design and consider evolutions of \DG networks. Thus, in considering related work, we look into three main related areas: Evolution of Complex Networks, \CNA studies applied to the Power Grid, and electrical engineering power lines design and adaptation.
%\footnote{dividere in tre sottosezioni? Per TSG poi cambierei l'ordine e metterei pre primo quello dei power engineers, per il tech report, cosi' va bene. OK, DONE}

\subsection*{Evolution of Complex Networks}

The problem of the evolution of networks by using \CNA techniques has been widely investigated. The focus is mainly to assess and analyze existing networks that have evolved over time and understand which principle is underlying such evolution. More in general, the evolution of networks is considered in defining abstract algorithms that can be used to build networks that mimic the same behavior of natural or synthetic networks. For a general introduction to \CNA generation techniques, we refer to the extensive works such as~\cite{Chakrabarti2006,Newman2003a}. 
The work of Liben-Novell and Kleinberg~\cite{Liben-Nowell07} discuss link prediction. The authors consider the evolution of the ArXiv library co-authorship network (i.e., a social network) and define several quantitative metrics to infer the future links to be formed in the network (i.e., new co-authored publications). The metrics that the authors use for predictions are based on evaluating the paths, the neighborhood of nodes, rank of the matrix representing the graph, and clustering. Compared to a random prediction of new links the methods used in the paper are far better: the random predictions is correct less than 1\% of the times, while on some topics of the ArXiv library the  predictions reach a correct outcome 50\% of the time. The ideas used in the paper are interesting and valuable, and the metrics used in the prediction of new connections between nodes prove valuable for the co-authorship social network. Even if at abstract level network are the same (i.e., graphs) it is difficult to apply these metrics and prediction strategies in a real infrastructure such as the Power Grid. In fact, many of the metrics considered in the paper are inspired by the underlying social aspects of the co-author relationship such as neighbors in common. This aspect is reinforced by the different results that the same metrics used on different topics of the ArXiv library provide sometimes far different results in the predictions following the different metrics. Given the small average node degree of the \PG networks, the absence of samples at different time stamps, it is difficult to apply the metrics and methods of the paper to Distribution Networks and expect meaningful and interesting results.

The survey of L\"{u}~\etal~\cite{lu11} provides a good overview on link prediction. The authors propose several methods and algorithms based on similarity properties of the nodes at local level (e.g., the neighborhood of a node), global similarity (i.e., over the whole network), and quasi-local that combines the best of the two approaches. They also propose other strategies such as probabilistic models to evaluate the likelihood of the formation of a link. One of the networks that is considered for link prediction in~\cite{lu11} is the Western U.S. \PG (data already used in~\cite{Watts03}) and in this case the methods that give more stress on the local interactions between nodes perform better in the prediction than other approaches. The authors also state that for the infrastructural networks the geography plays an important role where the long distant connections are discouraged and very rare. The paper describes three main application of link prediction in the field of Complex Network Analysis: reconstruction of networks, evolution mechanism of networks and classification of partially labeled networks. We consider these prediction mechanisms valuable for networks with rich neighborhood structures where differences and intersection between neighborhoods are meaningful (e.g., social networks) or where the discovery of links is particularly costly or long in time and efforts (e.g., protein interactions). The current \PG networks have low average node degree and the structure is usually close to radial, especially at \MLV level, thus we consider the predictions less interesting. However, these techniques and link predictions might be more applicable in the further evolution and prediction of new links when the networks are more rich in connectivity. In addition, some techniques for link prediction would benefit from additional information  available to characterize the nodes, i.e., the number of prosumers that are connected to a node (i.e., substation) and the available production power.

The concept of network evolution of a technological infrastructure is considered in~\cite{piraveenan10} where a model for the evolution of the Internet Autonomous System (AS) is provided. As most works (e.g.,~\cite{dorogo02,barabasi02}), this one focuses on creating a model of the existing evolution of the Internet rather than proposing new ways of evolving networks. The interesting aspect on the modeling of the Internet is the better results of the proposed model Parallel Addition and Rewiring Growth (PARG) compared to other models in the literature to capture specific features of the AS topology. In particular, the model is able to capture the dissortativeness of hubs (i.e., the most connected nodes in the network) and linear local assortativeness of nodes. The model proposed proves successful, however one must recognize that the AS connection are mainly logical and are due more to business and contractual agreements than physical interconnections, therefore missing part of the problem that is essential in the case of the \PG topology.

\subsection*{\CNA Applied to the \PG}

Complex Network Analysis works take into account the \PG at the High Voltage level usually to analyze the structure of the network without considering in detail the physical properties of the power lines. In our previous work~\cite{pag:preprint1102}, we have analyzed several scientific papers that investigate \PG properties using \CNA approach. There are two main categories: 1) understand the intrinsic property of  Grid topologies and compare them to other types of networks assessing the existence of properties such as small-world or scale-free~\cite{Amaral2000,Watts03,Watts98,Barabasi1999}; 2) better understand the behavior of the network when failures occur (i.e., edge or node removal) and analyze the topological causes that bring to black-out spread and cascading failures of power lines~\cite{Corominas-murtra2008,alb:str04,Crucitti04}. Few studies in the \CNA landscape consider the possibility of using the insight gained through the analysis to help the design in terms of the reliability of the Power Grid. Examples are the study of the Italian \HV \G~\cite{Crucitti2005} and the study of improvement by line addition in Italian, French and Spanish Grids~\cite{Rosato2007}. Also Holmgren~\cite{holmgren06} uses the \CNA to understand which \G improvement strategies are most beneficial showing the different improvement of typical \CNA metrics (e.g., path length, average degree, clustering coefficient, network connectivity) although in a very simple small graph (less than 10 nodes) when different edges and nodes are added to the network. Broader is the work of Mei~\textit{et al.}~\cite{Mei11} where a self-evolution process of the \HV \G is studied with \CNA methodologies. The model for \PG expansion considers an evolution of the network where power plants and substations are connected in a ``local-world'' topology through new transmission lines; overall the \PG reaches in its evolution  the \sw topology after few-steps of the expansion  process. Concerning the topic of \SG with focus on the \HV network, Wang~\textit{et al.}~\cite{Wang2010,wang08} study the \PG to understand the kind of communication system needed to support the decentralized control required by the the new \PG applying \CNA techniques. The analyses aim at generating samples using random topologies based on uniform and Poisson probability distributions and a random topology with small-world network features. The simulation results are compared to the real samples of U.S. \PG and synthetic reference models belonging to the IEEE literature. These works also investigate the property of the physical impedance to assign to the generated Grid samples.
\CNA is not generally used as a design tool to propose adaptation strategies of current \MLV samples for the future \SG as we use in this paper where we also assess the benefits in terms of economical improvement.

\subsection*{Power Engineering Approach for Power Lines Design and Adaptation}
%Power Grid new part about reconfiguration problems

A power engineering problem that resembles our present proposal is that of Distribution Grid topological reconfiguration. The early works on the topic of Merlin and Back~\cite{merlin1975search} considered the problem with heuristic techniques and algorithms (e.g., branch and bound)
%\footnote{branch and bound per me e' algoritmica e non operation research, o vuoi dire qualcos'altro? SORRY, AVENDOLO STUDIATO SOLO IN RICERCA OPERATIVA ERO CONVINTO CHE APPARTENESSE A QUELL AREA} 
 to reconfigure the switches of the distribution infrastructure. Later works of Cinvilar~\etal~\cite{civanlar88} provide computationally tractable methods to modify the topology of the \DG by opening/closing switches to reduce losses. The authors provide a simplified formula to estimate the reduction in losses between the situation before and after the reconfiguration in the topology. Baran~\etal~\cite{baran89} have also the idea of minimizing the losses and guaranteeing the load balancing. Basically, the problem  deals with finding the minimal spanning tree, since the radial configuration needs to be preserved, that minimizes the objective function (i.e., system losses) while satisfying the constraints on voltage, capacity of lines and transformers, and reliability. For this problem, Baran~\etal~use a simplification of the power flow equations to compute the power flow in the network to be optimized.

More recently, with the new interest in the \DG due to the \SG and distributed generation opportunities to be placed in the Distribution layer of the Power Grid, new interest has grown concerning the reconfiguration problem together with islanded Grids and micro-generation plants.
%\footnote{questo e' interessante per noi per future research!}
The work of Ramesh~\etal~\cite{ramesh09} focuses on the minimization of losses in the Distribution Grid. The authors provide several options that have been simulated and tested  on the field to reduce losses on real and reactive power. The three solutions proposed concern three areas: distributed generation, capacitor placement and restructuring of feeders. The first proposal consist in placing local energy generation closer to the end users. In certain buses of the simulated IEEE 37 Bus provides loss reduction of up to 9 MW. The second solution proposes the installation at optimal locations of capacitors. The third proposed technique concerns the restructuring of the topology of the network. This last aspect is closer to the topic of the current paper. The solution proposed in~\cite{ramesh09} is basically a reconfiguration of the opening and closing of the switches of the distribution network to maintain the network radial and satisfy its load requirements. This approach although valuable from the practical point of view, lacks the vision in imagining the future energy scenario and there is no investigation of which topology is best or according which principles the new pieces of infrastructure (e.g., cables) should be interconnected.
The aspect of network reconfiguration of power networks to achieve a minimization of losses is the topic of~\cite{nath11}. The authors state that the only way to improve efficiency is by altering the topology. They focus on two aspects: (i) finding an optimal switching scheme of the switches connecting the lines and minimize losses, (ii) adding lines. The case study used is a IEEE 14 Bus where three more lines are added. The authors do not give any details why only three lines are considered and the motivation to choose specific nodes to attach the new lines. In addition, the only argument provided concerns the economic aspect of adding only the three lines, but no quantitative evidence is provided. The idea of intervening on topology to improve the efficiency of the \DG is as we consider in this paper, but we provide quantitative values, metrics and an economic analysis to decide which and how many lines need to be added in the Distribution Grid. We also consider another key ingredient that in the various works on reconfiguration is missing that is the added robustness that a more connected \G may provide.
In~\cite{shariat12} the problem of network reconfiguration in a \SG environment is addressed. The idea of the authors is to reconfigure the topology of the network by operating switches in order to minimize the overload in the branches of the network. The scenario they consider takes also into account the higher penetration rate that distributed generation will have in the future. The numerical evaluation of the proposed genetic algorithm to reconfigure the network  is realized on a simple 33 Bus network where the system configuration problem is solved to minimize the objective function (i.e., losses) and without violating the voltage and current constraints on the lines. Even in that work the importance of topology is claimed and the benefits in terms of reduced losses are shown. However, no additional lines or investment of the \DG are proposed and the benefit come just from the reconfiguration; no motivation or order of magnitude of the avoided investment are provided and the effects of topological changes available in the switching are anyway limited. 
The work of Xiaodan \etal~\cite{xiaodan09} is in the spirit of studying the reconfiguration in a micro-grid environment with micro-generation plant based on renewables. Basically, the problem is considered as two optimization problems:  one related to the determining the capacity of each island of the \G that has micro-generation capabilities, and the other is a problem of reconfiguration of the \DG with the objective of minimizing the power losses. The optimization algorithm is validated against two test Grids: the IEEE 33 Bus and the PG\&E 69 node system.
The reconfiguration for the system in an island and micro-generation uses techniques and approaches similar to traditional Distribution Systems.
%\footnote{con sta frase ti fai amici gli autori, :-) La puoi dire un tantino piu' morbida?  MIGLIORATA} 
The problem that is considered is an optimization problem and the authors constrain the \DG to be always operated in a radial configuration.
Usually, the reconfiguration problem boils down to the definition of an objective function to minimize the losses of the system and to establish constraints to satisfy the load. The function is then solved resorting to some heuristic.
Our approach in the present work is novel: we do not deal with the congestion problem and our configuration of the network does not take into account how the switches are operated. In fact, in our dataset, we do not have the information of the state of the switches and we consider the full topology of the network for the \MLV samples, that is, the paths in the physical samples are already the best (i.e., shortest) achievable in the optimal operation of the switches (i.e., enabling all paths).

Power system engineers do resort to graph theory principles~\cite{wall79,crawford75}, though CNA is not typically considered as a design tool. The traditional techniques involve the individuation of an objective function representing the cost of the power flow along a certain line which is then subject to physical and energy balance. This problem translates in an operation research problem. These models are applied both for the \HV planning~\cite{garver70,lee74} and the \MLV~\cite{wall79,crawford75} since long time. Not only operation research, but also expert systems~\cite{machias89} have been developed to help in the process of designing grounding stations based on physical requirements as well as heuristic approaches from engineering experience. 
%Usually a problem faced in power system is the expansion of the Grid due to increase in electricity usage. 
The substation grounding issue is approached as an optimization problem of construction and conductor costs subject to the constraints of technical and safety parameters, its solution is investigated through a random walk search algorithm~\cite{gilbert11}. In~\cite{fratkin96}, a pragmatic approach using sensitivity analysis is applied to a linear model of load flow related to various overloading situations and a contingency analysis (N-1 and N-2 redundancy conditions) is performed with different grades of uncertainty in medium and long term scenarios. 
In practice, the planning and expansion problem is even more complex since it implies power plants, transmission lines, substations and Distribution Grid. In~\cite{grigsby07} all these aspects are assessed separately and several challenges appear. For instance, in the planning of a \HV overhead transmission line, specific clearance code must be followed and load is not the only driver, but also topography and weather/climatic conditions (above all wind and ice) play an import role. For substation planning, the authors of~\cite{grigsby07} emphasize, in addition to the need for upgrading the Grid (e.g., load growth, system stability) and budgeting aspects, the multidisciplinary aspects which involve environmental, civil, electrical, and communications engineering. 
%All these aspects together relate to three main aspects: economical evaluation, technical evaluation and community acceptance. 
A more general approach proposed in~\cite{grigsby07} to deal with power system planning might be regarded as a multi-objective (e.g., economics, environment, feasibility, safety) decision problem thus requiring the tools typical of decision analysis~\cite{keeney82}.

Most of the works cited take into account mainly the \HV end of the \G while not least important is the Distribution \G especially in the vision of the future electrical system as proposed in this work where the end-user plays a vital role. The integrated planning of \MLV networks is tackled by Paiva \etal~\cite{paiva05} who emphasize the need of considering the two networks together to obtain a sensible optimal planning. The problem is modeled as a mixed integer-linear programming one, considering an objective function for investment, maintenance, operation and losses costs that need to be minimized satisfying the constraints of energy balance and equipment physical limits.
Even more challenges to the Electrical system planning are posed by the change in the energy landscape with several companies running different aspects of the business (generation, transmission, distribution). In addition, by accommodating more players in the wholesale market, transmission expansion should follow (as it is already for generation) a market based approach i.e., the demand forces of the market and its forecast should trigger the expansion of the Grid~\cite{bresesti03}. 
The same consideration regarding the need of a different approach in planning in a deregulated market are expressed in~\cite{shariati08} where an optimization of an objective function in the market environment is applied. % environment should take into account three main aspects: 1) Locational Marginal Pricing (LMP) since each point in the \G is characterized by different congestion situation and losses that influence the price of electricity; 2) infrastructure investment construction costs; 3) security costs to account the cost enhancement to face situations of contingency for the designed transmission line. Also in an energy market with several players the expansion of the transmission \G could be beneficial to certain players, but not for all, thus creating potential conflict situations. These aspects are investigated in~\cite{jun10} to realize an optimal expansion plan with Game Theory principles. 

\section{Conclusions}\label{sec:conclusion}

The \SG promises to revolutionize the energy sector providing the \G with more efficiency, more real-time information and more renewable power sources. In such vision, the end-users will be able to produce their own energy and exchange the surplus in a completely free market. Based on the experience that we have accrued in our previous study of distribution \G evolution considering Complex Network models, we have gone forward considering several types of possible evolution of networks by increasing connectivity. We have used physical samples from the Distribution \G of the Northern Netherlands and we have applied six distinct strategies. We have performed two types of analysis of the results. First, a pure topological analysis considering a set of specifically defined metrics where random evolution, assortative high degree and least distance evolution strategies resulted as best satisfying the desiderata of the metrics. In our second set of analysis, we have also taken into account the physical properties of the cables, therefore realizing a weighted graph analysis, and considering the cost of evolving the networks assessing the cost of cables. In this weighted analysis, we have seen that the random and assortative high degree are the best strategies in a pure topological analysis, but they have costs that are extremely high and unrealistic to realize in practice. In addition, the weighted analysis gave us the possibility to investigate the elements of the \PG that influence the cost of electricity distribution (i.e., \G losses and \G reliability) and the results of this analysis suggest that evolving the network by adding connections between the nodes with smallest distance is beneficial and provides in many cases better results compared to the other evolution strategies. Therefore, with an investment about 25\% of the actual costs in cables already on the ground the \DG can improve consistently in reducing transportation costs.

In this work we have for the first time realized a study on how to evolve the Distribution Grid by using the principles of Complex Network Analysis. We have studied the feasibility of this approach taking into account the costs and benefits for realizing different evolution approaches. This study sets the basis for a multi-flow study of the Distribution Grid with enhanced prosumer contribution to local energy generation. Moreover, it is a step forward in realizing a decision support system tailored for the Smart Grid. Our future efforts will be devoted to the further development of this decision support system.

%{\color{red}In this work we have gone even one step closer in completing the steps depicted in Figure~\ref{fig:engProc} to realize a high level decision support tool in the design the \DG for the Smart Grid. What we leave as future work is the interaction with energy distribution providers to investigate the targets in reduction of electricity distribution costs and the evaluation of real budgets for \DG evolution; these two elements will enable to complete the missing piece of the decision support system.}
%\footnote{Non mi piace ques'utlimo pezzo di testo in rosso. Perche' non rimpiazzarlo dicendo che per la prima volta abbiamo realizzato uno studio dell'evoluzione della smart grid basandoci su CNA e analizzato le feasibilitiy, dunque ponendo le fondamenta per uno studio della grid multi flow, e di un decision support system taylored per la smart grid. Roba che e' difficile da capire e complicata, specie se sei un power engineer testa di legno :-) CAMBIATO, MEGLIO?}

\bibliographystyle{siam}
\bibliography{andreaNewBib}

\newpage

\appendix
\renewcommand\thesection{Appendix \Alph{section}}

\section{Network topological requirements}\label{sec:metrics}

In~\cite{pag:preprint12,PaganiEvol2013} we proposed a number of metrics useful for analyzing synthetic network topologies for the \PG having in mind decentralized energy trading. We recall them here since we use them to measure the adequacy of the proposed evolution strategies of the physical samples of the Dutch Distribution Grid.

%\subsection*{Qualitative requirements}
%
%The main qualitative requirement we envision for the future Distribution Network relies on the modularity of the network topology. In the power system domain, the modularity is invoked as a solution that provides benefits reducing uncertainties in energy demand forecasting and costs for energy generation plants as well as risks of technological and regulatory obsolescence~\cite{lovins02,hoff97}. Modularity is usually required not only in the energy sector, but more generally in the design and creation of product or organizations~\cite{gershenson03}. It is also a principle that is promoted in innovation of complex systems~\cite{ethiraj04} for the benefits it provides in terms of reduced design and development time, adaptation and recombination.  We assess the modularity of a network as the ability of building the network using a self-similar recurrent approach and having a repetition of a kind of pattern in its structure.

\subsection*{Quantitative requirements}

As a global statistical tool, quantitative requirements are even more useful as they give a precise indication of network properties. 
Here are the relevant ones when considering efficiency, resilience and robustness of a power system.
\begin{itemize}
\item {\em Characteristic Path Length (CPL) lower or equal to the natural logarithm of order of graph:} $CPL \leq ln(N)$. This requirement represents having a general limited path when moving from one node to another. In the Grid this provides for a network with limited losses in the paths used to transfer energy from one node to another.
\item {\em Clustering Coefficient (CC) which is 5 times higher than a corresponding random graph with same order and size}: $CC \geq 5\times CC_{RG}$. Watts and Strogatz~\cite{Watts98} show that \sw networks have clustering coefficient such that $CC \gg CC_{RG}$. Here we require a similar condition, although less strong by putting a constant value of 5. This requirement is proposed in order to guarantee a local clustering among nodes since it is more likely that energy exchanges occur at a very local scale (e.g., neighborhood) when small-scale distributed energy resources are highly implemented. 
\item {\em Betweenness-related requirements:}
\begin{itemize}
\item {\em A low value for average betweenness in terms of order of the graph} $\overline{\upsilon}=\frac{\overline{\sigma}}{N}$, where $\overline{\sigma}$ is the average betweenness of the graph and $N$ is the \textit{order} of the graph. For the Internet V\'{a}zquez \etal~\cite{Vazquez2002} have found for this metric  $\overline{\upsilon}\approx 2.5$. Internet has proved successful to tolerate failures and attacks~\cite{Cohen2000,Albert2000},
therefore we require a similar value for this metric for the future Grid.
\item {\em A coefficient of variation for betweenness} $c_v=\frac{s}{\overline{x}} < 1$ where $s$ is the sample standard deviation and $\overline{x}$ is the sample mean of betweenness. Usually distribution with $c_v < 1$ are known as low-variance ones.
\end{itemize}
The above two requirements are generally considered to provide network resilience by limiting the number of critical nodes that have a high number of minimal paths traversing them. These properties provide distributions of shortest paths which are more uniform among all nodes.
\item {\em An index for robustness such that} $Rob_N \geq 0.45$. Robustness is evaluated with a random removal strategy and a node degree-based removal strategy by computing the average of the \textit{order} of the maximal connected component (MCC) of the graph between the two situations when the 20\% of the nodes of the original graph are removed~\cite{PaganiAielloTSG2011}. It can be written as $Rob_N = \frac{|MCC_{Random20\%}|+|MCC_{NodeDegree20\%}|}{2}$. Such a requirement is about double the value observed for current \MV and 33\% more for \LV samples~\cite{PaganiAielloTSG2011}.
\item {\em A measure of the cost related to the redundancy of paths available in the network:} $APL_{10^{th}} \leq 2\times CPL$. With this metric we consider the cost of having redundant paths available between nodes. In particular, we evaluate the 10$^{th}$ shortest path (i.e., the shortest path when the nine best ones are not considered) by covering a random sample of the nodes in the network (40\% of the nodes whose half represents source nodes and the other half represents destination nodes). The values for the paths considered are then averaged. In the case where there are less than ten paths available, the worst case path between the two nodes is considered. This last condition gives not completely significant values when applied to  networks with small connectivity (i.e., absence of redundant paths).
\end{itemize}
\begin{center}
\begin{table}[h!]
\begin{center}
\begin{footnotesize}
    \begin{tabular}{ | l || c | c | c |}
    \hline
    \textbf{Metric} & \textbf{Efficiency} & \textbf{Resilience} & \textbf{Robustness} \\ \hline\hline
    CPL & \checkmark &  &  \\ \hline
    CC & \checkmark & &\\ \hline
    Avg. Betweenness &  & \checkmark & \\ \hline
    Betw. Coeff. of Variation &  & \checkmark & \\ \hline
    $Rob_N$&&& \checkmark \\ \hline
    $APL_{10^{th}}$ & \checkmark &\checkmark & \\
    \hline
    \end{tabular}
\caption{Metrics classification related to properties delivered to the network.}\label{tab:metricsClass}
\end{footnotesize}
\end{center}
\end{table}
\end{center}
The above quantitative metrics can be categorized into three macro categories with respect to how they affect a Power Grid: efficiency in the transfer of energy, resilience in providing alternative path if part of the network is compromised/congested and robustness to failures for network connectivity. Table~\ref{tab:metricsClass} summarized the property each metric assesses.

\end{document}